\documentclass[aps,prd,10pt,notitlepage,nofootinbib,superscriptaddress,showkeys,showpacs]{revtex4-1}

\usepackage{amsfonts, amsmath, amssymb, amsthm}

\usepackage{color}

\usepackage{graphicx}

\usepackage[caption=false]{subfig}

\usepackage[left = 2.0cm, right = 2.0cm, top = 2.5cm, bottom = 2.5cm]{geometry}

\newcommand{\be}{\begin{equation}}
\newcommand{\ee}{\end{equation}}

\newcommand{\tr}{\textrm{Tr}}

\newcommand{\cM}{\mathcal{M}}
\newcommand{\cB}{\mathcal{B}}

\newcommand{\cC}{\mathcal{C}}
\newcommand{\cE}{\mathcal{E}}
\newcommand{\cV}{\mathcal{V}}
\newcommand{\cG}{\mathcal{G}}

\newcommand{\bJ}{\mathbb{J}}
\newcommand{\bT}{\mathbb{T}}

\newtheorem{lemma}{Lemma}

\newcommand{\extd}{\mathrm{d}}

\DeclareMathOperator{\NNLO}{NNLO}
\DeclareMathOperator{\NLO}{NLO}
\DeclareMathOperator{\LO}{LO}
\DeclareMathOperator{\DS}{DS}

\begin{document}

\title{\Large \bf The double scaling limit of random tensor models}

\author{{\bf Valentin Bonzom}}\email{bonzom@lipn.univ-paris13.fr}
\affiliation{LIPN, UMR CNRS 7030, Institut Galil\'ee, Universit\'e Paris 13, Sorbonne Paris Cit\'e,
99, avenue Jean-Baptiste Cl\'ement, 93430 Villetaneuse, France, EU}

\author{{\bf Razvan Gurau}}\email{rgurau@cpht.ecolepolytechnique.fr}
\affiliation{CPHT, UMR CNRS 7644 \'Ecole Polytechnique,\\
91128 Palaiseau cedex, France, EU \\ and Perimeter Institute for Theoretical Physics, \\
31 Caroline st. N, N2L 2Y5, Waterloo, ON, Canada}

\author{{\bf James P.~Ryan}}\email{james.ryan@aei.mpg.de}
\affiliation{Max Planck Institute f\"{u}r Gravitationsphysik,\\ Am M\"{u}hlenberg 1, 14476 Golm, Germany, EU}

\author{{\bf Adrian Tanasa}}\email{tanasa@lipn.univ-paris13.fr}
\affiliation{LIPN, UMR CNRS 7030, Institut Galil\'ee, Universit\'e Paris 13, Sorbonne Paris Cit\'e,
99, avenue Jean-Baptiste Cl\'ement, 93430 Villetaneuse, France, EU \\
and Horia Hulubei National Institute for Physics and Nuclear Physics\\
PO Box MG-6, 077125 Magurele, Romania, EU
}

\date{\small\today}

\begin{abstract}
\noindent Tensor models generalize matrix models and generate colored triangulations of pseudo-manifolds in dimensions $D\geq 3$. The free energies of some models have been recently shown to admit a double scaling limit, i.e. large tensor size $N$ while tuning to criticality, which turns out to be summable in dimension less than six. This double scaling limit is here extended to arbitrary models. This is done by means of the Schwinger--Dyson
equations, which generalize the loop equations of random matrix models, coupled to a double scale analysis of the cumulants.
\end{abstract}

\medskip


\maketitle

\section{Introduction}

Random tensor models represent a natural generalization, to dimensions greater than two, of the celebrated random matrix models.
Thus, one can view these tensor models as an appealing new approach for a fundamental theory of quantum gravity (see the review paper \cite{review-vincent}).

The current revival of interest in the study of tensor models came from the definition of colored tensor models \cite{colored} and,
shortly thereafter, from the implementation of a $1/N$--expansion within these models \cite{largeN,largeN1,largeN2,largeN3, Uncoloring}. In the case of matrix models, the
$1/N$--expansion is controlled by the genus of the corresponding Feynman ribbon graphs. In dimensions greater than two, the role of the
genus is played by the degree of the tensor graphs. The degree of a tensor graph is defined to be the sum of the genera of certain well--chosen ribbon subgraphs (called the jackets of the tensor graph \cite{Heegaard}). At large--$N$, matrix models are dominated by planar graphs (tiling the two--dimensional sphere
${\cal S}^2$), while the dominant graphs for tensor models are the so-called melonic graphs. In $D$ dimensions, these correspond to particular
triangulations of the $D$--dimensional sphere ${\cal S}^D$ (see also \cite{GR} for an extensive review and \cite{Uncoloring} for a shorter one). The continuum limit of melonic graphs turns out to be the continuous random branched polymer \cite{GR2}.

To escape this universality class and further explore tensor models, the next-to-leading order of the $1/N$--expansion has been identified in \cite{ColoredNLO} where new structures are shown to emerge. Recently, this has led to a double scaling limit of random tensor models, exhibited for both the colored model and the so-called uncolored model with
a {\it quartic interaction}. The idea of the double scaling limit is to take $N$ large while sending the model to the continuum, so as to consistently retain Feynman graphs from arbitrary orders of the $1/N$--expansion. It is worth noticing that the double scaling limit in tensor models differs markedly from that in matrix models. In the matrix model case, while the resulting series \lq\lq consistently\rq\rq\ sums over topologies, it is divergent.  In the tensor model cases mentioned above, however, the double scaling limit 
leads to  a summable series in dimensions $D=3,4,5$. This imply that a 
reiteration of the double scaling limit procedure might be possible, which at criticality could ultimately lead to a genuinely new continuous random space.

The colored model double scaling limit was obtained in \cite{ColoredSchemes}, using a purely combinatorial approach; the uncolored model double scaling limit was obtained in \cite{DSQuartic}, using an appropriate intermediate field method. Both results therefore rely on a thorough analysis of each term of the series (in the coupling constant) associated to each order in the initial large--$N$ expansion. However, the method of \cite{DSQuartic} is not known to be applicable beyond the case of quartic interactions. To improve on these results and possibly go beyond, it is reasonable to look for a more effective way to reach the double scaling regime, which we argue here could be the Schwinger-Dyson equations.

\medskip

Taking some inspiration from matrix models (and later, string field theory), while there exist various ways to solve them, one interesting technique is the use of their Schwinger--Dyson equations (SDEs, often known as loop equations in this context) \cite{Fukuma,Dijkgraaf, Bouwknegt, Dijkgraaf2, Aganagic}. They allow one to probe the
correlators at all orders in the $1/N$--expansion and have unraveled some fascinating structures (e.g. integrability) that become transparent in the topological recursion \cite{eynard}. Interestingly,  topological recursion was initially developed as an intrinsic method to solve the SDEs. It is therefore natural to ask whether the SDEs could also be used to solve tensor models.

As in the case of matrix models, the SDEs of tensor models translate into differential constraints satisfied by
the partition function, constraints which have been shown, in the large--$N$ limit, to close a Lie algebra indexed by colored rooted $D-$ary trees.
This provides a natural generalization of the Virasoro constraints in arbitrary dimensions \cite{TreeAlgebra}. The SDEs and the associated algebra at all orders
in $1/N$ were then completed in \cite{BubbleAlgebra}, which extend the Virasoro generators to operators labeled by regular edge-colored graphs. From the computational perspective, the tensor SDEs can be solved at large--$N$, \cite{SDE}, to give a new proof that large random tensors are Gaussian, with the covariance being the full two-point function.

\medskip

In this paper we derive the double scaling limit of random tensor model using the SDEs. It is worth emphasizing that our method allows to obtain this
double scaling limit not only for tensor models with quartic interactions, thereby reproducing the result of \cite{DSQuartic}, but also for tensor models with a general melonic interactions. This is a result going beyond what is already known from the literature, and more importantly which would seem rather intricate to derive without the SDEs. Indeed, the method of \cite{DSQuartic} relies on a bijection that is only known to exist in the case of quartic interactions.

The organization of the paper is the following. In the next section we give a brief review of random tensor models and of the two relevant SDEs which will be used. Section \ref{sec:SDNLO} is dedicated to our analysis of the SDEs beyond the $1/N$ limit (leading order (LO), next-to-leading-order (NLO) and even another sub--leading order in the case of the quartic model). Those results are essential to achieve the double scaling limit of the SDEs in section \ref{ssec:DSSD}. The main result is the doubly--scaled 2-point function for a model with generic melonic interactions (symmetrized on their colors). However several assumptions on the large--$N$ scaling of cumulants are made along the way. They are proved by means of combinatorial methods in sections \ref{ssec:proofdsl}, \ref{ssec:FromQuarticToGeneric} and in appendix \ref{app:4pdom}. The reader mostly interested in solving the SDEs may skip those parts and simply consider their conclusions as ansatz which allow to extract particular solutions of the SDEs.

\section{Brief review of random tensor models and Schwinger-Dyson equations}

\subsection{The framework of random tensor theory}

Observables in random tensor theory are generalizations of trace-invariants in matrix models. They are generated by polynomials in the tensor
entries $\bT_{a_1\dotsb a_D}$ and $\overline{\bT}_{a_1\dotsb a_D}$ ($a_1,\dotsc,a_D=1,\dotsc,N$), which are invariant under transformations of the following form \cite{Uncoloring,Universality}:
\begin{equation}
\label{UnitaryTransfo}
\bT'_{a_1\dotsb a_D} = \sum_{b_1,\dotsc,b_D} U^{(1)}_{a_1 b_1}\, \dotsm \, U^{(D)}_{a_D b_D}\ \bT_{b_1 \dotsb b_D}\;,
\end{equation}
where $(U^{(1)},\dotsc,U^{(D)})$ is a $D$-uple of independent unitary matrices.
The complex conjugated tensor $\overline{\bT}$ transforms in a similar fashion.  Since different unitary transformations are applied
to the different indices, invariants can be obtained by \emph{contracting} (that is, by identifying and summing) indices pairwise only
when they have the same position between 1 and $D$. Invariance also requires all indices to be contracted.

It emerges that the generating polynomials can be labeled by connected, regular, bipartite graphs of degree $D$, whose edges have a color label drawn from $\{1,\dotsc,D\}$ such that the $D$ edges incident to a vertex have distinct colors. Such graphs are called \emph{bubbles}. The correspondence between polynomials and bubbles is tabulated in Table \ref{table:GraphicalRules}.

\begin{table}
  \begin{tabular}{c|c}
    Polynomials & Bubbles\\
    \hline
    $\bT_{a_1\dotsb a_D}$ & White vertex $\begin{array}{c}\includegraphics[scale=.65]{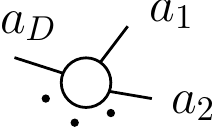}\end{array}$ \\
    \hline
    $\overline{\bT}_{a_1\dotsb a_D}$ & Black vertex $\begin{array}{c}\includegraphics[scale=.65]{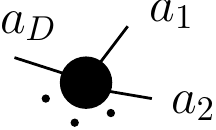}\end{array}$ \\
    \hline
    Contraction & Edge with color label $i$ \\
    $\sum_{a_i} \bT_{a_1\dotsb a_i \dotsb a_D} \overline{\bT}_{b_1\dotsb a_i \dotsb b_D}$ &
    $\begin{array}{c}\includegraphics[scale=.65]{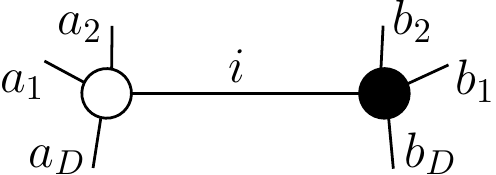}\end{array}$
  \end{tabular}
  \caption{\label{table:GraphicalRules}}
\end{table}

Let us denote bubbles by $\cB$, the vertex set of $\cB$ by $\cV(\cB)$ of cardinality $|\cV(\cB)|=2p(\cB)$ and the set of edges of color $c$ of $\cB$ by $\cE_c(\cB)$, of cardinality $|\cE_c(\cB)|=p(\cB)$. The vertex set of $\cB$ is bipartite. We generically denote a white vertex of $\cB$ by $v$, and a black vertex by $\bar v$ and an
edge of color $c$ by $e_c$. We furthermore use the shorthand notation $\vec a$ for the $D$-uple of indices $(a_1,\dotsc,a_D)$ (hence $\vec a^{v} = (a_1^v,\dotsc,a_D^v)$) with $a_i=1,\dotsc,N$ for $i\in \{1,\dotsc,D\}$.

The invariant polynomial associated to $\cB$ is denoted $\tr_{\cB}(\bar \bT, \bT)$ and it writes:
\begin{equation}
 \tr_{\cB}(\bar \bT, \bT) = \sum_{\{\vec a^{v}, \vec b^{\bar v}\}_{v,\bar{v}}} \left(  \prod_{\bar v \in \cV(\cB) } \bar \bT_{\vec b^{\bar v}} \right)
 \left( \prod_{v\in \cV(\cB)} \bT_{\vec a^{v}} \right) \delta^{\cB}_{\vec a^v, \vec b^{\bar v}} \; , \qquad
    \delta^{\cB}_{\vec a^v, \vec b^{\bar v}} = \prod_{c=1}^D \prod_{e^c=(  x , \bar y ) \in \cE^c(\cB)} \delta_{a_c^{x} b_c^{\bar y} } \; ,
 \end{equation}
that is for each white vertex $v$ (resp. black vertex $\bar v$) of $\cB$ we take a tensor $ \bT_{\vec a^{v}} $ (resp. $\bT_{\vec b^{\bar v} } $),
and for each edge of color $c$ we contract the indices of color $c$ of the tensors associated to its end vertices.
The operator $ \delta^{\cB}_{\vec a^v, \vec b^{\bar v}}  $ is called the \emph{trace-invariant operator} associated to $\cB$.

\begin{figure}
\subfloat[The 2-vertex bubble $\cB_2$.]{\begin{tabular}{c}\includegraphics[scale=.55]{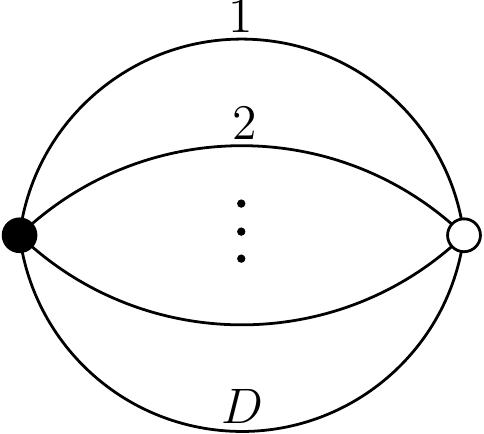}\end{tabular} \label{fig:2PointBubble}}
\hspace{1.5cm}
\subfloat[The 4-vertex bubble $\cB_{4,\{c\} }$.]{\begin{tabular}{c}\includegraphics[scale=.6]{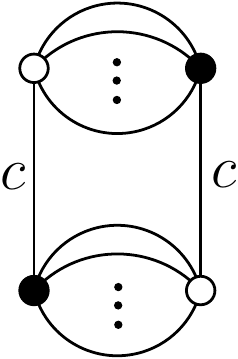}\end{tabular} \label{fig:4PointBubble}}
\hspace{1.5cm}
\subfloat[A 10-vertex bubble at $D=4$.]{\begin{tabular}{c}\includegraphics[scale=.5]{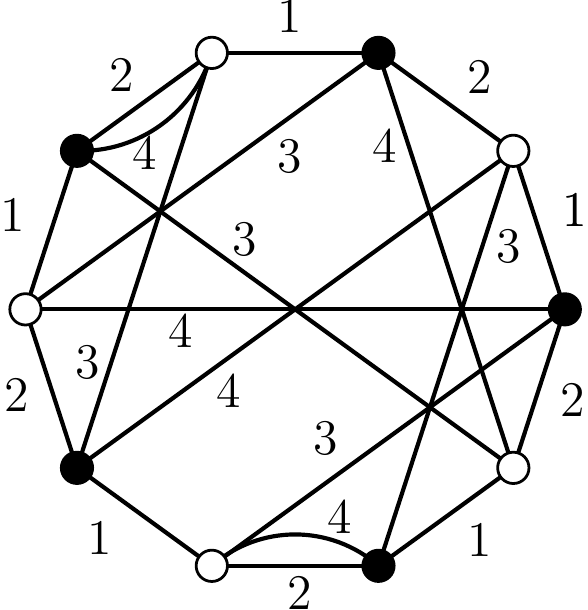}\end{tabular} \label{fig:10VertexGraph}}
\caption{ \label{fig:Bubbles} Some examples of bubbles, that is, connected, bipartite, regular graphs of degree $D$ with colored edges. The dots indicate multiple edges.}
\end{figure}

There are a number of bubbles that will play an important role later on.
Firstly, there is the unique 2--vertex bubble $\cB_2$, displayed in figure \ref{fig:2PointBubble}. Secondly, there are the 4--vertex bubbles illustrated in figure \ref{fig:4PointBubble}.  They are 1--particle--irreducible yet 2--particle--reducible. They are labeled $\cB_{4,\{c\} }$ where $c\in\{1,\dotsc,D\}$ is the color of the edges that, when cut, disconnect the graph.  Another (less important) example of a bubble is given in figure \ref{fig:10VertexGraph}.

Let $I$ be a finite set and $\{\cB_i\}_{i\in I}$ be a set of bubbles such that $\cB_i$ has $p_i\geq2$ black vertices. We shall denote the set of
corresponding invariant polynomials by $\{\tr_{\cB_i}\}_{i\in I}$.  The \emph{action} is an invariant function of $\bT$ and $\overline{\bT}$:
\begin{equation}
  \label{Action}
  S(\bT,\overline{\bT}) = \tr_{\cB_2}(\bT,\overline{\bT}) - \sum_{i\in I} \frac{z^{p_i-1}}{p_i} t_i\,\tr_{\cB_i}(\bT,\overline{\bT})\;,
\end{equation}
where the parameters $z$ and $\{t_i\}_{i\in I}$ are called the couplings. The partition function of a generic ``single trace'' tensor model
is:
\begin{equation}
  \label{eq:partition}
  Z(N,z,\{t_i\}) = \int   [\extd\bar \bT \extd \bT ] \ e^{-N^{D-1} S}\; , \qquad
   [\extd\bar \bT \extd \bT ] = \left( \prod_{\vec a} N^{D-1} \frac{\extd \bT_{\vec a}\, \extd
    \overline \bT_{\vec a}} {2\pi i}\right) .
\end{equation}

\subsection{Bubble observables}

Among all the observables one can build out of a tensor and its complex conjugate, the invariant observables labeled by bubbles $\cB$ play a distinguished role. Their expectations are:
\begin{equation}
  \label{eq:moments}
  \langle \tr_{\cB} (\bT,\overline{\bT}) \rangle = \frac{1}{Z(N,\{t_i\})} \int d\bT\,d\overline{\bT}\ e^{-N^{D-1} S}\ \tr_{\cB} (\bT,\overline{\bT})\;,
\end{equation}
and are functions of $N$, $z$ and $\{t_i\}_{i\in I}$.

The above integrals can be computed via their Feynman expansions, that is, as perturbative expansions in the couplings.
The Feynman expansion is organized with respect to Feynman graphs. These graphs result from first Taylor expanding the
exponentials $e^{ N^{D-1} \frac{z^{p_i-1}}{ p_i}  t_i \tr_{\cB_i}(\bT,\overline{\bT}) } $, and commuting the sums with the remaining Gaussian integral\footnote{It is well known that the resulting series is not summable but only Borel summable, however such subtleties are beyond the scope of our work.}. We thus obtain a sum over terms, each term being a Gaussian integral of a product of trace-invariants. Each Gaussian integral is computed using Wick's theorem, as a sum over pairings of $\bT$s with $\overline{\bT}$s contracted with the covariance. Graphically, a pairing connects a black vertex ($\bT$) to a white vertex ($\overline{\bT}$) via an edge to which the fictitious color $0$ is attributed.

\begin{figure}
  \includegraphics[scale=.4]{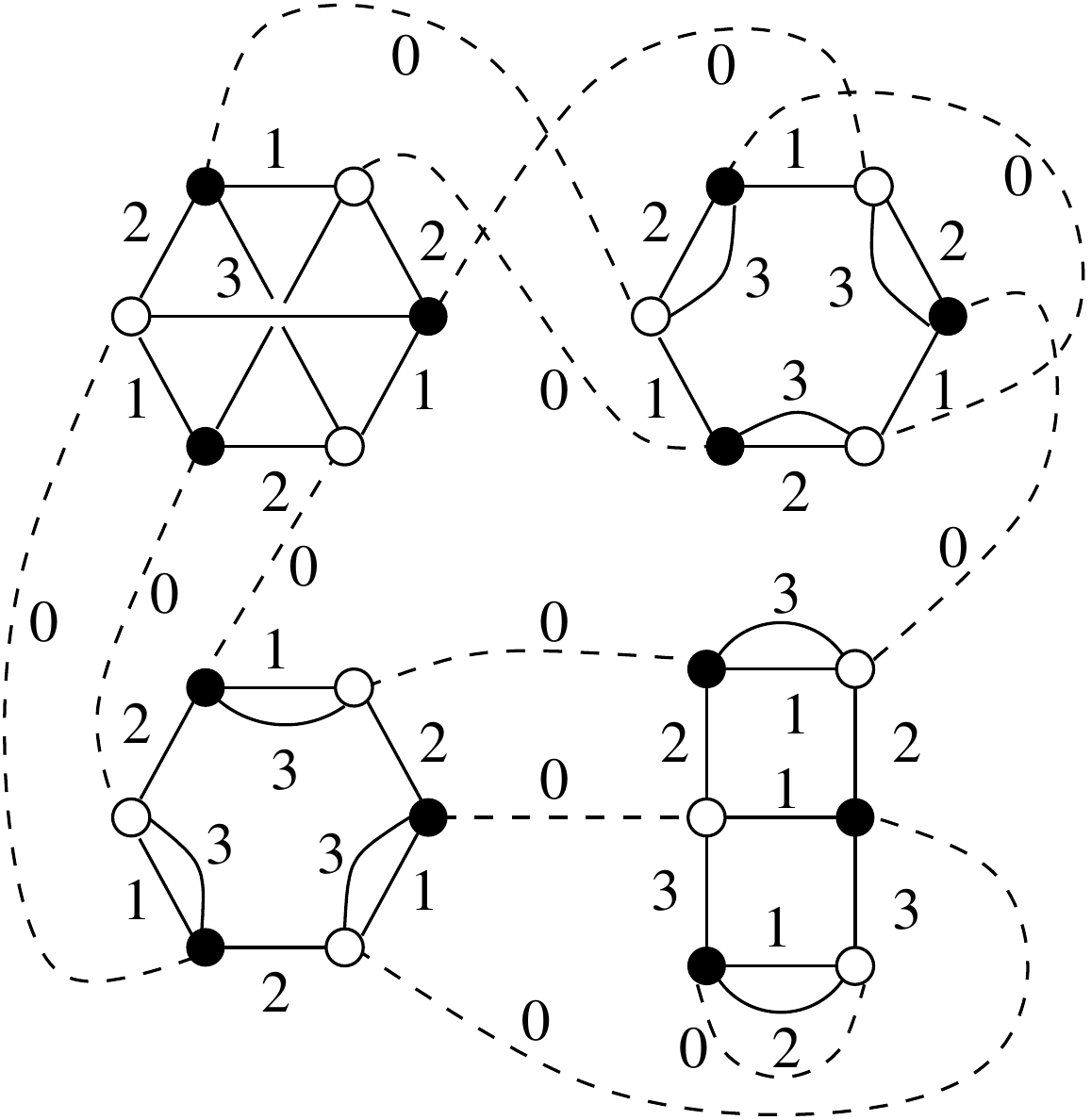}
  \caption{\label{fig:tensobsgraph} An example of a $(D+1)$--colored Feynman graph (with $D = 4$). The bubbles (solid edges)
  have colors 1, 2 and 3, while the propagators (dashed edges) are assigned the color 0.}
\end{figure}

The Feynman graphs of the partition function are therefore regular bipartite edge-colored graphs of degree $D+1$ (the colors of the bubbles plus the color 0). An example is shown in figure \ref{fig:tensobsgraph}. The Feynman graphs contributing to the expectation of $\tr_{\cB} (\bT,\overline{\bT})$ for a connected bubble $\cB$ are the connected regular bipartite edge-colored graphs of degree $D+1$ built from the set of bubbles $\{\cB_i\}_{i\in I}$ and containing the bubble $\cB$ as a marked sub--graph.

The Feynman amplitude associated to a closed connected Feynman graph $\mathcal{G}$ with no marked bubble is (up to
some symmetry factor) easily found. Each bubble $\cB_i$, $i\in I$, contained in $\cG$ brings a factor $z^{p_i-1} N^{D-1}$ and its associated trace-invariant. Each edge of color 0 brings $N^{-(D-1)}$ times $\prod_{c=1}^D \delta_{a_c^v b_c^{\bar{v}}}$, which identifies the index $a_c^v$ on the white vertex adjacent to the edge with the index $b_c^{\bar{v}}$ on the black vertex adjacent to the edge. A \emph{face with colors $cd$} is defined as a connected component of the graph obtained from $\cG$ by removing all edges of colors different of $c$ and $d$. Contracting the Kronecker deltas of the propagators with the trace-invariants coming from the bubbles, it follows that the indices are identified along the faces of color $0c$ of $\cG$. One gets a free sum per face, hence a factor $N$. We denote $b_i$ the number of bubbles of type $\cB_i$ of $\mathcal{G}$, and $b = \sum_{i\in I} b_i$ is the total number of bubbles of $\cG$. Moreover $p=\sum_{i\in I} p_i b_i$ is the total number of black vertices of $\cG$ and it also counts the edges of color 0 of $\cG$. The amplitude then reads:
\begin{equation}
  \label{eq:amps}
  A_{\mathcal{G}} = N^{\sum_{c=1}^D f_{0c} - (D-1)(v-b)}\, z^{v-b}\, \prod_{i\in I} (t_i)^{b_i}\;.
\end{equation}
At $D=2$, the $(D+1)$--colored graphs $\mathcal{G}$ are also ribbon graphs and the exponent of $N$ in the amplitude reduces to $2-2g$ where $g$ is the genus of the ribbon graph.

The amplitude of a Feynman graph in the expansion of the expectation of $\tr_\cB$ is similar, expect that the marked sub--graph $\cB$ does not bring any power of $z$ and $N$. The exponent of $N$ for the graphs entering an expectation is bounded and leads to a $1/N$--expansion of expectations \cite{Uncoloring,Universality} of the following form:
\begin{equation}
 \frac{1}{N}  \langle \tr_{\cB} (\bT,\overline{\bT}) \rangle =   \sum_{\omega\geq 0} N^{-\omega}\,A_{\omega,\cB}(z,\{t_i\})\;.
\end{equation}
Furthermore, the large--$N$ limit is Gaussian:
\begin{equation} \label{GaussianLargeNLimit}
 \frac{1}{N}\langle \tr_{\cB} (\bT,\overline{\bT}) \rangle = N^{ -\omega^*_{\cB}}\, \alpha_{\cB}\, [T(z,\{t_i\})]^{v} + \mathcal{O}(N^{-\omega^*_{\cB}})\;.
\end{equation}
Here, $\omega^*_{\cB}$ is the minimal value of $\omega$ for which $A_{\omega,\cB}(z,\{t_i\})\neq 0$, $\alpha_{\cB}$ counts the number of leading
order Wick contractions on $\cB$, $v$ is the number of black vertices of $\cB$ and $T(z,\{t_i\})$ is the large--$N$ limit of the 2--point function:
\begin{equation} \label{LargeN2Pt}
T(z,\{t_i\}) \equiv \lim_{N\to\infty} \frac1N\,\langle \tr_{\cB_2}(\bT,\overline{\bT}) \rangle =
\lim_{N\to\infty}  K\Big(\cB_2 ; N,z,\{t_i\}  \Big)  = K\Big(\cB_2 ; z,\{t_i\}  \Big) \;.
\end{equation}
(Here we have introduced the notation $K\Bigl(\cB_2;N,z,\{t_i\}\Bigr)$ which refers to the 2-point cumulant and is equivalent to the full 2-point function. It will be generalized to cumulants of higher orders in the Section \ref{sec:cumulants}.)

The equations \eqref{GaussianLargeNLimit} and \eqref{LargeN2Pt} constitute an illustration
of the \emph{Universality Theorem for large random tensors} equipped with joint distributions that are invariant under \eqref{UnitaryTransfo}.
The details about this theorem can be found in \cite{Universality}, where it was proven using mostly the combinatorics of cumulants. In the context
of random tensor models defined by an action like \eqref{Action}, it is also possible to get to this universal behavior through the Schwinger-Dyson
equations, as done in \cite{SDE}.\footnote{Although universality was shown in \cite{SDE} only for the so-called melonic polynomials, the same method
applies to the expectations of non-melonic polynomials.}

On the one hand, it is a difficult task to find $\omega_{\cB}^*$ and $\alpha_{\cB}$ in general. On the other hand, there is one important class for which
they are known~: the polynomials $\tr_{\cB}$ for which $\omega_\cB^*=0$. Members of this class are called \emph{melonic} polynomials and correspond
to \emph{melonic} bubbles. The structure of melonic bubbles is quite simple and defined recursively. The basic building blocks are the so--called
$(D-1)$--dipoles. A \emph{$(D-1)$--dipole of color $c$} is comprised of two vertices, connected by $D-1$ edges \emph{not} carrying the color $c$,
and two open edges of color $c$:
\begin{equation}\nonumber
  \begin{array}{c}
    \includegraphics[scale=.5]{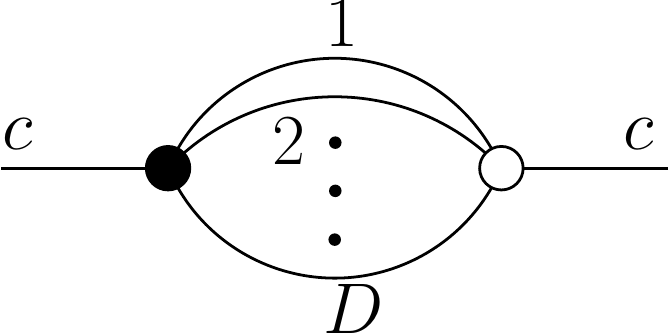}
  \end{array}
\end{equation}
One starts with the bubble $\cB_2$ and performs a \emph{$(D-1)$-dipole insertion}, where the edge of color $c$ is removed and replaced by the
$(D-1)$-dipole of the same color. This results in a bubble with 4 vertices, which is actually $\cB_{4,\{c\} }$ (Figure \ref{fig:4PointBubble}).
Such a $(D-1)$-dipole insertion can then be performed on an arbitrary edge of the new graph. In this manner, the bubble grows according to a
tree structure that records the history of dipole insertions. Melonic bubbles have the following properties:
\begin{itemize}
  \item[--] They are exactly the bubbles for which $\omega_\cB^*=0$.

  \item[--] The combinatorial coefficient is $\alpha_\cB=1$. Indeed, vertices in a melonic bubble come in canonical pairs (the vertices
  of a pair are those associated to a dipole insertion). $\alpha_\cB=1$ means that there is a single family of contributions coming from
  Wick contracting the two vertices of each pair.

  \item[--] The large--$N$ 2--point function $T$ is non--trivial, that is $T\neq1$, if and only if the set of bubbles $\{\cB_i\}_{i\in I}$
  contains a melonic bubble (different from $\cB_2$). The function $T$ can then be found by means of the Schwinger--Dyson equations, as we explain below.
\end{itemize}

\subsection{The Schwinger--Dyson equations}

The Schwinger--Dyson equations are the quantum equations of motion, governing the behavior of the expectations. A detailed presentation of
these equations can be found in \cite{BubbleAlgebra} and a focus on the melonic sub--algebra in \cite{TreeAlgebra}. They can be solved at
leading order to recover the Universality Theorem, as well as to find the equation satisfied by the large--$N$ 2--point function \cite{SDE}.
Here, we shall only need a couple of Schwinger--Dyson equations and not the full tower of equations derived in
\cite{BubbleAlgebra}. The first equation we use is the simplest one, coming from the identity:
\begin{equation}
\label{eq:sd1}
\sum_{a_1,\dotsc,a_D} \frac{1}{Z(N,\{t_i\})} \int d\bT\,d\overline{\bT}\ \frac{\partial}{\partial \bT_{a_1\dotsb a_D}}
\left( \bT_{a_1\dotsb a_D}\,e^{-N^{D-1} S}\right) = 0\;.
\end{equation}
Performing the derivatives explicitly, it is easy to see that it leads to:
\begin{equation}
\label{1stSDE}
N - \langle \tr_{\cB_2}(\bT,\overline{\bT}) \rangle + \sum_{i\in I} z^{p_i-1}\,t_i\,\langle \tr_{\cB_i}(\bT,\overline{\bT}) \rangle = 0\;.
\end{equation}
Here $p_i$ once again denotes the number of black vertices of $\cB_i$.
The second equation comes from the identity:\footnote{The overall factor $1/2$ has been introduced to counter the fact that the action of
the derivative (w.r.t.~$\overline T$) on $\tr_{\cB_{4,\{c\}}}(\bT,\overline{\bT})$ results in two identical terms.}
\begin{equation}
  \label{eq:sd2}
\sum_{a_1,\dotsc,a_D} \frac{1}{Z(N,\{t_i\})} \int d\bT\,d\overline{\bT}\ \frac{\partial}{\partial \bT_{a_1\dotsb a_D}}
\left( \frac12 \frac{\partial\, \tr_{\cB_{4,\{c\}}}(\bT,\overline{\bT})}{\partial \overline{\bT}_{a_1 \dotsb a_D}}\,e^{-N^{D-1} S}\right) = 0\;.
\end{equation}
 We shall recast this expression bit by bit. When acting on the first factor the derivative w.r.t.~$\bT$ gives:
 \begin{equation}
   \label{eq:sd2term1}
\sum_{a_1,\dotsc,a_D} \frac{1}{Z(N,\{t_i\})} \int d\bT\,d\overline{\bT}\ \frac12
\frac{\partial^2\, \tr_{\cB_{4,\{c\}}}(\bT,\overline{\bT})}{\partial \bT_{a_1\dotsb a_D}\,\partial \overline{\bT}_{a_1 \dotsb a_D}}\,e^{-N^{D-1} S}
= N^{D-1} \langle \tr_{\cB_2}(\bT, \overline{\bT})\rangle + N \langle \tr_{\cB_2}(\bT, \overline{\bT})\rangle\;.
\end{equation}
Meanwhile, we split its effect on the second factor into two parts. Operating on the quadratic part of the action produces, thanks to
$\frac{\partial\,\tr_{\cB_2}(\bT, \overline{\bT})}{\partial \bT_{a_1\dotsb a_D}} = \overline{\bT}_{a_1\dotsb a_D}$:
\begin{equation}
  \label{eq:sd2term2}
\sum_{a_1,\dotsc,a_D} \frac{1}{Z(N,\{t_i\})} \int d\bT\,d\overline{\bT}\ \frac12
\frac{\partial\, \tr_{\cB_{4,\{c\}}}(\bT,\overline{\bT})}{\partial \overline{\bT}_{a_1 \dotsb a_D}}
 \left(  - \frac{\partial\,\tr_{\cB_2}(\bT, \overline{\bT})}{\partial \bT_{a_1\dotsb a_D}} \right)
 \ e^{-N^{D-1} S} = - N^{D-1}\,\langle \tr_{\cB_{4,\{c\}}}(\bT,\overline{\bT}) \rangle\;,
\end{equation}
leading to:
\begin{multline}
  \label{2ndSDEFormal}
\Bigl(1+\frac1{N^{D-2}}\Bigr) \langle \tr_{\cB_2}(\bT,\overline{\bT}) \rangle - \langle \tr_{\cB_{4,\{c\}}}(\bT,\overline{\bT}) \rangle \\
+\sum_{a_1,\dotsc,a_D} \int \frac{d\bT\,d\overline{\bT}}{Z(N,\{t_i\})}\frac12
\frac{\partial\, \tr_{\cB_{4,\{c\}}}(\bT,\overline{\bT})}{\partial \overline{\bT}_{a_1 \dotsb a_D}}\,\frac{\partial}{\partial \bT_{a_1\dotsb a_D}}
\left(\sum_{i\in I} \frac{z^{p_i-1}}{p_i}\,t_i\,\tr_{\cB_i}(\bT,\overline{\bT}) \right) e^{-N^{D-1} S} = 0\;.
\end{multline}
In order to conclude we must evaluate the final contribution in eq. \eqref{2ndSDEFormal}.
Let us explain some of its components in detail.  The first factor
$\partial\, \tr_{\cB_{4,\{c\}}}(\bT,\overline{\bT})/\partial \overline{\bT}_{a_1 \dotsb a_D}$ is of order
two in $\bT$ and one in $\overline{\bT}$. The graphical rules of Table \ref{table:GraphicalRules} still apply in this context,
meaning that the resulting (non-invariant) polynomial has a graphical representative:
\begin{equation} \label{OpenB4}
\frac12 \frac{\partial\, \tr_{\cB_{4,\{c\}}}(\bT,\overline{\bT})}{\partial \overline{\bT}_{a_1 \dotsb a_D}} = \begin{array}{c}
\includegraphics[scale=.5]{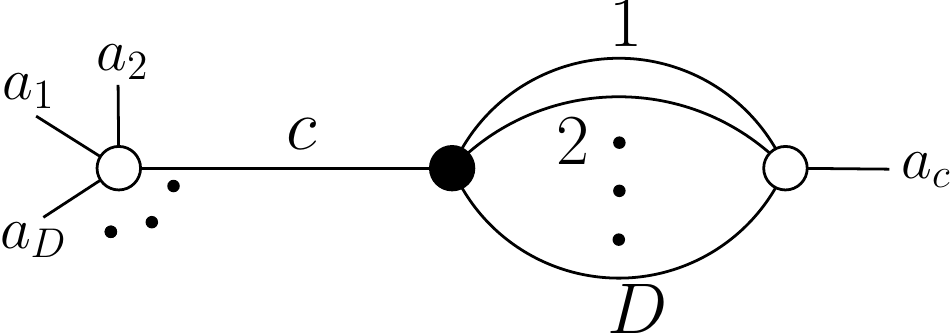}\end{array}\;.
\end{equation}
As expected, the graph is open, with all open edges emerging from white vertices.  This means that the polynomial transforms like $\bT$ in \eqref{UnitaryTransfo}.

The factor $\partial\,\tr_{\cB_i}(\bT,\overline{\bT})/\partial \bT_{a_1\dotsb a_D}$ is a polynomial (which transforms like $\overline{\bT}$)
obtained from  $\tr_{\cB_i}(\bT, \overline{\bT})$ by summing over all the ways to remove a $\bT$. Graphically, this gives to a sum over a
collection of open graphs, each of which corresponds to a distinct way to excise a white vertex from $\cB_i$.

Now notice that if one removes a white vertex $V$ from an arbitrary bubble $\cB$ and inserts the right hand side of \eqref{OpenB4}, one
produces a new bubble that is $\cB$ with a $(D-1)$-dipole inserted on the edge of color $c$ incident to $V$. This is precisely the process
operating in the last term of \eqref{2ndSDEFormal}. When it is contracted with the left hand side of \eqref{OpenB4}, it means that the
right hand side of \eqref{OpenB4} is glued back instead of the removed white vertex. That yields a sum over all the ways to insert a $(D-1)$-dipole
on an edge $e_c$ of color $c$ in $\cB_i$. When the insertion occurs on the edge $e_c$, we denote the resulting bubble $\cB_i+(e_c)$.
The Schwinger-Dyson equation \eqref{2ndSDEFormal} therefore reads:
\begin{equation}
  \label{2ndSDE}
\Bigl(1+\frac1{N^{D-2}}\Bigr) \langle \tr_{\cB_2}(\bT,\overline{\bT}) \rangle - \langle \tr_{\cB_{4,\{c\}}}(\bT,\overline{\bT})
\rangle + \sum_{i\in I} \frac{z^{p_i-1}}{p_i}\,t_i \sum_{e_c\in \cB_i} \langle \tr_{\cB_i+(e_c)}(\bT, \overline{\bT}) \rangle =0\;.
\end{equation}
There are $D$ such equations, one for each value of the color $c\in\{1,\dotsc,D\}$.

\subsection{Navigating the following sections}

In light of the technical nature of Sections \ref{sec:SDNLO} and \ref{sec:DS}, the main points of the argument are presented here. 
\begin{itemize}
  \item[--]
    To begin Section \ref{sec:SDNLO}, we present a quick recapitulation of the large--$N$ limit.  In particular, combining its Gaussian universality \eqref{GaussianLargeNLimit} with the Schwinger--Dyson equations allows on to solve for the leading order contribution $T(z,\{t_i\})$, to the $1/N$--expansion of the 2--point function. Only the melonic subsector survives and,  in the quartic model, one finds:
\begin{equation*}
  T(z) = \frac{1 - \sqrt{1 - 4Dz}}{2Dz}\;.
\end{equation*}

\item[--] Beyond leading order, non--Gaussian contributions creep into the mix.  Having catalogued the pertinent examples, we examine the next--to--leading order in detail. At NLO, one needs only to utilize the two SD equations \eqref{1stSDE} and \eqref{2ndSDE} to obtain a pair of coupled equations, linear in $K_2^{\NLO}(z,\{t_i\})$ and $K_{4,\bullet}^{\LO}(z,\{t_i\})$. Respectively, these are the NLO contribution to the 2--point function and the leading order contribution to the connected 4--point function, based on the graphs of species $\mathcal{B}_{4,\{c\}}$. One solves these equations to arrive at the result (in the case of the quartic model):
\begin{align}
  \nonumber
K_2^{\NLO}(z) &= \frac1{\sqrt{1-4Dz}}\,\frac{Dz\,T(z)^2}{1-z\,T(z)^2},\crcr
\nonumber
K_{4,{\bullet}}^{\LO}(z) &= \frac{z\,T(z)^4}{1-z\,T(z)^2} \; ,
\end{align}

\item[--] Specializing to the quartic model allows one to most easily probe deeper into the SD equations, \eqref{1stSDE} and \eqref{2ndSDE}, and retrieve information about $K_{2}^{\NNLO}(z)$ and $K_{4,\emptyset}^{\LO}(z)$. The latter is the leading order contribution to the connected 4--point function based on the graph $\mathcal{B}_{4,\emptyset}$ (as described next to Figure \ref{fig:4pointbub1}). Again, solving the associated coupled equations yields our next result:
\begin{align}
  \nonumber
K_2^{\NNLO}(z) &= \frac1{1-4Dz}\,\frac{D(D-1)\,z^2\,[T(z)]^3}{1-z\,[T(z)]^2},
\\
\nonumber
K^{\LO}_{4,\emptyset}(z) & = \frac{1}{\sqrt{1-4Dz}}\,\frac{D(D-1)\,z^2\,[T(z)]^5}{1-z\,[T(z)]^2}.
\end{align}

\item[--] In Section \ref{ssec:DSSD}, we change the parameter set from $(N, z, \{t_i\})$ to $(N,x,\{t_i\})$ where $x  = N^{D-2}(z - \frac{1}{4D})$, dubbed the double--scaling parameter. While this choice for $x$ is, at the outset, an ansatz, its validity is unequivocally confirmed by subsequent analysis. In principle, one can now send $N\rightarrow \infty$ and $z\rightarrow\frac{1}{4D}$, keeping $x$ fixed.  This is the double scaling limit. Moreover, it allows for a new expansion of the cumulants around the melonic sector evaluated at criticality.  Once substituted into the SD equation \eqref{1stSDE}, one obtains an equation for $K^{DS}(x,\{t_i\})$, the dominant contribution to the 2--point function in this new expansion, \emph{provided} \textit{i}) certain higher order correction terms truly remain sub--dominant as one tunes to criticality in $z$ and \textit{ii}) the contributing series are actually summable (which is the case for $D<6$).  
In this instance, one arrives at our main result, the limiting double scaled behaviour of the 2--point function. In the quartic case, this takes the form:
  \begin{equation*}
    \nonumber
K^{\DS}(x) = 4\sqrt{D}\ \sqrt{x - \frac{1}{4\,(D-1)}}\;.
\end{equation*}

\item[--] In Section \ref{ssec:proofdsl}, we prove that the assumption of sub--dominance, vital for the results of the preceding section, does in fact hold. In Section \ref{ssec:FromQuarticToGeneric}, we present an argument for the universality of such a double scaling limit within the subclass of tensor models with melonic interaction terms.

\end{itemize}

\section{The Schwinger-Dyson equations beyond the large--$N$--limit} \label{sec:SDNLO}

\subsection{The leading order}

Plugging the result \eqref{GaussianLargeNLimit} of the Universality Theorem into the SD equations enables one to obtain a closed
the system of equations, since all expectations factorize as products of the large--$N$ 2--point function $T(z,\{t_i\})$. Only the
melonic expectations survive and all SD equations ultimately reduce to the same algebraic equation \cite{SDE}:
\begin{equation} \label{SDLO}
1- T(z,\{t_i\}) + \sum_{i\in I} z^{p_i-1}\,t_i\,[T(z,\{t_i\})]^{p_i} = 0\;.
\end{equation}
Together with the initial condition $T_{|z=0}=1$, this equation determines $T$ as long
as the derivative of \eqref{SDLO} with respect to $T$ does not vanish. Examining the equation \eqref{SDLO}, one sees
that for generic couplings $\{t_i\}_{i\in I}$, $T$ has a square--root singularity at a finite value of $z$ called the critical coupling $z_c$:
\begin{equation} \label{CriticalT}
T(z,\{t_i\}) \simeq T_c + T'_c\,\bigl(1-z/z_c\bigr)^{1/2}\;,
\end{equation}
where $T_c$, $z_c$ and $T'_c$ are functions of $\{t_i\}$. The critical values $T_c$ and $z_c $
are determined by eq. \eqref{SDLO} supplemented with the criticality condition that (minus) the derivative of \eqref{SDLO}
with respect to $T$:
\begin{equation} \label{SingularityOfS}
  C(z,\{t_i\}) \equiv 1 - \sum_{i\in I} t_i\,p_i\,[z\,T(z,\{t_i\})]^{p_i-1}  \;,
\end{equation}
vanishes.
Notice that $C(z,\{t_i\})$ not only controls how far we are from criticality, but it is also singular at criticality.
Indeed, plugging the expansion \eqref{CriticalT} into $C(z,\{t_i\})$, we see that around $z_c$:
\begin{equation}
C(z,\{t_i\}) = -\Bigl(\sum_{i\in I} t_i p_i(p_i-1) z_c^{p_i-1} T_c^{p_i-2}\Bigr)\,T'_c\,\bigl(1-z/z_c\bigr)^{1/2} = c(\{t_i\})\,\bigl(1-z/z_c\bigr)^{1/2}.
\end{equation}

\emph{Quartic case.} The above scenario is easily illustrated when $I=\{1,\dotsc,D\}$ and the interaction part of the action is
defined as the set of quartic bubbles $\{\cB_i\}_{i\in I} = \{\cB_{4,\{c\}}\}_{c=1,\dotsc,D}$ with the same global coupling $z$,
i.e. $t_i=1$ for all $i\in I$.  Thus, the model is symmetric with respect to the colors $\{1,\dotsc,D\}$, the number of black
vertices for each interaction bubble satisfies $p_i=2$ ($i\in I$), and the sums over $i\in I$ reduce to multiplication by the factor $D$.
The equations \eqref{SDLO} specializes to:
\begin{equation}
  \label{eq:SDLO4}
1-T(z) + Dz\,T(z)^2 = 0\;,
\end{equation}
whose physical solution is:
\begin{equation}
  \label{eq:SDLO4S}
T(z) = \frac{1-\sqrt{1-4Dz}}{2Dz}\;.
\end{equation}
The critical point is $z_c = 1/(4D), T_c =2$, which satisfies the criticality condition, $1-2Dz_cT_c=0$. Moreover, close to criticality:
\begin{equation}
  \label{eq:Criticality4}
C(z) = 1 - 2Dz\,T =  \sqrt{1-4Dz}\;,
\end{equation}
is singular as expected from \eqref{SingularityOfS}.

\subsection{Moments and Cumulants} \label{sec:cumulants}

The expectations of arbitrary polynomials in the tensor entries can be computed as derivatives of the moment generating function
\begin{equation}
 Z(\bJ, \overline \bJ) = \int
     \left( \prod_{\vec a} N^{D-1} \frac{\extd \bT_{\vec a}\, \extd
    \overline \bT_{\vec a}} {2\pi i}\right) e^{-N^{D-1}S(\bT,\overline \bT) + \tr_{\cB_2}(\bT,  \overline \bJ) + \tr_{\cB_2}(\bJ,  \overline \bT)} \; .
\end{equation}
It turns out that it is more convenient to work with the generating function of connected moments,
or \emph{cumulants}:
  \begin{equation}
    \label{eq:cumdef}
    W(\bJ, \overline \bJ) =  \ln Z(\bJ, \overline \bJ)
    \quad\textrm{with}\quad
    W(\bJ, \overline \bJ) = \sum_{p = 0}^\infty \frac{1}{(p!)^2} \;  W^{(2p)}_{\vec a^1 \dotsc \vec a^p, \vec b^1 \dotsc \vec b^p} \Big(N, z, \{t_i\} \Big)\;
    \bJ_{\vec a^1} \dotsm \bJ_{\vec a^p} \overline \bJ_{\vec b^1} \dotsm \overline \bJ_{\vec b^p}\;.
  \end{equation}
The normalization factor $1/(p!)^2$ is conventional and accounts for the invariance of the cumulant of order $2p$ (or $2p$ point cumulant)
$W^{(2p)}_{\vec a^1 \dotsc \vec a^p, \vec b^1 \dotsc \vec b^p} \Big(N, z,\{t_i\} \Big)$ under independent permutations $\sigma$ and $\tau$ of its indices:
 \begin{equation}
   \label{eq:sdbasic}
   W^{(2p)}_{\vec a^1 \dotsc \vec a^p, \vec b^1 \dotsc \vec b^p}\Big(N, z,\{t_i\} \Big) =
   W^{(2p)}_{\vec a^{\sigma(1)} \dotsc \vec a^{\sigma(p)}, \vec b^{\tau(1)} \dotsc \vec b^{\tau(p)}}\Big(N, z,\{t_i\} \Big) \; .
 \end{equation}

The generating function $W(\bJ, \overline \bJ)$ is itself invariant under the unitary transformations \eqref{UnitaryTransfo}. It follows that it admits an expansion in invariants labeled by (non necessarily connected) bubbles. Denoting $p(\cB)$ the number of black vertices of the bubble $\cB$, the $2p$-point cumulant admits an expansion:
\begin{equation}
   W^{(2p)}_{\vec a^1 \dotsc \vec a^p, \vec b^1 \dotsc \vec b^p}\Big(N, z,\{t_i\} \Big)
   = \sum_{\cB, p(\cB) = p } \bar\delta^{\cB}_{\vec a^1 \dotsc \vec a^p,\vec b^1\dotsc \vec b^p} \; W \Big(\cB ; N,z,\{t_i\}  \Big)  \; ,
\end{equation}
where $\bar\delta^{\cB}_{\vec a^1 \dotsc \vec a^p,\vec b^1\dotsc \vec b^p}$ is the trace-invariant operator associated to $\cB$, symmetrized over its indices.

\begin{figure}
\includegraphics[scale=.4]{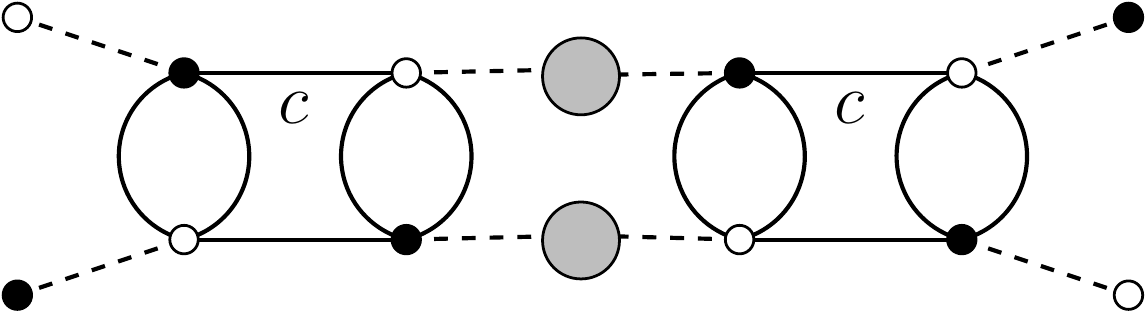}
\caption{\label{fig:4PointExample} This shows a 4-point graph where the grey blobs represent arbitrary 2-point insertions, which do not affect the boundary graph.}
\end{figure}

The graphical interpretation is as follows. The $2p$-point cumulants are sums over connected Feynman graphs $\cG$ with $2p$ external lines, i.e. $(D+1)$ edge-colored graphs with $2p$ edges of color $0$ hooked to univalent external vertices, and such that all their subgraphs with colors $\{1,\dotsc,D\}$ are from the set $\{\cB_i\}_{i\in I}$. A typical example is presented in figure \ref{fig:4PointExample}.

The \emph{faces} of color $0c$ of a Feynman graph (i.e. the subgraphs with colors $0,c$) fall in two categories. Either they are \emph{internal faces}, that is closed cycles of edges of colors $0$ and $c$, or they are \emph{external faces}, that is chains of edges of colors $0$ and $c$. The external faces start and end at external vertices and there is a single face of color $0c$ which start/end at an external vertex (there are $p$ external faces of each color $c\in\{1,\dotsc,D\}$ in $\cG$). The univalent external vertices have $D$-uples of indices, $\vec{a}^v$ for white vertices and $\vec{b}^{\bar{v}}$ for the black ones. The index of color $c$ at an external white vertex is identified along the corresponding external face with the index of the same color at the black vertex on the other end of the chain.

Due to the unitary invariance, the pattern of index identification along the external faces of $\cG$ can be encoded in a $D$-colored graph called the \emph{boundary graph} of $\cG$ and denoted $\partial \cG$. It only depends on the external faces and is obtained by only keeping the univalent external vertices of $\cG$ and drawing an edge of color $c$ for every external face of colors $0c$ connecting two external vertices. For instance, the figure \ref{fig:4PointExample} shows a 4-point graph whose boundary graph is $\cB_{4,\{c\}}$. From the boundary graph $\partial \cG$, one gets the trace--invariant operator $\bar\delta^{\partial\cG}_{\vec a^1 \dotsc \vec a^p,\vec b^1\dotsc \vec b^p}$ which contains the whole index dependence of the Feynman graph. The function $W\Big(\cB;N,z,\{t_i\}\Big)$ is then the sum of the amplitudes of all graphs $\cG$ contributing with the fixed boundary graph $\partial\cG=\cB$.

The scaling with $N$ of each of the terms in the sum above has been studied in \cite{Universality}. Let us define the rescaled contribution of an invariant:
\begin{equation}\label{eq:propbound}
K \Big( \cB;N,z,\{t_i\} \Big) \equiv \frac{ W \Big(\cB ; N,z,\{t_i\}  \Big)}{ N^{  D - 2(D-1) p(\cB) -\rho(\cB) }  } \; ,
\end{equation}
where $\rho(\cB)$ denotes the number of connected components of $\cB$. In the sense of perturbation theory,
for any tensor model, these quantities are bounded for all $N$ and admit a finite large--$N$ limit \cite{Universality}:
\begin{equation}
\lim_{N\to \infty} K \Big( \cB;N,z,\{t_i\} \Big) =  K \Big( \cB,z,\{t_i\} \Big) \; .
\end{equation}

In the sequel the 2-- and 4--point cumulants $W^{(2)}_{\vec a^1, \vec b^1}$ and $W^{(4)}_{\vec a^1\vec a^2,\vec b^1 \vec b^2}$ will play a distinguished role. $W^{(2)}_{\vec a^1, \vec b^1}$ is proportional to the trace-invariant $\delta^{\cB_2}$ since there is a single bubble on two vertices, $\cB_2$.

\begin{figure}[htb]
\centering
\includegraphics[scale = 0.8]{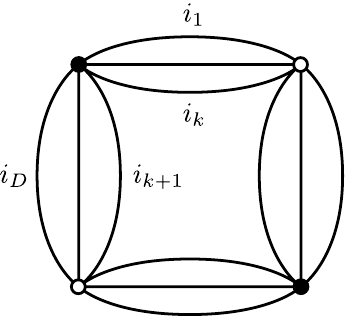}
\caption{\label{fig:4pointbub1} The bubble graph $\cB_{4,  \{c_{k+1},\dots, c_D\} }$.}
\end{figure}

At order $4$, we have several bubbles $\cB_{4,\cC}$, one for each choice of a subset $\cC\subset \{1,\dotsc,D\}$ of cardinality $0 \le |\cC| \le D/2$ (it is important to keep in mind that the expansion includes non-connected bubbles, here the union of two copies of $\cB_2$). The four vertices of $\cB_{4,\cC}$ are divided into two pairs. The two vertices in a pair are connected by edges of colors $\{1,\dotsc,D\} \setminus \cC$, while the edges with  color in $\cC$ connect the pairs in between them. An example is presented in figure \ref{fig:4pointbub1}.

Remark that $\cB_{4,\emptyset}$ represents two copies of $\cB_2$, while the bubbles $\cB_{4,\{c\}}$ have already been presented in figure \ref{fig:4PointBubble}. The trace--invariant operators associated to $\cB_2$ and $\cB_{4,\cC}$ are
\begin{align}
 & \bar \delta^{\cB_2}_{\vec a^1, \vec b^1} = \prod_{c=1}^D \delta_{a^1_c b^1_c}   \;, \\
& \bar \delta^{\cB_{4,\cC}}_{\vec a^1 \vec a^2, \vec b^1 \vec b^2 } = \left( \prod_{c\notin \cC} \delta_{a_c^1 b_c^1} \right)
 \left( \prod_{c\notin \cC} \delta_{a_c^2 b_c^2} \right) \left( \prod_{c\in \cC}  \delta_{a_c^1 b_c^2}
  \delta_{a_c^2 b_c^1} \right) + \left( \prod_{c\notin \cC} \delta_{a_c^2 b_c^1} \right)
 \left( \prod_{c\notin \cC} \delta_{a_c^1 b_c^2} \right) \left( \prod_{c\in \cC}  \delta_{a_c^2 b_c^2}
  \delta_{a_c^1 b_c^1} \right)
  \; ,
\end{align}
where, according to our previous discussion we use the symmetrized trace-invariant operator for $\cB_{4,\cC}$. The 2- and 4-point cumulants write therefore as:
\begin{align}\label{eq:twofourpoint}
&  W^{(2)}_{\vec a_1, \vec b_1}   = \delta^{\cB_2}_{\vec a^1 \vec b^1} \; \; W \Big(\cB_2 ; N,z,\{t_i\}  \Big)  \; ,\\
&  W^{(4)}_{\vec a_1\vec a_2,\vec b_1\vec b_2}  = \sum_{\cC \subset \{1,\dotsc,D\} }
     \delta^{\cB_{4,\cC}}_{\vec a^1 \vec a^2; \vec b^1 \vec b^2 }  \; \;  W \Big(\cB_{4,\cC} ; N,z,\{t_i\}  \Big) \; .
\end{align}

\subsection{Gaussian and non-Gaussian contributions}

The Universality Theorem asserts that the large--$N$ limit of a random tensor model is Gaussian with covariance $T(z,\{t_i\})$. At next--to--leading orders, non--Gaussian contributions eventually show up. Loosely speaking, a non--Gaussian contribution to an expectation $\langle \tr_{\cB}(\bT, \overline{\bT}) \rangle$ is any Feynman graph $\mathcal{G}$ in the expansion of the expectation which would not appear in a Gaussian distribution whatever the covariance.

Let us denote $\pi$ a partition of the set of vertices $\cV(\cB)$ of the bubble $\cB$ into (disjoint) bipartite subsets, $\cV(\cB)=\bigcup_{\alpha} \cV_\alpha$, with $\cV_\alpha=\{v^{(\alpha)}_1,\dotsc,v^{(\alpha)}_{p_\alpha}, \bar v^{(\alpha)}_1,\dotsc,\bar v^{(\alpha)}_{p_\alpha}\}$ of cardinality $2p_\alpha$ (hence $\sum_\alpha p_\alpha = p(\cB)$).
The expectation of a bubble observable expands in cumulants as:
\begin{align}\label{eq:momentexpansion}
& \langle \tr_{\cB}(\bT, \overline{\bT}) \rangle =
\sum_{\{\vec a^{v}, \vec b^{\bar v}\}_{v,\bar v}}  \delta^{\cB}_{\vec a^v, \vec b^{\bar v}}
\left \langle
\left(  \prod_{\bar v \in \cV(\cB) } \bar \bT_{\vec b^{\bar v}} \right)
 \left( \prod_{v\in \cV(\cB)} \bT_{\vec a^{v}} \right) \right \rangle \\
 & = \sum_{\{\vec a^{v}, \vec b^{\bar v}\}_{v,\bar v}}  \delta^{\cB}_{\vec a^v, \vec b^{\bar v}}
 \left(\sum_{ \pi  } \prod_{\cV_\alpha \in \pi}
W^{(2p_\alpha)}_{\vec a^{v^{(\alpha)}_1} \dotsc \vec a^{v^{(\alpha)}_{p_\alpha}}, \vec b^{ \bar v^{(\alpha)}_1 } \dotsc \vec b^{\bar v^{(\alpha)}_{p_\alpha}}} \Big(N, z,\{t_i\} \Big)  \right)\; .
\end{align}
This means that we can classify the Feynman graphs contributing to the expectation using partitions of the vertex set $\cV(\cB)$. Each partition $\pi$ gives rise to a family of graphs $\cG$ such that upon cutting off the edges of color 0 which connect the marked sub--graph $\cB$ to the rest of $\cG$, one gets as many connected components as parts in $\pi$ and these connected components have $2p_\alpha$ external edges which were connected to the vertices of $\cV_\alpha$ in $\cG$.

We call a \emph{Gaussian contribution} a partition $\pi$ such that $p_\alpha=1, \;  \forall \alpha$, and a
\emph{non--Gaussian contribution} any other partition. This means that a Gaussian contribution is exactly a graph obtained by attaching to $\cB$ only 2-point graphs.

The case $p(\cB)=1$ is quite trivial, by definition. Indeed there is a single partition $\pi$ which is the vertex set of $\cB_2$ and $\langle \tr_{\cB_2}(\bT, \overline{\bT}) \rangle/N = K(\cB_2;N,z,\{t_i\})$. Recall that $K \Big(\cB_2 ; N,z,\{t_i\}  \Big)$ is bounded for all $N$ and has a finite limit, namely $T(z,\{t_i\}) \equiv  K \Big(\cB_2 ; z,\{t_i\}  \Big) $, when $N$ goes to infinity. We can therefore start the $1/N$--expansion of the 2-point function as
\be \label{2ptFunctionExpansionNLO}
\frac1N \langle \tr_{\cB_2}(\bT, \overline{\bT}) \rangle = K \Big(\cB_2 ; N,z,\{t_i\}\Big) = T(z,\{t_i\}) + \frac1{N^{D-2}} \Big(K^{\NLO} \Big(\cB_2 ,z,\{t_i\}  \Big)  + \mathcal{O}(1/N) \Big)\;,
\ee

Next, consider the case $p(\cB)=2$. Labeling the four vertices $v_1, v_2, \bar{v}_1, \bar{v}_2$, three partitions into bipartite subsets are obtained. The partitions $\{\{v_1,\bar{v}_1\}, \{v_2, \bar{v}_2\}\}$ and $\{\{v_1,\bar{v}_2\}, \{v_2, \bar{v}_1\}\}$ are both Gaussian, since the parts have cardinality two, meaning only 2-point functions are attached to $\cB$. The third partition is the vertex set itself, corresponding to attaching to $\cB$ a 4-point cumulant. Graphically, this leads to the following exact expansion on $\cB_{4,\{c\}}$,
\begin{equation*}
\frac1N \left\langle \begin{array}{c} \includegraphics[scale=.55]{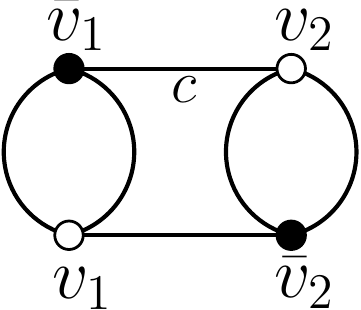} \end{array}\right\rangle = \begin{array}{c} \includegraphics[scale=.5]{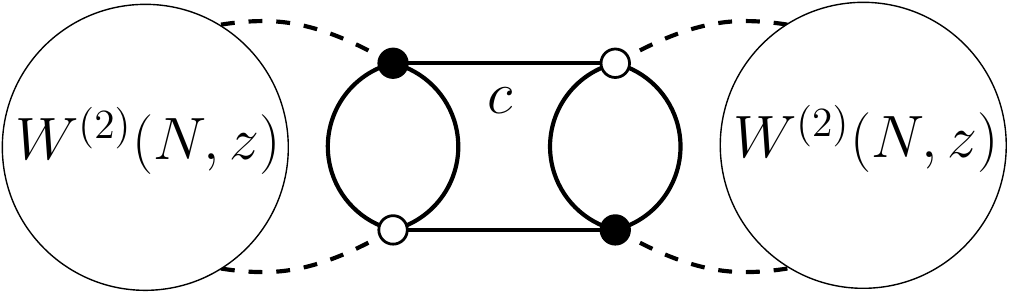} \end{array} + \begin{array}{c} \includegraphics[scale=.5]{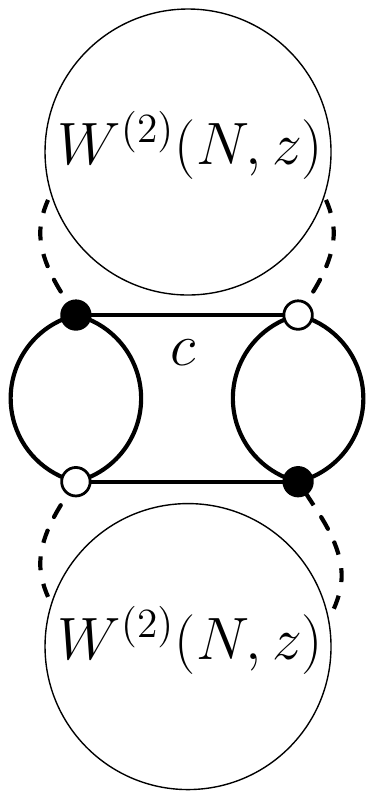} \end{array} + \begin{array}{c} \includegraphics[scale=.5]{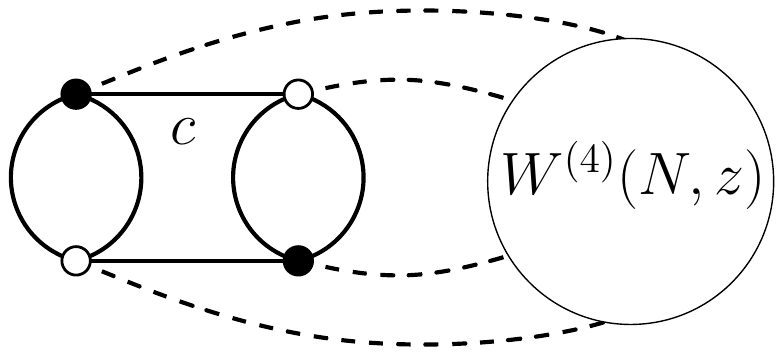} \end{array}
\end{equation*}
Clearly, the parts of a partition corresponding to a graph can be visualized by cutting the edges of color 0 around the bubble $\cB_{4,\{c\}}$.

Let us consider in more detail the expectation of $\cB_{4,\{c\}}$. We have
\begin{align}
\frac{1}{N} &\langle \tr_{\cB_{4,\{c\} }}(\bT, \overline{\bT}) \rangle =
\frac{1}{N} \sum_{\substack{\vec a^1,\vec a^2\\ \vec b^1,\vec b^2}} \Bigl(
   \delta_{a^1_c b^2_c } \delta_{a^2_c b^1_c}
\prod_{c_1\neq c} \delta_{a^1_{c_1} b^1_{c_1} } \delta_{a^2_{c_1} b^2_{c_1} } \Bigr) \langle \bT_{\vec a^1} \bT_{\vec a^2}
\overline{\bT}_{\vec b^1}  \overline{\bT}_{\vec b^2}\rangle  \\
& \begin{aligned} =
\frac{1}{N} \sum_{\substack{\vec a^1,\vec a^2\\ \vec b^1,\vec b^2}} \Bigl(
   \delta_{a^1_c b^2_c } \delta_{a^2_c b^1_c}
\prod_{c_1\neq c} \delta_{a^1_{c_1} b^1_{c_1} } \delta_{a^2_{c_1} b^2_{c_1} } \Bigr) &
 \Bigg[ W^{(2)}_{ \vec a^1, \vec b^1}\Big(N, z,\{t_i\} \Big) W^{(2)}_{ \vec a^2, \vec b^2} \Big(N, z,\{t_i\} \Big) \\
&+  W^{(2)}_{ \vec a^1, \vec b^2}\Big(N, z,\{t_i\} \Big) W^{(2)}_{ \vec a^2, \vec b^1}\Big(N, z,\{t_i\} \Big) +
 W^{(4)}_{ \vec a^1 \vec a^2, \vec b^1 \vec b^2 }\Big(N, z,\{t_i\} \Big)
 \Bigg] . \end{aligned}
\end{align}
The two products of 2-point cumulants are the two Gaussian contributions.

\begin{figure}
\includegraphics[scale=.5]{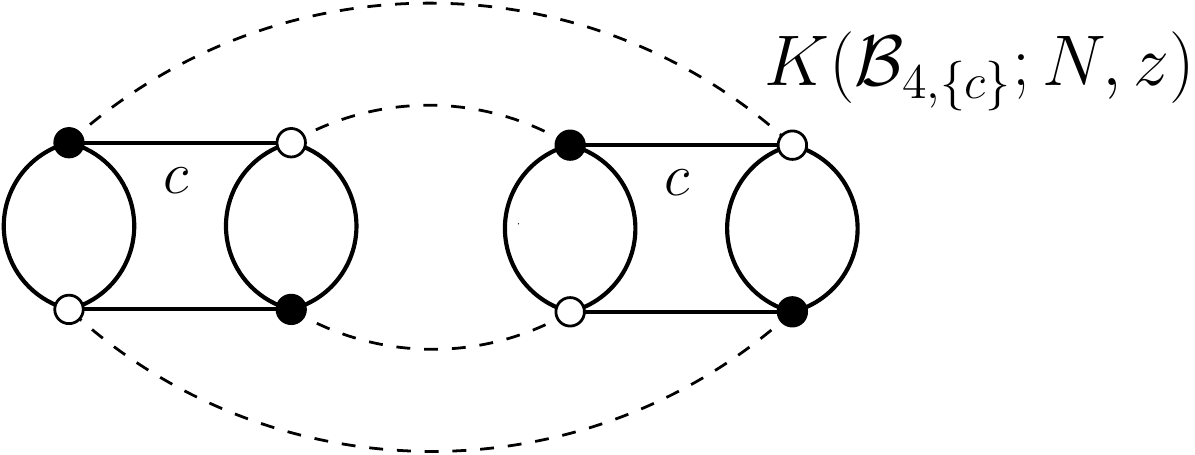}
\caption{\label{fig:contractionG4} The dominant non-Gaussian contribution to the expectation of $\cB_{4,\{c\}}$, where the rightmost bubble represents the boundary graph of $K(\cB_{4,\{c\}};N,z,\{t_i\})$. It is necessary to explicitly draw this boundary graph in order to describe how it is connected to the observable. There are indeed two ways to connect it, corresponding to the two terms proportional to $K(\cB_{4,\{c\}};N,z,\{t_i\})$ in eq. \eqref{4ptBubbleExpansion}: the dominant one, which comes with a factor $1/N^{D-2}$ and a sub--dominant one which comes with a scaling $N^{-D-1}/N^{D-2}$.}
\end{figure}

We treat the Gaussian and non--Gaussian contributions above separately. Using eq. \eqref{eq:twofourpoint}, the Gaussian contributions write:
\begin{equation}
\begin{aligned}
&\begin{array}{c} \includegraphics[scale=.4]{4PointBubbleGaussianLO.pdf} \end{array} + \begin{array}{c} \includegraphics[scale=.4]{4PointBubbleGaussianNLO.pdf} \end{array} = \frac{1}{N} \big(N^{2D-1 } + N^{D+1} \big)   \bigg[ N^{-(D-1)} K \Big(\cB_2 ; N,z,\{t_i\}  \Big) \bigg]^2 \\
&= \Bigl(1+\frac{1}{N^{D-2}}\Bigr)   \bigg[ K \Big(\cB_2 ; N,z,\{t_i\}  \Big) \bigg]^2 = \begin{array}{c} \includegraphics[scale=.4]{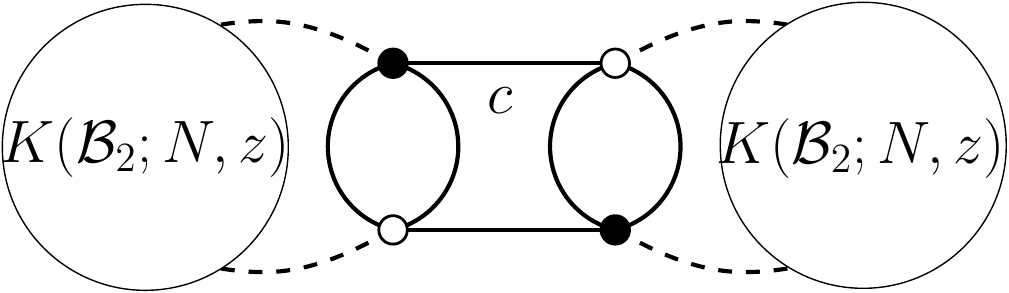} \end{array} + \frac1{N^{D-2}} \begin{array}{c} \includegraphics[scale=.4]{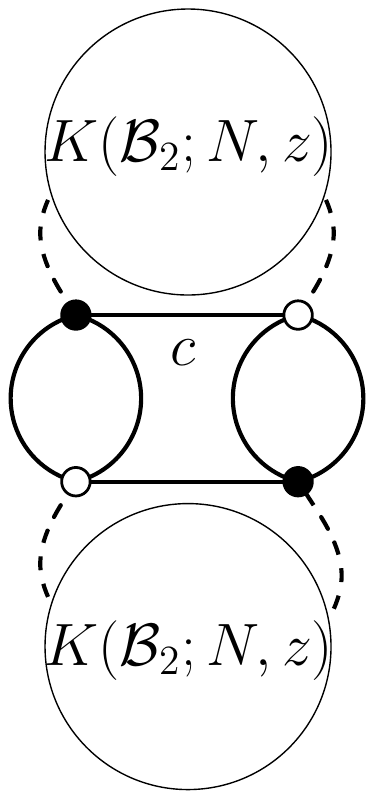} \end{array}\; .
\end{aligned}
\end{equation}
The non-Gaussian contributions:
\begin{equation}\label{eq:nongauss1}
\begin{array}{c} \includegraphics[scale=.5]{4PointBubble4pt.pdf} \end{array} =
 \frac{1}{N} \sum_{a,b} \Bigl(
   \delta_{a^1_c b^2_c } \delta_{a^2_c b^1_c}\prod_{c_1\neq c} \delta_{a^1_{c_1} b^1_{c_1} } \delta_{a^2_{c_1} b^2_{c_1} } \Bigr)
   W^{(4)}_{ \vec a^1 \vec a^2, \vec b^1 \vec b^2 }\Big(N, z,\{t_i\} \Big) \; ,
\end{equation}
are shown in appendix \ref{app:4pdom} to yield:
\begin{align}
 \begin{array}{c} \includegraphics[scale=.5]{4PointBubble4pt.pdf} \end{array} = & \Bigl( \frac{1}{N^{2(D-1) } } + \frac{1}{N^D} \Bigr)K \Big(\cB_{4,\emptyset} ; N,z,\{t_i\}  \Big) \crcr
 & + \sum_{\cC \in \{1,\dotsc,D\} , c\in \cC , |\cC| \le D/2}\Bigl( \frac{1}{N^{D-2}} N^{1-|\cC|} +\frac{1}{N^{D-2}} N^{ -D + |\cC|-1} \Bigr)    K \Big(\cB_{4,\cC} ; N,z,\{t_i\}  \Big) \crcr
&  +\sum_{\cC \in \{1,\dotsc,D\} , c\notin \cC, |\cC|\le D/2 } \Bigl( \frac{1}{N^{D-2}} N^{-D + 1 + |\cC|} + \frac{1}{N^{D-2}} N^{-1-|\cC|} \Bigr) K \Big(\cB_{4,\cC } ; N,z,\{t_i\}  \Big) \;.
\end{align}
We conclude that these terms contribute at most at order $N^{D-2}$. Furthermore, at order $N^{D-2}$ only the term with $\cC = \{c\}$ represented in Figure \ref{fig:contractionG4} contributes.

Gathering the above pieces brings up the following cumulant expansion,
\begin{equation} \label{4ptBubbleExpansion}
  \begin{aligned}
\frac1N \langle \tr_{\cB_{4,\{c\} }}(\bT, \overline{\bT}) \rangle &= \Bigl(1+\frac{1}{N^{D-2}}\Bigr)  \bigg[ K \Big(\cB_2 ; N,z,\{t_i\}  \Big) \bigg]^2 + \Bigl( \frac1{N^{D-2}} + \frac{1}{N^{2D-1}}\Bigr)   K \Big(\cB_{4,\{c\} } ; N,z,\{t_i\}  \Big) \\
    & + \Bigl( \frac{1}{N^{2(D-1) } } + \frac{1}{N^D} \Bigr)K \Big(\cB_{4,\emptyset} ; N,z,\{t_i\}  \Big)\\
    & + \sum_{\cC \in \{1,\dotsc,D\} , c\in \cC , 1 < |\cC| \le D/2} \Bigl( \frac{1}{N^{D-2}} N^{1-|\cC|} +\frac{1}{N^{D-2}} N^{ -D + |\cC|-1} \Bigr)    K \Big(\cB_{4,\cC} ; N,z,\{t_i\}  \Big)  \\
    &+  \sum_{\cC \in \{1,\dotsc,D\} , c\notin \cC , 1\le |\cC| \le D/2 } \Bigl( \frac{1}{N^{D-2}} N^{-D + 1 + |\cC|} + \frac{1}{N^{D-2}} N^{-1-|\cC|} \Bigr) K \Big(\cB_{4,\cC } ; N,z,\{t_i\}  \Big) \; .
  \end{aligned}
\end{equation}
This equation is exact. Since we are only interested in the LO and NLO of the expectation, we only keep the first line in the form
\be \label{4ptCumExpNLO}
\frac1N \langle \tr_{\cB_{4,\{c\} }}(\bT, \overline{\bT}) \rangle = \Bigl(1+\frac{1}{N^{D-2}}\Bigr)  \bigg[ K \Big(\cB_2 ; N,z,\{t_i\}  \Big) \bigg]^2 + \frac1{N^{D-2}} K \Big(\cB_{4,\{c\} } ; N,z,\{t_i\}  \Big) + \mathcal{O}(1/N^{D-1}).
\ee

Truncating at order $\frac{1}{N^{D-2}}$ and using \eqref{2ptFunctionExpansionNLO}, we obtain:
\begin{equation} \label{4PtExpNLO}
\frac1N\,\langle \tr_{\cB_{4,\{c\} }}(\bT, \overline{\bT})\rangle = T^2 + \frac1{N^{D-2}}\Bigl(2T\,K^{\NLO} \Big(\cB_2 ,z,\{t_i\}  \Big)  + T^2 +
K \Big(\cB_{4,\{c\} } ; z,\{t_i\}   \Big)  + \mathcal{O}(1/N)\Bigr)\;.
\end{equation}

We will use from now on the shorthand notations $K^{\NLO} \Big(\cB_2 ,z,\{t_i\}  \Big)  \equiv K_2^{\NLO} $, and $K \Big(\cB_{4,\{c\} } ; z,\{t_i\}   \Big) \equiv K_{4,\{c\}}^{\LO} $. In appendix \ref{app:4pdom} we prove that the same kind of expansion can be performed on the expectations of the polynomials associated to arbitrary melonic bubbles:
\begin{multline}
\frac1N\,\langle \tr_{\cB_{i}}(\bT, \overline{\bT})\rangle = \Bigl(1+\frac{\alpha_i}{N^{D-2}}\Bigr)  \bigg[ K \Big(\cB_2 ; N,z,\{t_i\}  \Big) \bigg]^{p_i} + \frac{\alpha'_i}{N^{D-2}} \bigg[K \Big(\cB_2 ; N,z,\{t_i\}  \Big) \bigg]^{p_i-2} K \Big(\cB_{4,\{c\} } ; N,z,\{t_i\}  \Big)\\
 + \mathcal{O}(1/N^{D-1}),
\end{multline}
reducing at NLO to:
\begin{equation} \label{BiExpNLO}
\frac1N\,\langle \tr_{\cB_{i}}(\bT, \overline{\bT})\rangle =
T^{p_i} + \frac1{N^{D-2}}\bigl(p_i\, T^{p_i-1}\, K_2^{\NLO} + \alpha_i\, T^{p_i} +
T^{p_i-2} \sum_{c=1,\dotsc,D}\alpha'_{i,c}\, K_{4,\{c\} }^{\LO} + \mathcal{O}(1/N)\bigr)\;,
\end{equation}
where $\alpha_i$ and $\alpha'_{i,c}$ are some coefficients which depend on $\cB_i$.

\emph{From now on, to simplify the analysis, we shall assume that the set of interaction bubbles $\{\cB_i\}_{i\in I}$
(as well as the set of corresponding couplings $\{t_i\}_{i\in I}$) is invariant under color relabeling.}  As a consequence,
there is a single function $K_{4,\bullet}^{\LO} \equiv K_{4,\{c\}}^{\LO}$ for all $c\in\{1,\dotsc,D\}$. We denote in \eqref{BiExpNLO} $\alpha'_i \equiv \sum \alpha'_{i,c}$.

The same kind of expansion holds for $\langle \tr_{\cB_i+(e_c)}(\bT, \overline{\bT})\rangle$, albeit with different numerical constants, $\alpha_{i,e_c}$ and $\alpha'_{i,e_c}$. It is convenient to sum directly over $e\in\cB_i$, that is, the edges of color $c$, upon which one may insert a $(D-1)$--dipole:
\begin{equation} \label{Bi+ExpNLO}
\sum_{e_c\in\cB_i} \frac1N\,\langle \tr_{\cB_{i}+(e_c)}(\bT, \overline{\bT})\rangle =
p_i\, T^{p_i+1} + \frac1{N^{D-2}}\bigg(p_i(p_i+1) T^{p_i}\, K_2^{\NLO} + \beta_{i,c}\, T^{p_i+1} + \beta'_{i,c}\, T^{p_i-1}\, K_{4,\bullet}^{\LO} + \mathcal{O}(1/N)\bigg)\;,
\end{equation}
with $\beta_{i,c} = \sum_{e} \alpha_{i,e_c}$ and $\beta'_{i,c} = \sum_e \alpha'_{i,e_c}$.

\emph{6-point bubbles.} In the quartic model of \cite{DSQuartic}, the SD equations involve the bubbles $\cB_{4,\{c'\}}+(e_c)$, that is, $\cB_{4,\{c'\}}$ with a melonic insertion of color $c$, which have six vertices and are of two types: either $c=c'$ or $c\neq c'$. If $c=c'$, we get a bubble with three $(D-1)$--dipole insertions of the same color. Setting for convenience $c'=1$ and $c=1$ or $c=2$ (depending on the case), we can write:
\begin{gather}
\label{6ptBubble1NLOExp}
\frac1N \left\langle \begin{array}{c} \includegraphics[scale=.3]{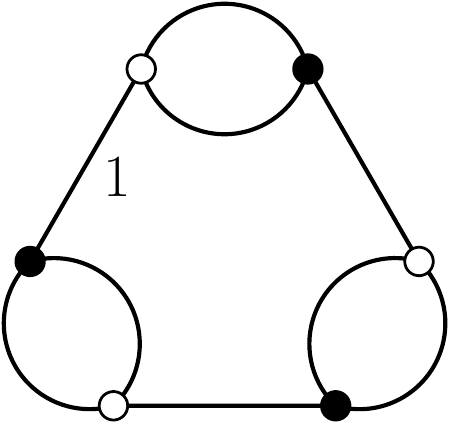} \end{array}
\right\rangle = \Bigl(1+\frac{3}{N^{D-2}}\Bigr) \Bigl[K \Big( \cB_2;N,z,\{t_i\}  \Big)\Bigr]^3 + \frac{3}{N^{D-2}}\Bigl[K \Bigl(\cB_2 ; N,z,\{t_i\} \Bigr)\, K \Bigl( \cB_{4,\bullet } ; N,z,\{t_i\} \Bigr) + \mathcal{O}(1/N)\Bigr], \\
\label{6ptBubble2NLOExp}
\frac1N \left\langle \begin{array}{c} \includegraphics[scale=.28]{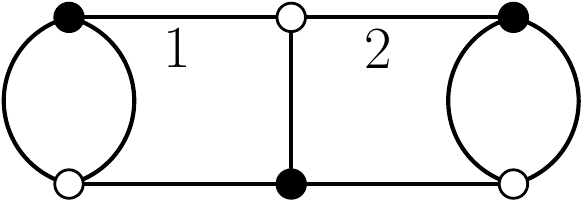} \end{array}  \right\rangle
= \Bigl(1+\frac{2}{N^{D-2}}\Bigr) \Bigl[K \Big(\cB_2; N,z,\{t_i\} \Big) \Bigr]^3 + \frac{2}{N^{D-2}}\Bigl[ K \Big(\cB_2; N,z,\{t_i\} \Big) K\Big(\cB_{4,\bullet } ; N,z,\{t_i\}\Big) + \mathcal{O}(1/N)\Bigr].
\end{gather}
In both equations, the $K(\cB_2; N,z,\{t_i\})^3$ term is the Gaussian contribution.  Its coefficient takes into account the single melonic contraction, as well as the $\alpha$ NLO Gaussian contractions.  The second term is the non--Gaussian contribution, where there are $\alpha'$ ways to introduce a 4--point contribution with the appropriate $1/N^{D-2}$ scaling.
For the first bubble $\alpha = \alpha' =3 $, since there are three contributions of the types:\footnote{Repeatedly rotating the graphs by $2\pi/3$ produces  another two graphs of each type.}
\begin{equation*}
\begin{array}{c} \includegraphics[scale=.4]{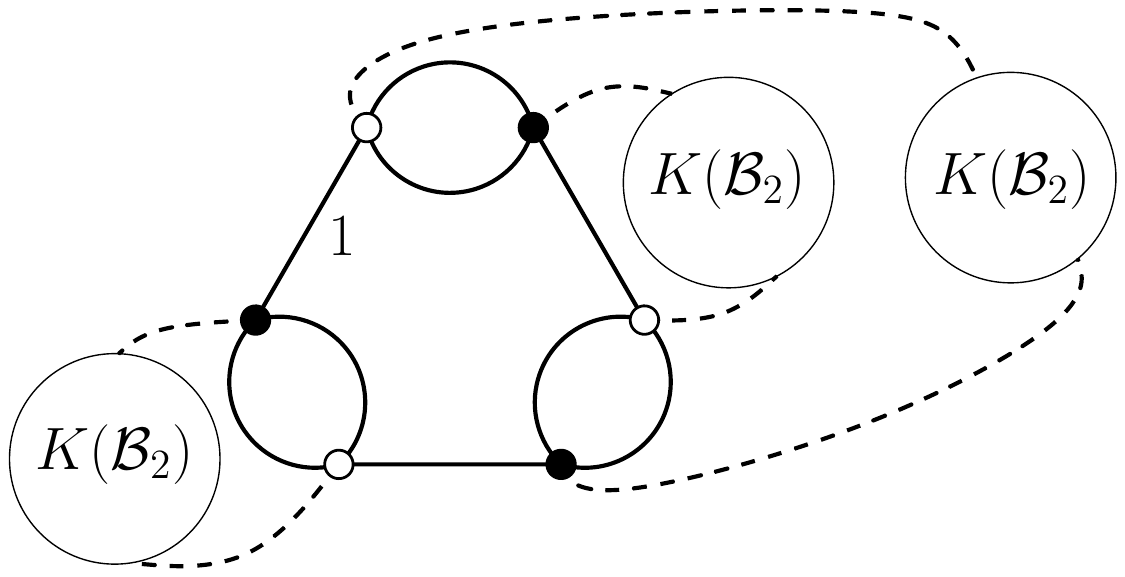} \end{array}\;,
\hspace{2cm}
\begin{array}{c} \includegraphics[scale=.4]{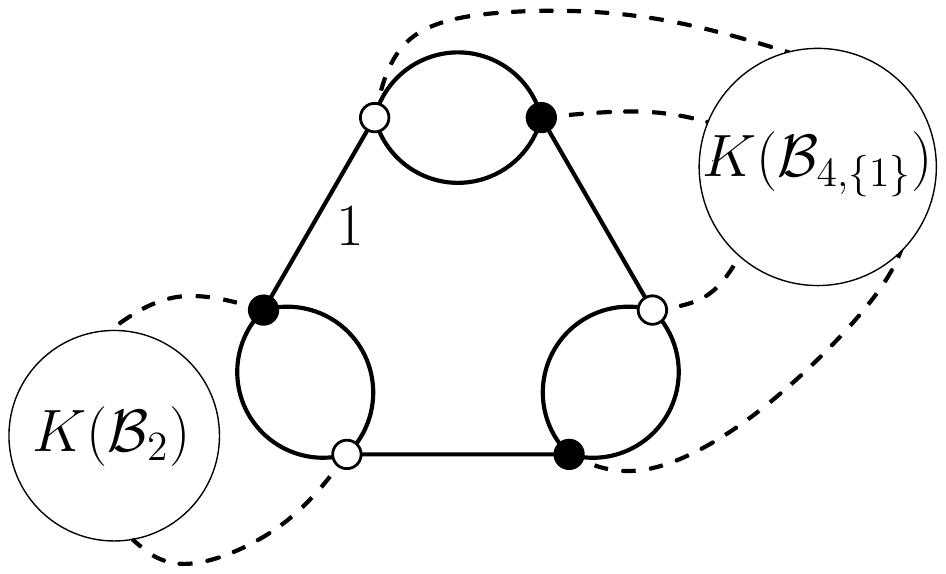} \end{array}\;.
\end{equation*}
For the second bubble $\alpha=2$, which corresponds to the two terms:
\begin{equation*}
\begin{array}{c} \includegraphics[scale=.4]{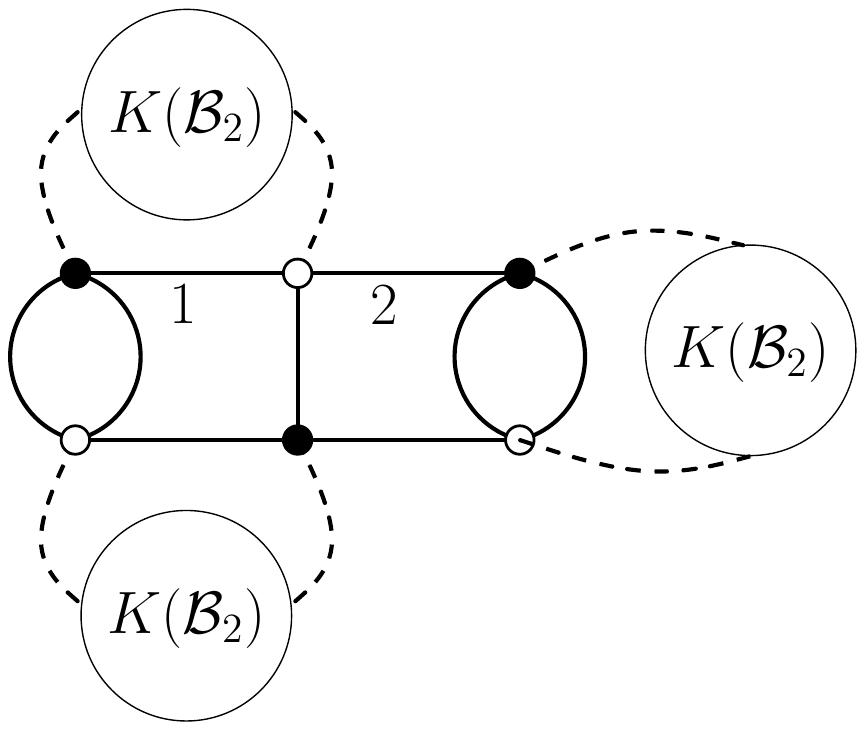} \end{array}\;,
\hspace{2cm}
\begin{array}{c} \includegraphics[scale=.4]{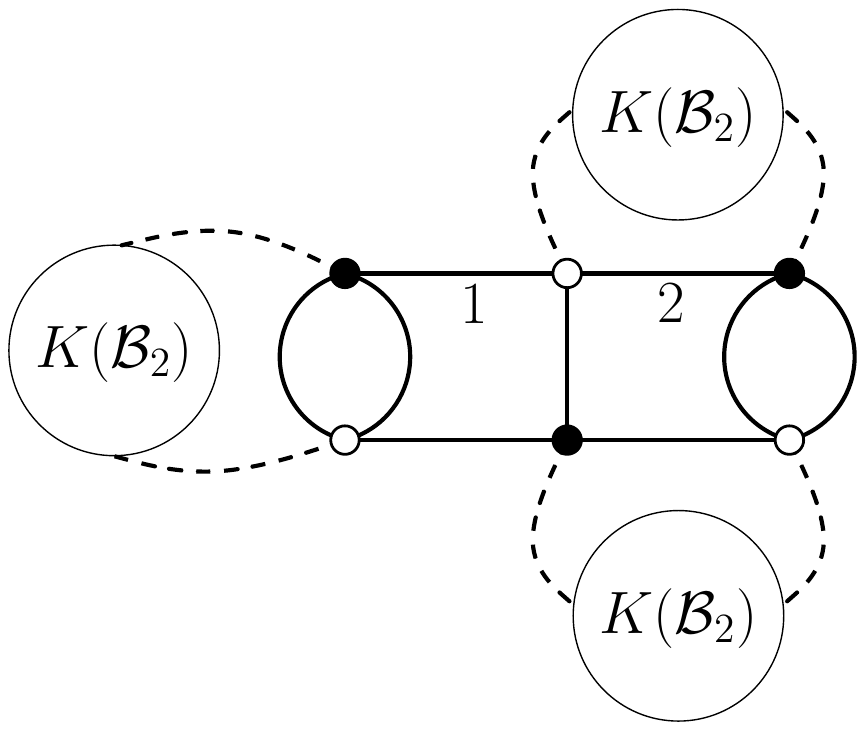} \end{array}\;,
\end{equation*}
and $\alpha'=2$, corresponding to the following two terms,
\begin{equation*}
\begin{array}{c} \includegraphics[scale=.4]{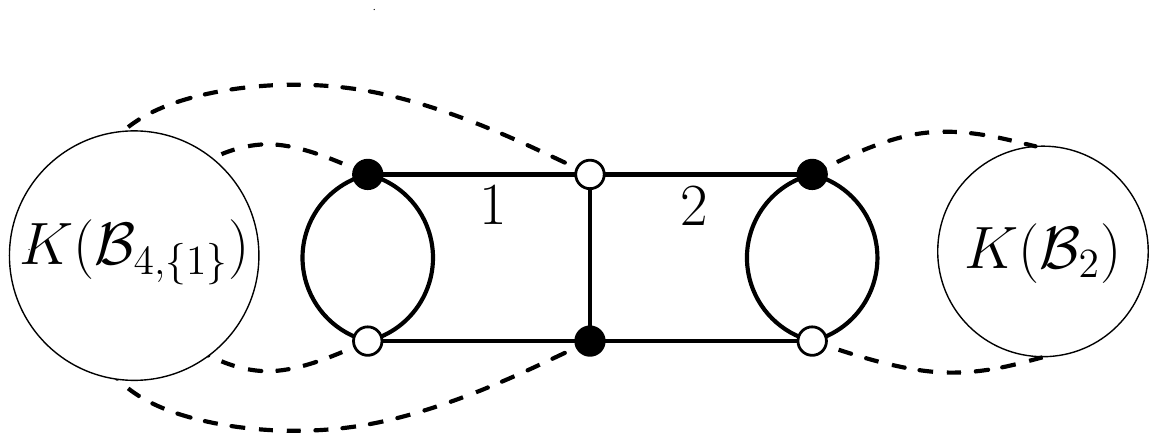} \end{array}\;,
\hspace{2cm}
\begin{array}{c} \includegraphics[scale=.4]{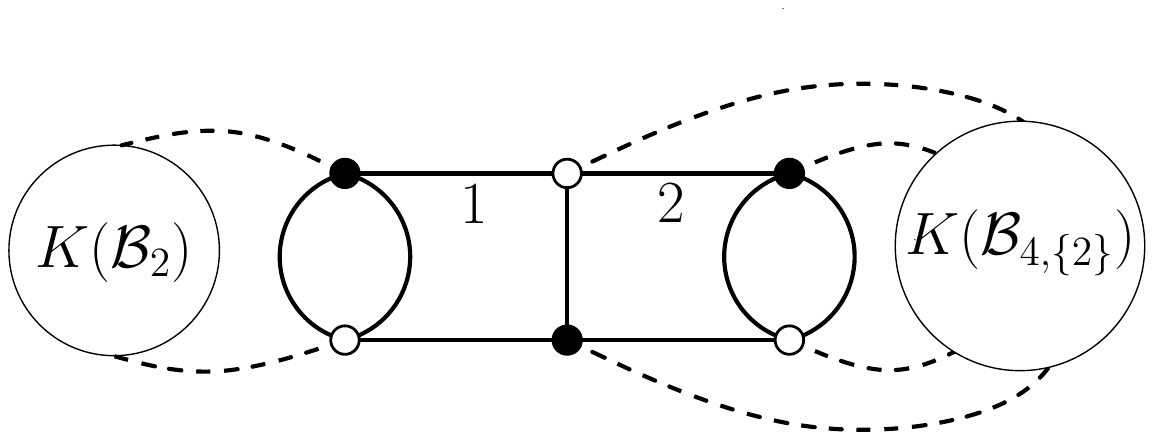} \end{array}\;.
\end{equation*}

\subsection{The Schwinger--Dyson equations at NLO}
\label{ssec:sdnlo}
We are now in position to solve the SD equations at NLO to extract $K_2^{\NLO}, K_{4,\bullet}^{\LO}$.
Plugging the expansions \eqref{2ptFunctionExpansionNLO}, \eqref{4PtExpNLO}, \eqref{BiExpNLO}, \eqref{Bi+ExpNLO}
into the SD equations \eqref{1stSDE} and \eqref{2ndSDE} yields a \emph{linear} system on $K_2^{\NLO}, K_{4,\bullet}^{\LO}$. The first SD equation \eqref{1stSDE} reads:
\begin{equation} \label{1stSDENLO}
\Bigl(-1+\sum_{i\in I} t_i p_i [zT]^{p_i-1}\Bigr) K_2^{\NLO} + \Bigl(\sum_{i\in I} t_i \alpha'_i z^{p_i-1} T^{p_i-2}\Bigr) K_{4,\bullet}^{\LO} = -\sum_{i\in I} t_i \alpha_i z^{p_i-1} T^{p_i}.
\end{equation}
Notice that the coefficient of $K_2^{\NLO}$ is the singular function $C(z,\{t_i\})$ defined in \eqref{SingularityOfS}, which vanishes at criticality.

The second SD equation, \eqref{2ndSDE}, is:
\begin{equation} \label{2ndSDENLO}
\Bigl(1-2\,T + \sum_{i\in I} t_i (p_i+1) z^{p_i-1} T^{p_i}\Bigr) K_2^{\NLO} + \Bigl(-1+\sum_{i\in I} t_i \frac{\beta'_{i,c}}{p_i} [z T]^{p_i-1}\Bigr) K_{4,\bullet}^{\LO}
= -T\,\Bigl(1-T+\sum_{i\in I} t_i \frac{\beta_{i,c}}{p_i} z^{p_i-1}\,T^{p_i}\Bigr)
\end{equation}
It should be noted that, although the numerical coefficients $\beta_{i,c}$ and $\beta'_{i,c}$ depend on both $i$ and $c$, the sum over $i\in I$ ensures that everything is symmetric
with respect to color relabeling, so that the equation is the same for all values of $c\in\{1,\dotsc,D\}$. We then set $c=1$ and denote
$\beta_i\equiv \beta_{i,c=1}$ and $\beta'_i \equiv \beta'_{i,c=1}$.

To simplify this equation, we use $1-T+\sum_{i\in I} t_i z^{p_i-1} T^{p_i}=0$, first to observe that the coefficient of $K_2^{\NLO}$ is again proportional to the singular function $C$:
\begin{equation}
1-2\,T + \sum_{i\in I} t_i (p_i+1) z^{p_i-1} T^{p_i} = -T\,C\;,
\end{equation}
and second to reduce the right hand side:
\begin{equation}
1-T+\sum_{i\in I} t_i \frac{\beta_{i}}{p_i}z^{p_i-1}\,T^{p_i} = -\sum_{i\in I} t_i\Bigl(1-\frac{\beta_i}{p_i}\Bigr) z^{p_i-1} T^{p_i}\;.
\end{equation}

After dividing the equation \eqref{2ndSDENLO} by $T$, \eqref{1stSDENLO} and \eqref{2ndSDENLO} can be cast into a linear system:
\begin{equation}
\begin{pmatrix} -C & \sum_{i\in I} t_i \alpha'_i z^{p_i-1} T^{p_i-2}\\ -C & \frac1{T}\bigl(-1+\sum_{i\in I} t_i \frac{\beta'_i}{p_i} [zT]^{p_i-1}\bigr) \end{pmatrix}
\begin{pmatrix} K_2^{\NLO} \\ K_{4,\bullet}^{\LO} \end{pmatrix} = \begin{pmatrix} -\sum_{i\in I} t_i \alpha_i z^{p_i-1} T^{p_i} \\ \sum_{i\in I} t_i \frac{p_i-\beta_i}{p_i} z^{p_i-1} T^{p_i} \end{pmatrix}\;,
\end{equation}
whose determinant is:
\begin{equation}
  \begin{aligned}
    \det \begin{pmatrix} -C & \sum_{i\in I} t_i \alpha'_i z^{p_i-1} T^{p_i-2}\\ -C & \frac1{T}\bigl(-1+\sum_{i\in I} t_i \frac{\beta'_i}{p_i} [zT]^{p_i-1}\bigr) \end{pmatrix}
    &= \frac{C}{T}\Bigl(1-\sum_{i\in I} t_i \Bigl(\frac{\beta'_i}{p_i}-\alpha'_i\Bigr) [zT]^{p_i-1}\Bigr)\\
    & = \frac{C}{T^2}\Bigl(1-\sum_{i\in I} t_i \Bigl(\frac{\beta'_i}{p_i}-(\alpha'_i+1)\Bigr) z^{p_i-1}T^{p_i}\Bigr)\;.
  \end{aligned}
\end{equation}
Inversion is then straightforward and leads to:
\begin{align} \label{G2NLO}
&K_2^{\NLO} = \frac{T}{C} \frac{\bigl(1-\sum_i t_i\beta'_i[zT]^{p_i-1}/p_i\bigr) \sum_i t_i \alpha_i z^{p_i-1} T^{p_i} - \sum_{i,j\in I} t_i t_j  \alpha'_i (v_j-\beta_j)z^{p_i+v_j-2}T^{p_i+v_j-1}/v_j}
{1-\sum_{i\in I} t_i \bigl(\beta'_i/p_i-(\alpha'_i+1)\bigr) z^{p_i-1} T^{p_i}}\;, \\
\label{G4LO}
&K_{4,\bullet}^{\LO} = \frac{\sum_i t_i \bigl(\beta_i/p_i-(\alpha_i+1)\bigr)\,z^{p_i-1}\,T^{p_i+2}}{1-\sum_{i\in I} t_i \bigl(\beta'_i/p_i-(\alpha'_i+1)\bigr) z^{p_i-1} T^{p_i}}\;.
\end{align}

\emph{The quartic case.} For $I=\{1,\dotsc,D\}$ and $\{\cB_i\}_{i\in I}= \{\cB_{4,\{c\} }\}_{c=1,\dotsc,D}$, the combinatorial coefficients are $\alpha  = \alpha'  =1$
and $\beta_c=\beta'_c$ with $\beta_1=6$ and $\beta_c=4$ for $c\neq1$. Specializing the above formula simply gives:
\begin{align}
K_2^{\NLO}(z) &= \frac1{\sqrt{1-4Dz}}\,\frac{Dz\,T(z)^2}{1-z\,T(z)^2},\crcr
K_{4,{\bullet}}^{\LO}(z) &= \frac{z\,T(z)^4}{1-z\,T(z)^2} \; ,
\end{align}
where we have used the explicit solution $T(z)= (1-\sqrt{1-4Dz})/(2Dz)$, $C=1-2DzT = \sqrt{1-4Dz}$.

The function $K_{4,\bullet}^{\LO}$  has a simple combinatorial interpretation as the chain of quartic bubbles transmitting a single color:
\begin{equation*}
\begin{array}{c} \includegraphics[scale=.4]{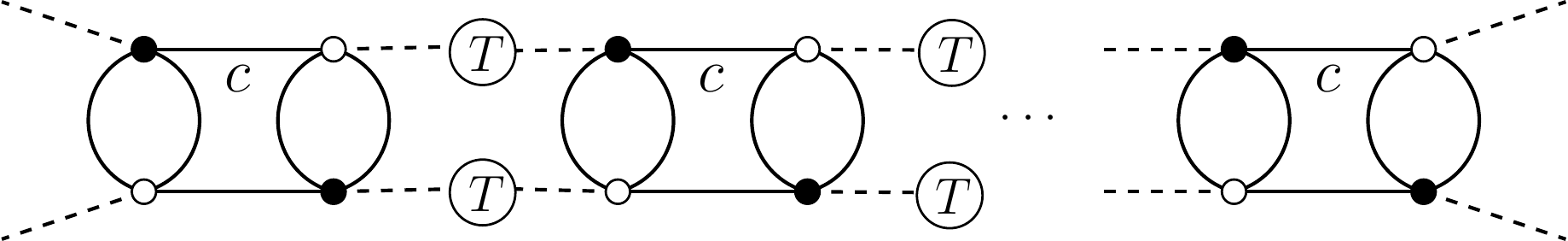} \end{array} = \frac{z\,T^4}{1-z\,T^2} = K_{4,\bullet}^{\LO}(z).
\end{equation*}
This makes the link with the results of \cite{DSQuartic, ColoredSchemes}, where these chains play a prominent role.

\subsection{The order $1/N^D$ in the quartic model} \label{sec:NNLO}

In the context of the quartic model, it is easy enough to go beyond the order $1/N^{D-2}$. Here, we solve the SD equations at the order $1/N^D$ to get the leading order of $K(\cB_{4,\emptyset};N,z)$. This result will be useful to derive the double scaling limit (and will actually be also derived in the section \ref{ssec:proofdsl}, using combinatorics).

Notice that $1/N^D$ is the NNLO when $D\geq 5$. For $D=3$, the NNLO is however $1/N^{2(D-2)}$, while those two orders coincide at $D=4$. Nonetheless, the SD equations hold order by order in the $1/N$ expansion, so $1/N^D$ not being the NNLO is not an issue. The only critical case is $D=4$, for which it would seem that looking at terms of order $1/N^D$ is not enough. It is however not true. The terms which scale like $1/N^4$ at $D=4$ can be unambiguously split into contributions which generalize to $D\neq 4$ with scaling $1/N^D$ and others with scaling $1/N^{2(D-2)}$. If the SD equations are satisfied by some functions of $D$ at the order $1/N^D$ for $D\neq 4$, they will still be satisfied at this order when $D=4$ by the same functions evaluated on $D=4$.

We here need to come back to the equation \eqref{4ptBubbleExpansion}, i.e. the cumulant expansion of the expectation $\langle \tr_{\cB_{4,\{c\}}}(\bT,\overline\bT)\rangle/N$ (fixed color $c$), so as to identify the terms which pop up at the order $1/N^D$. We find directly the term $K(\cB_{4,\emptyset};N,z)/N^D$ which shows that the disconnected boundary graph is now involved. We also find the term $\frac{N^{-1-|\cC|}}{N^{D-2}} K(\cB_{4,\cC};N,z)$ with $c\not\in \cC$, which for $|\cC|=1$ reduces to $\frac{1}{N^D} K(\cB_{4,\{c'\}};N,z)$ with $\cC=\{c'\}$. We can now extend the explicit terms of \eqref{4ptCumExpNLO} to
\begin{multline}
\frac1N \langle \tr_{\cB_{4,\{c\} }}(\bT, \overline{\bT}) \rangle = \Bigl(1+\frac{1}{N^{D-2}}\Bigr)  \bigg[ K \Big(\cB_2 ; N,z,\{t_i\}  \Big) \bigg]^2 + \frac1{N^{D-2}} K \Big(\cB_{4,\{c\} } ; N,z,\{t_i\}  \Big) \\
+ \frac1{N^D}\biggl[K\Bigl(\cB_{4,\emptyset};N,z\Bigr) + \sum_{c'\neq c} K\Bigl(\cB_{4,\{c'\}};N,z\Bigr)+\mathcal{O}(1/N)\biggr].
\end{multline}
Due to the color symmetry, all the terms $K\Bigl(\cB_{4,\{c'\}};N,z\Bigr)$ are equal and their LO is therefore $K^{\LO}_{4,\bullet}(z)$. We introduce $K_2^{\NNLO}(z)$ the restriction of the 2-point function at the order $1/N^D$, and $K^{\LO}_{4,\emptyset}(z)$ the leading order of $K\Bigl(\cB_{4,\emptyset};N,z\Bigr)$. Identifying those terms at order $1/N^D$, we find:
\be
\frac1N \langle \tr_{\cB_{4,\{c\} }}(\bT, \overline{\bT}) \rangle|_{1/N^D} = 2T(z)\,K_2^{\NNLO}(z) + (D-1)K^{\LO}_{4,\bullet}(z) + K^{\LO}_{4,\emptyset}(z).
\ee

Similar cumulant expansions exist for larger melonic bubbles. For the quartic model, we need to extend the expansions \eqref{6ptBubble1NLOExp}, \eqref{6ptBubble2NLOExp} of the two 6-point bubbles. We skip the details which generalize the appendix \ref{app:4pdom} to the order $1/N^D$. At the end of the day, one gets
\begin{multline}
\frac1N \left\langle \begin{array}{c} \includegraphics[scale=.3]{6PointBubble1.pdf} \end{array}
\right\rangle = \Bigl(1+\frac{3}{N^{D-2}}\Bigr) \Bigl[K \Big( \cB_2;N,z,\{t_i\}  \Big)\Bigr]^3 + \frac{3}{N^{D-2}} K \Bigl(\cB_2 ; N,z,\{t_i\} \Bigr)\, K \Bigl( \cB_{4,\bullet } ; N,z,\{t_i\} \Bigr) \\
+\frac{3}{N^D} K \Bigl(\cB_2 ; N,z,\{t_i\} \Bigr) \Bigl[ (D-1)\,K \Bigl( \cB_{4,\bullet } ; N,z,\{t_i\} \Bigr) + K \Bigl(\cB_{4,\emptyset} ; N,z,\{t_i\} \Bigr) + \mathcal{O}(1/N)\Bigr],
\end{multline}
and
\begin{multline}
\frac1N \left\langle \begin{array}{c} \includegraphics[scale=.3]{6PointBubble2.pdf} \end{array}  \right\rangle
= \Bigl(1+\frac{2}{N^{D-2}}+\frac{1}{N^D}\Bigr) \Bigl[K \Big(\cB_2; N,z,\{t_i\} \Big) \Bigr]^3 + \frac{2}{N^{D-2}} K \Big(\cB_2; N,z,\{t_i\} \Big)\, K\Big(\cB_{4,\bullet } ; N,z,\{t_i\}\Big) \\
+ \frac{1}{N^D}K \Big(\cB_2; N,z,\{t_i\} \Big) \Bigl[\bigl(2 (D-1)+D\bigr)\,K \Bigl( \cB_{4,\bullet } ; N,z,\{t_i\} \Bigr) + 3 K \Bigl(\cB_{4,\emptyset} ; N,z,\{t_i\} \Bigr) + \mathcal{O}(1/N)\Bigr].
\end{multline}

This way the SD equation \eqref{1stSDE} gives at order $1/N^D$
\be
\bigl(-1+2Dz\,T(z)\bigr)\,K_2^{\NNLO}(z) + Dz\,K^{\LO}_{4,\emptyset}(z) = -D(D-1)z\,K^{\LO}_{4,\bullet}(z),
\ee
while the SD equation \eqref{2ndSDE} gives
\be
\bigl(1-2T(z)+3Dz\,T(z)^2\bigr)\,K_2^{\NNLO}(z) + \bigr(-1+3Dz\,T(z)\bigr)\,K^{\LO}_{4,\emptyset}(z) = (D-1)\Bigl(1-(3D+1)z\,T(z)\Bigr)\,K^{\LO}_{4,\bullet}(z)-(D-1)z\,T(z)^3.
\ee
Substracting 2 times the first equation to the second, we get the linear system
\be
\begin{pmatrix} -1+2Dz\,T & Dz \\ 1-Dz\,T^2 & -1+Dz\,T \end{pmatrix} \begin{pmatrix} K_2^{\NNLO}(z) \\ K^{\LO}_{4,\emptyset}(z) \end{pmatrix} = \begin{pmatrix} -D(D-1)z K^{\LO}_{4,\bullet}(z) \\ (D-1)K^{\LO}_{4,\bullet}(z)-(D+1)(D-1)z T K^{\LO}_{4,\bullet}(z)-(D-1)z T^3 \end{pmatrix}
\ee
where $K^{\LO}_{4,\bullet}$ acts as a source. The determinant is found to be
\be
\det \begin{pmatrix} -1+2Dz\,T(z) & Dz \\ 1-Dz\,T(z)^2 & -1+Dz\,T(z) \end{pmatrix}  = \biggl[\frac{1-Dz\,T(z)^2}{T(z)}\biggr]^2 = 1-4Dz = 1-z/z_c,
\ee
which leads to
\begin{align}
K_2^{\NNLO}(z) &= \frac{D(D-1)\,z^2\,[T(z)]^5}{(1-Dz\,[T(z)]^2)^2\,(1-z\,[T(z)]^2)} = \frac1{1-4Dz}\,\frac{D(D-1)\,z^2\,[T(z)]^3}{1-z\,[T(z)]^2},
\\
K^{\LO}_{4,\emptyset}(z) &= \frac{D(D-1)\,z^2\,[T(z)]^6}{(1-Dz\,[T(z)]^2)\,(1-z\,[T(z)]^2)} = \frac{1}{\sqrt{1-4Dz}}\,\frac{D(D-1)\,z^2\,[T(z)]^5}{1-z\,[T(z)]^2}.
\end{align}
Whereas $K^{\LO}_{4,\bullet}(z)$ was found to be the (sum of all) monocolored chains, $K^{\LO}_{4,\emptyset}(z)$ instead corresponds to the (sum of all) strictly multicolored chains, i.e. the chains of quartic bubbles where there is at least one change of transmitted color. Indeed, there are obtained by considering all chains with arbitrary colors and then substracting the $D$ monocolored chains,
\be
\frac{Dz\,T^4}{1-Dz\,T^2} - \frac{Dz\,T^4}{1-z\,T^2} = \frac{D(D-1)\,z^2\,T^6}{(1-Dz\,T^2)\,(1-z\,T^2)} = K^{\LO}_{4,\emptyset}(z).
\ee
All those combinatorial interpretations within the quartic model will be clear in the section \ref{ssec:proofdsl} which makes use of a powerful combinatorial representation (which however only works for the quartic model).

\section{The double scaling limit} \label{sec:DS}

\subsection{Double scaling limit in the SD equation}
\label{ssec:DSSD}

We now look for a different scaling limit, which can be reached by sending both $N\to\infty$ and $z$ to its critical value $z_c$, at a certain rate such that a
combination (which we denote $x$), of $z_c-z$ and $N$ (to be determined) is held fixed.
The system of SD equations that we have solved so far is linear in the $1/N$ perturbations: only linear functions arise beyond the LO.
To find an interesting limit, the SD equations must become \emph{non--linear} in the perturbations, at least in the $1/N$ corrections to the 2--point function.

The plan is to expand the 2--point function in a different manner, after which we shall test the viability of this ansatz using the simplest SD equation \eqref{1stSDE}:
\begin{equation*}
  0 = 1 - \frac1N \langle \tr_{\cB_2}(\bT,\overline\bT)\rangle + \frac1N\sum_{i\in I}z^{p_i-1}t_i\langle\tr_{\cB_i}(\bT,\overline\bT)\rangle\;.
\end{equation*}
A sign that non--linear perturbations are possible stems from the fact that the expectation $\langle \tr_{\cB_i}(\bT, \overline{\bT})\rangle$ is of order $p_i$
in the 2--point function:
\begin{equation}
\frac1N \langle \tr_{\cB_i}(\bT, \overline{\bT})\rangle - \text{non--Gaussian parts} = \Bigl(1+\frac{\alpha_i}{N^{D-2}} +
\mathcal{O}(1/N^D)\Bigr)\,\left[K \Big(\cB_2; N,z,\{t_i\} \Big) \right]^{p_i}\;.
\end{equation}
Therefore, if we make the following generic ansatz for the double--scaled 2--point function:
\begin{equation} \label{1/N^a}
K \Big(\cB_2; N,x,\{t_i\} \Big) = T(z_c,\{t_i\}) + \frac1{N^{a}}K^{\DS}(x,\{t_i\}) + \dotsm
\end{equation}
when $N\to\infty, z\to z_c$, we see that a quadratic term in $ K^{\DS}(x,\{t_i\})$  arises from the expansion:
\begin{align}
&\frac1N \langle \tr_{\cB_i}(\bT, \overline{\bT})\rangle - \text{non--Gaussian parts} \crcr
&= T_c^{p_i} + \frac{1}{N^a} p_i T_c^{p_i-1} K^{\DS}(x)
+ \frac1{N^{D-2}} \alpha_i T_c^{p_i} + \frac1{N^{2a}} \frac{p_i(p_i-1)}{2}\,T_c^{p_i-2} \bigl[K^{\DS}(x)\bigr]^2 + \dotsm \; ,
\end{align}
where $T_c\equiv T(z_c,\{t_i\}) $, at order $1/N^{2a}$.
It follows that a nontrivial double scaling equation can be obtained for $a = \frac{D-2}{2}$ (when $N^{-2a}=N^{-(D-2)}$)
since then the expectation \emph{does} have other contributions at this order, which can provide a source for the quadratic term $[K^{\DS}(x)]^2$.

A reasonable ansatz is thus to start with \eqref{1/N^a} with $a=(D-2)/2$, completed up to the order $1/N^{D-2}$ to which we want to solve the SD equation. We introduce the function $K^{\DS,\NLO}(N,x,\{t_i\})$ which contains all the orders between $1/N^{\frac{D-2}{2}}$ and $1/N^{D-2}$ and write:
\begin{equation} \label{DSAnsatz}
K \Big(\cB_2; N,x,\{t_i\} \Big) = T(z_c,\{t_i\}) + \frac{K^{\DS}(x,\{t_i\})}{N^{\frac{D-2}{2}}} + K^{\DS,\NLO}(N,x,\{t_i\}) + \frac{K^{\DS,(D-2)}(x,\{t_i\})}{N^{D-2}} + \mathcal{O}(N^{-D+1}),
\end{equation}
and we expect $K^{\DS}(x)$ to be determined by the SD equation \eqref{1stSDE} at order $1/N^{D-2}$.

In order to determine the fixed coupling $x$ as a function of $z_c-z$ and $N$, the ansatz \eqref{DSAnsatz}
must be compared with the expansion of the 2--point function to NLO, in the vicinity of the critical coupling $z_c$:
\begin{equation*}
K \Big(\cB_2; N,x,\{t_i\} \Big) \underset{z\sim z_c}{=} T(z,\{t_i\}) + \frac1{N^{D-2}\ \sqrt{1-z/z_c}}\,(\dotsb) + \mathcal{O}(N^{-D+1})\;,
\end{equation*}
where the ellipses represent the currently irrelevant contributions to the function $K_2^{\NLO}$ given in \eqref{G2NLO}
which take a finite value when $z\to z_c$.
The singular factor $1/C$ of $K_2^{\NLO}$, which behaves as $1/\sqrt{1-z/z_c}$, has been explicitly factorized.
While $1/N^{D-2}$ goes to zero as $N\to\infty$, the singularity of $1/C$ as $z$ approaches its critical value makes
it possible to find a double scaling limit. Comparing with eq. \eqref{DSAnsatz} it appears that the only way to reach $a=(D-2)/2$ while $z\to z_c$ is to choose:
\begin{equation}
z = z_c - \frac{x}{N^{D-2}},
\end{equation}
where $x$ is held fixed in the double scaling limit $z\to z_c,\, N\to\infty$.

Substituting the ansatz \eqref{DSAnsatz} into the expansion \eqref{BiExpNLO} of the expectation of $\tr_{\cB_i}$ leads to:
\begin{multline}
\frac1N \langle \tr_{\cB_i}(\bT, \overline{\bT}) \rangle =
\Bigl(1+\frac{\alpha_i}{N^{D-2}}\Bigr) \biggl[T(z_c,\{t_i\}) + \frac{K^{\DS}(x,\{t_i\})}{N^{\frac{D-2}{2}}} + K^{\DS,\NLO}(N,x,\{t_i\}) + \frac{K^{\DS,(D-2)}(x,\{t_i\})}{N^{D-2}}\biggr]^{p_i} \\
+ \frac1{N^{D-2}}\,\bigl[T(z_c,\{t_i\})\bigr]^{p_i-2}\,K^{\LO}_{4,\bullet}\Big{|}_{ z\to z_c } + \dotsm
\end{multline}

Importantly, $K_{4,\bullet}^{\LO}$ is finite at criticality, since it does not have the singular factor $1/C$. Therefore, it can be safely evaluated
at $z_c$ (it already comes at order $1/N^{D-2}$ so there is no need to expand around $z_c$ which would generate $1/N$ corrections).
At this stage, however, it is not clear that the ellipses in the above equation contain only terms that can be neglected in the double scaling limit.
It could indeed be the case that corrections to the 4--point function, as well as non--Gaussian contributions of higher orders, diverge at criticality,
and furthermore, that they come with sufficiently many powers of $1/\sqrt{1-z/z_c}$ to counter--balance their $1/N$ suppression. In this way,
those terms could contribute in the double scaling limit. Solving this issue requires a precise analysis that will be performed in the next section, \ref{ssec:proofdsl}.

For the time being, we simply give the result of the analysis, in the context of the \emph{quartic} model. We need to know the behavior of the various 4-point cumulants in the eq. \eqref{4ptBubbleExpansion}. In the standard $1/N$--expansion:
\begin{align}\label{eq:4cumulz}
& K \Big(\cB_{4,\{c\} } ; N,z   \Big) = \frac{z\,T(z)^4}{1-z\,T(z)^2} + \mathcal{O}\left( \frac{1}{N^{D-2}} \right) \; ,\crcr
& K \Big(\cB_{4, \emptyset } ; N,z\Big) =
\frac{1}{\sqrt{1-4Dz}}\,\frac{D(D-1)\,z^2\,[T(z)]^5}{1-z\,[T(z)]^2} + \mathcal{O}\left( \frac{1}{N^{D-2}} \right) \; ,\crcr
&  K \Big(\cB_{4, \cC } ; N,z\Big)\Big{|}_{|\cC|\ge 2} =  \mathcal{O}\left( \frac{1}{N^{D-2}} \right) \; .
\end{align}
Both the first and second lines have been obtained by solving the SD equations in the section \ref{sec:SDNLO} (the estimates on their rests follow combinatorial arguments). The third line is a scaling argument. In the following section \ref{ssec:proofdsl}, those results will actually be re-derived on the way to showing that the cumulants in the double scaling limit become:
\begin{align}\label{eq:4cumulx}
&   K \Big( \cB_{4,\{c\} }; N,x  \Big) = f^{(0)}(\cB_{4, \{c\} };N,x)  +  N^{- \frac{D-2}{2}} f^{ (-\frac{D-2}{2} ) }(\cB_{4, \{c\} };N,x) \; ,\crcr
&  K \Big( \cB_{4,\emptyset}; N,x  \Big) = N^{\frac{D-2}{2}} f^{(\frac{D-2}{2})}(\cB_{4,\emptyset};N,x) \;, \crcr
&  K \Big( \cB_{4,\cC}; N,x  \Big) = \mathcal{O} \Bigl( \frac{1}{N^{D-2}} \Bigr) \; .
\end{align}
Importantly, the functions $f^{(q)}(\cB;N,x)$ are \emph{bounded for all $N$} (have a finite value as $N\to\infty$). Note that the sub--leading corrections to $ K \Big( \cB_{4,\{ c\} }; N,x  \Big) $ contribute at order $N^{-\frac{D-2}{2}}$, that is sooner than expected in the $1/N$--expansion.

As a result, $K \Big( \cB_{4,\emptyset}; N,x  \Big) $ is enhanced in the double scaling regime by a factor $N^{\frac{D-2}{2}}$. This can be understood already from the $1/N$--expansion \eqref{eq:4cumulz}. Its LO indeed contains the singular factor $1/C(z)=1/\sqrt{1-4Dz}$ which as a function of $x$ and $N$ becomes
\begin{equation*}
\frac1{C(z(N,x))} = \frac{N^{\frac{D-2}{2}}}{\sqrt{x}}\;.
\end{equation*}
The divergence of the LO of $K \Big( \cB_{4,\emptyset}; N,x  \Big) $ as $z$ approaches $z_c$ is traded for a large $N$ enhancement.

Therefore the non--Gaussian contributions to the expectation of $\tr_{\cB_{4,\{c\}}}$ read:
\begin{multline}
\frac1{N^{D-2}} K \Big( \cB_{4,\{c\} }; N,x  \Big) + \frac1{N^D} K \Big( \cB_{4,\emptyset}; N,x  \Big) +\mathcal{O}(N^{-D-1}) \\
= \frac1{N^{D-2}} \Bigl(\underbrace{f^{(0)}(\cB_{4, \{c\} };N,x)}_{\mathcal{O}(1)} + \underbrace{\frac{N^{D-2} N^{\frac{D-2}{2}}}{N^{D}} f^{(\frac{D-2}{2})}(\cB_{4,\emptyset};N,x)}_{\mathcal{O}\bigl(N^{\frac{D-6}{2}}\bigr)} + \mathcal{O}(1/N)\Bigr)\;.
\end{multline}

From this analysis, we conclude that in the double scaling limit, the dominant 4-point cumulant in the quartic model is dominated by the leading order of:
\begin{itemize}
  \item[--] $f^{(0)}(\cB_{4,\{c\}};N,x)$ when $D<6$,
  \item[--] both $f^{(0)}(\cB_{4,\{c\}};N,x)$ and $f^{(\frac{D-2}{2})}(\cB_{4,\emptyset};N,x)$ when $D=6$,
  \item[--] $f^{(\frac{D-2}{2})}(\cB_{4,\emptyset};N,x)$ when $D>6$.
\end{itemize}
This way, we recover the $D=6$ barrier found in \cite{DSQuartic, ColoredSchemes}. Since the doubly scaled 2-point function is not-summable for $D\geq 6$, we hereafter focus on the case $D<6$. Moreover, we will argue in the section \ref{ssec:FromQuarticToGeneric} that this conclusion holds for any model with melonic bubbles in the action (symmetrized on the colors) and not only the quartic model.

Plugging the expansion \eqref{DSAnsatz} for the 2-point function into the SD equation \eqref{1stSDE} gives (some terms contributing beyond $1/N^{D-2}$ have already been neglected):
\begin{multline}
1-\biggl(T_c + \frac{K^{\DS}(x,\{t_i\})}{N^{\frac{D-2}{2}}} + K^{\DS,\NLO}(N,x,\{t_i\}) + \frac{K^{\DS,(D-2)}(x,\{t_i\})}{N^{D-2}}\biggr) \\
+ \sum_{i\in I} t_i \Bigl(z_c-\frac{x}{N^{D-2}}\Bigr)^{p_i-1}
\biggl(T_c^{p_i} + \frac1{N^{\frac{D-2}{2}}}p_i T_c^{p_i-1} K^{\DS}(x,\{t_i\}) + p_i T_c^{p_i-1} K^{\DS,\NLO}(N,x,\{t_i\}) \\
+ \frac1{N^{D-2}} \Bigl(\alpha_i T_c^{p_i} + p_i T_c^{p_i-1} K^{\DS,(D-2)}(x,\{t_i\}) + \frac{p_i(p_i-1)}{2} T_c^{p_i-2} \bigl[K^{\DS}(x,\{t_i\})\bigr]^2 + \alpha'_i T_c^{p_i-2} K^{\LO}_{4,\bullet}(z_c, \{t_i\})\Bigr) \biggr) = 0\;,
\end{multline}
which we shall evaluate order--by--order in the $1/N$--expansion up to $1/N^{D-2}$. The leading order equation is trivially satisfied as the evaluation of $1-T+\sum_i t_i z^{p_i-1}T^{p_i}$ at criticality. At order $N^{-\frac{D-2}{2}}$, we get
\begin{equation}
  \Bigl(-1 + \sum_{i\in I} t_i p_i [z_c\,T_c]^{p_i-1} \Bigr)\,K^{\DS}(x, \{t_i\}) = 0\;.
\end{equation}
The quantity into brackets is the function $C$ evaluated at criticality and so vanishes. Thus, it provides no new information. Then we need to take care of all the terms between the order $1/N^{\frac{D-2}{2}}$ and $1/N^{D-2}$. Because they are \emph{strictly bounded} by $N^{-(D-2)/2}$, the functions $K^{\DS,\NLO}$ and $K^{\DS,(D-2)}$ only appear \emph{linearly} in the above expansion (their square would only contribute beyond $1/N^{D-2}$). Therefore, they both come in factor of the first derivative of the LO equation, i.e. the function $C(z,\{t_i\})$, evaluated at criticality which is zero.

Remarkably, we are left at order $1/N^{D-2}$ with an equation on the function $K^{\DS}$ only:
\begin{multline} \label{2PtDoublyScaled}
\Bigl(\sum_{i\in I} t_i \frac{p_i(p_i-1)}{2} z_c^{p_i-1} T_c^{p_i-2}\Bigr) [K^{\DS}(x,\{t_i\})]^2 - x \Bigl(\sum_{i\in I} t_i (p_i-1) z_c^{p_i-2} T_c^{p_i}\Bigr)\\
+ \sum_{i\in I} t_i z_c^{p_i-1} \bigl(\alpha_i T_c^{p_i} + \alpha'_i T_c^{p_i-2} K^{\LO}_{4,\bullet}(z_c, \{t_i\})\Bigr) = 0\;,
\end{multline}
which directly gives its expression as a function of $x,\{t_i\}$. This is the main result of the article.

\emph{The quartic case.} Specializing this calculation to the quartic case directly reproduces the doubly scaled 2--point function found in \cite{DSQuartic}:
\begin{equation}
[K^{\DS}(x)]^2 = \Bigl(\frac{x}{z_c}-1\Bigr)\,T_c^2 - K^{\LO}_{4,\bullet}(z_c)\;.
\end{equation}
Using $z_c=1/4D$, we find $T_c=2$ and $K^{\LO}_{4,\bullet}(z_c) = 4/(D-1)$ and finally:
\begin{equation}
K^{\DS}(x) = 4\sqrt{D}\ \sqrt{x - \frac{1}{4\,(D-1)}}\;.
\end{equation}



\subsection{Proof of the double scaling limit}
\label{ssec:proofdsl}

Our task in this section is to establish the equation \eqref{eq:4cumulx} (we will actually derive \eqref{eq:4cumulz} on our way).

At the end of the day, we do not know how to derive these equalities solely via the SD equations. We shall utilize in this section a different analysis, devised in \cite{beyondpert} (and utilized first in \cite{DSQuartic} for the 2--point function), to probe these sub--leading terms.

In this proof, we shall restrict to the quartic model described earlier and use a universality argument, presented in the next section, to extend the proof to the generic model.  Our main reason for doing so is that the universality argument is succinct yet powerful. As one will see, the following analysis is quite involved and tailored for the quartic model only. To prove these results for generic models directly is an arduous task, deserving its own paper.

The quartic model is defined by the action and the generating function:
\begin{align*}
&  S_{(\rm{quart} )}(\bT,\overline{\bT}) = \tr_{\cB_2}(\bT,\overline{\bT}) - \sum_{c=1}^D \frac{z}{2} \,\tr_{ \cB_{  4,\{c\} } }(\bT,\overline{\bT})\;, \crcr
&  Z_{(\rm{quart} )}(\bJ, \overline \bJ) = \int
     \left( \prod_{\vec a} N^{D-1} \frac{\extd \bT_{\vec a}\, \extd
    \overline \bT_{\vec a}} {2\pi i}\right) e^{-N^{D-1}S_{(\rm{quart} )}(\bT,\overline \bT) + \tr_{\cB_2}(\bT,  \overline \bJ) + \tr_{\cB_2}(\bJ,  \overline \bT)} \; .
\end{align*}

\begin{figure}
\subfloat[A Feynman graph with two external legs.]{\begin{tabular}{c}\includegraphics[width=5cm]{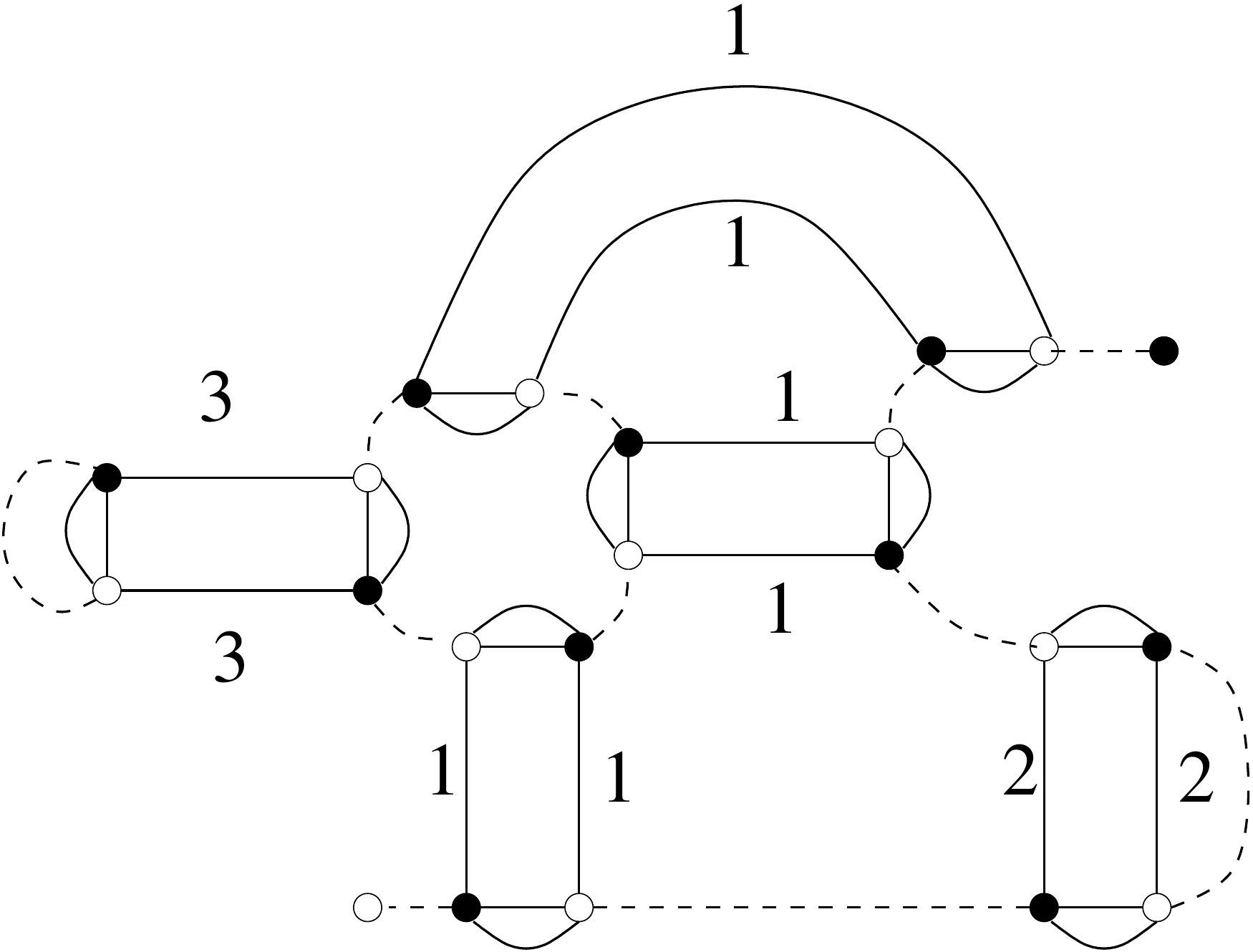}\end{tabular} \label{fig:Feynman}}
\hspace{1.5cm}
\subfloat[The associated edge colored map with a (dashed) cilium.]{\begin{tabular}{c}\includegraphics[width = 3cm]{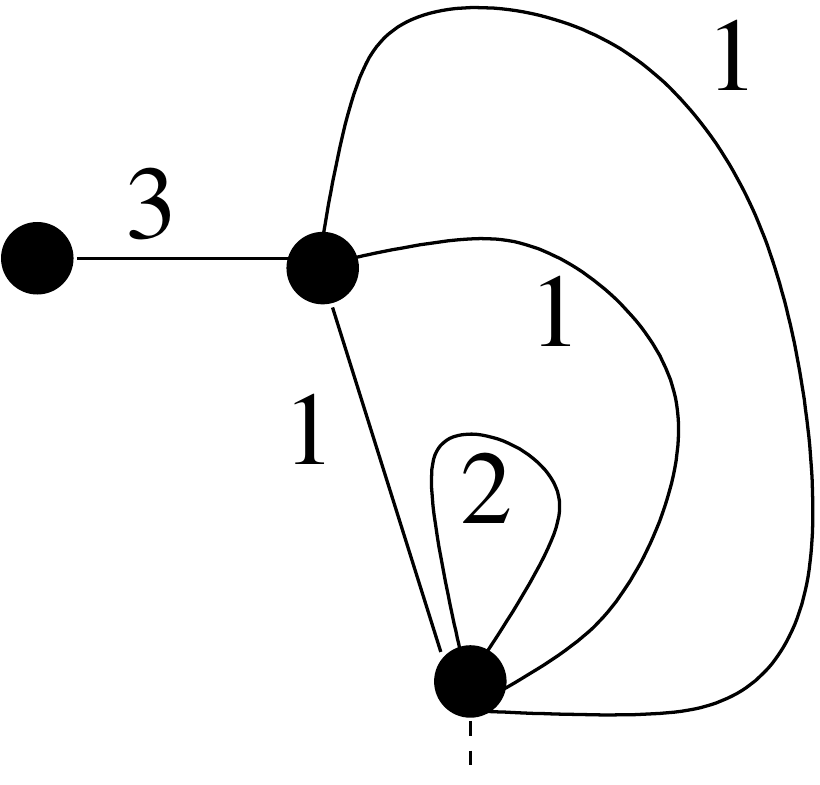}\end{tabular} \label{fig:edgecolmap}}
\caption{ \label{fig:Feytomap} An example of the bijection between Feynman graphs with $2p$ external legs and edge--colored maps with $p$ cilia.}
\end{figure}
The $2p$-point cumulants are sums over connected $(D+1)$-colored graphs with $2p$ edges of color 0 adjacent to $2p$ vertices of degree 1 called the external vertices. Moreover the connected components of the sub--graphs with colors $1,\dotsc,D$ are bubbles $\cB_{4,\{c\}}$.

We recall that the amplitude of such a Feynman graph $\cG$ is easily evaluated: each subgraph $\cB_{4,\{c\}}$ brings
a $N^{D-1}$ scaling factor and a trace-invariant operator. Each edge of color $0$ brings a $\frac{1}{N^{D-1}} \prod_{c=1}^D \delta_{a_c b_c} $ factor. It follows that the indices are identified along the faces of color $0c$ of $\cG$.
The indices corresponding to the internal faces are summed and bring a factor $N$ each. The indices corresponding to the external faces of $\cG$ reconstitute the trace-invariant operator associated to $\partial \cG$, $\delta^{\partial \cG}_{ \vec a^v, \vec b^v}$. Denoting $E^0(\cG)$ the number of edges of color $0$ of $\cG$ (including the external edges), $B(\cG)$ the number of subgraphs of colors $\{1,\dotsc,D\}$ of $\cG$, and $F^{0c}_{\rm{int}}(\cG)$ the number of internal faces of color $0c$ of $\cG$, the $2p$-point cumulant writes:
\begin{equation}
  W^{(2p)}_{\vec a^1 \dotsc \vec a^p, \vec b^1 \dotsc \vec b^p}\Big(N, z,\{t_i\} \Big) = \sum_{n\ge 0 }
   \frac{1}{n!} z^n \sum_{ \genfrac{}{}{0pt}{}{\cG, p(\partial \cG)=p}{B(\cG)=n} } \frac{N^{B(\cG)(D-1)} N^{ \sum_{c=1}^D F^{0c}_{\rm{int}}(\cG) }}{N^{(D-1)E^0(\cG)} } \;\; \delta^{\partial \cG}_{ \vec a^v, \vec b^v} \; ,
\end{equation}
where the sum runs over graphs $\cG$ with labeled sub--graphs $\cB_{4,\{c\}}$. The contribution of an invariant $\cB$
to the $2p$-point cumulant is obtained by restricting to graphs whose boundary is $\cB$, that is $\partial \cG = \cB$:
\begin{equation}\label{eq:amplitudeeee}
 W\Big( \cB; N,z \Big) = \sum_{n\ge 0 }
   \frac{1}{n!} z^n \sum_{ \genfrac{}{}{0pt}{}{\cG,\partial \cG = \cB }{B(\cG)=n; } } \frac{N^{B(\cG)(D-1)} N^{ \sum_{c=1}^D  F^{0c}_{\rm{int}}(\cG) }}{N^{(D-1)E^0(\cG)} } \; ,
\end{equation}

The graphs and amplitudes of this model can be recast in terms of an \emph{intermediate field representation}, the details of which can be found in \cite{beyondpert}. Although somewhat lengthy to introduce, this representation clarifies greatly the $1/N$--expansion. The intermediate field representation can be obtained by introducing Hubbard-Stratonovich intermediate fields, integrating out $\bT, \overline{\bT}$ and deriving the new Feynman rules of the theory. Here we do not need this full machinery, but we will take advantage of the fact that the graphs of this intermediate field representation are in a one-to-one correspondence with the Feynman graphs of the tensor model.

\subsubsection{The intermediate field representation}

We will call \emph{effective graphs} the graphs of the intermediate field representation. The mapping to the Feynman graphs of our tensor model is exemplified in the figure \ref{fig:edgecolmap}. The effective graphs are simply obtained from the regular edge--colored graphs by contracting all the edges of color $0$ and all the edges of color $c'\neq c$ in each $\cB_{4,\{c\}}$ to constitute \emph{effective vertices} (e--vertices for short), while associating an \emph{effective edge} of color $c$ (e--edges of color $c$ for short) to the couple of edges of color $c$ in each $\cB_{4,\{c\}}$.

\begin{figure}[htb]
  \centering
  \includegraphics[scale = 1]{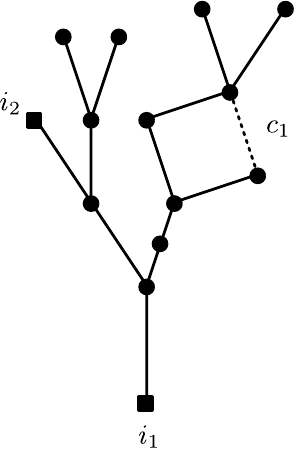}
  \caption{\label{fig:planar} An effective graph with two marked vertices $\{i_1,i_2\}$.
   The e--edges have a color, as stressed out by the dashed edge with color $c_1$.}
\end{figure}

The external edges of color $0$ will then be partitioned into pairs associated to some of the e--vertices. We decorate those e--vertices by a mark, or a \emph{cilium}, to signal such a couple. An e--vertex can have at most one cilium. This mapping is obviously bijective. A typical example of a contribution to the 4-point cumulant is presented in figure \ref{fig:planar}.

We use boxes to represent the marked vertices when there is no ambiguity as for the cilium positions around the vertices. Note indeed that, in this intermediate field representation, the order of the e--edges adjacent to an e--vertex is specified. It means that the effective graphs are in fact combinatorial maps (i.e. graphs with ascribed order of the edges at a vertex) with edges colored $\{1,\dotsc,D\}$. We call a \emph{corner} the piece of an e--vertex comprised between two consecutive e--edges. Note that a cilium is incident to two corners (or to a unique corner, if the graph has one ciliated vertex and no edges). We will denote the maps thus obtained by $\cM$.

Every $\cM$ has $D$ canonical sub--maps $\cM_c$ obtained by deleting all the edges of color $c'\neq c$ in $\cM$. All the vertices of $\cM$ belong to $\cM_c$. The sub--maps $\cM_c$ have a well defined notion of \emph{faces}. They fall in two categories: the \emph{internal faces} of $\cM_c$ are the circuits obtained by going along the e--edges (of color $c$) and along the corners of the e--vertices of $\cM_c$, while the \emph{external faces} of $\cM_c$ are the open paths obtained by going along the corners and the e--edges (of color $c$) of $\cM_c$ from one cilium to another. By convention, all the faces are oriented clockwise. We define the \emph{faces of color $c$} of $\cM$ as the faces of $\cM_c$. Note that some of the faces can be reduced to a single corner on an isolated vertex.

All the elements present in the formula \eqref{eq:amplitudeeee} are faithfully represented within effective graphs:
\begin{itemize}
 \item each subgraph $\cB_{4,\{c\}}$ of $\cG$ corresponds to an e--edge of color $c$ of $\cM$,
 \item each edge of color $0$ of $\cG$ corresponds to a corner of $\cM$,
 \item each (internal or external) face of color $0c$ of $\cG$ corresponds to an (internal or external)
 face of color $c$ of $\cM$.
\end{itemize}

The boundary graph $\partial\cG$ can be reconstructed from $\cM$. To do so, one draws a black and a white vertex for every cilium of $\cM$ and for each external face of color $c$ of $\cM$ going from a source cilium to a target cilium, one connects the white vertex corresponding to the source cilium with the black vertex of the target cilium by an edge of color $c$. We denote $\partial \cM$ the boundary graph of the effective map $\cM$.

Let us denote $E^c(\cM)$, $ V(\cM)$ and $ F^{c}_{\rm { int} } (\cM)$ the numbers of e--edges of color $c$, e--vertices and \emph{internal} faces of color $c$ of $\cM$. Furthermore, let us define $E(\cM) = \sum_{c=1}^D E^c(\cM)$ and $\sum_{c=1}^DF^{c}_{\rm { int} } (\cM) =F_{\rm{int}}(\cM)$. The numbers of corners of $\cM$ is $p(\cB) + 2  E (\cM)$.  The number of cilia of $\cM$ is half the number of vertices of $\cB=\partial \cM$. The equation \eqref{eq:amplitudeeee} becomes:
\be
\begin{aligned}
  W\Big( \cB; N,z \Big) &= \sum_{v\ge 0 }
   \frac{1}{v!} \sum_{ \genfrac{}{}{0pt}{}{\cM, \partial \cM = \cB} { V(\cM) =v } }  z^{ E(\cM)}  \frac{N^{ E (\cM)(D-1)}
    N^{   F _{\rm{int}}(\cM) }}{N^{(D-1)[p(\cB) + 2  E (\cM)]} } \\
    &= N^{-(D-1) p(\cB)} \sum_{v\ge 0 }
   \frac{1}{v!} \sum_{ \genfrac{}{}{0pt}{}{\cM, \partial \cM = \cB} { V(\cM) =v } }  z^{ E(\cM)}   N^{ - E (\cM)(D-1) +  F _{\rm{int}}(\cM) }
\end{aligned}
\ee
where the sum runs over edge--colored maps with labeled vertices such that $\partial\cM = \cB$.

\begin{lemma}\label{lem:bun}
 We have the bound:
 \begin{equation}
   - E (\cM)(D-1) +  F _{\rm{int}} (\cM)  \le D - (D-1) p(\cB) -\rho(\cB)  - (D-2) \Big[   E (\cM) -V(\cM) +1 \Big] \; .
 \end{equation}
\end{lemma}

{\bf Proof.} The external faces are open paths. They can be represented as cycles by adding to our drawings \emph{external strands}. For each external face of color $c$, we connect its source and target cilia by an external strand of color $c$ (which can for instance be represented as a dashed edge). By convention we orient the external strands form the target cilium to the source cilium. The external faces now become cycles, obtained by going between the cilia
along the corners and e--edges of the graph and closing the path into a cycle by following the external strands.

Note that the external strands encode the boundary $\partial \cM$ of $\cM$. For each cilium we draw a black and a white vertex and for each strand of color $c$ we connect the white vertex of its target cilium with the black vertex of its source cilium. Henceforth we use this as the definition of the boundary graph of a map.

We denote by $F^{c}_{\rm{ext}} (  \cM) $ the number of external faces of color $c$ of the map $\cM$, and $\sum_{c=1}^D F^{c}_{\rm{ext}} (  \cM) =  F _{\rm{ext}} (  \cM) $.
Initially, $\cM$ has exactly $Dp(\cB)$ external faces, hence a total of:
\[
 Dp(\cB) + F_{\rm{int}} (\cM)  \;,
\]
faces either internal or external. We are interested in finding a bound on this total number of faces.

An e--edge belongs to either one or two faces (internal or external). By deleting an e--edge and merging the corners of the two e--vertices to which it is hooked, the total number of faces of the map can not increase by more than $1$. Remark that, while in the initial map every external face contained exactly one external strand, by deleting an edge we can create external faces containing several external strands. However, as the deletion does not affect the connectivity of the external strands, the latter still encode the boundary of the initial map $\partial\cM$.

We choose a spanning tree in $\cM$ and iteratively delete the e--edges in its complement. We denote the map obtained at the end of this procedure (which is a tree decorated with external strands)
by $\cM^{(0)}$ and we have:
\begin{equation*}
 Dp(\cB) +   F_{\rm{int}} (\cM)  \le   F _{\rm{ext}} (\cM^{(0)}) +  F _{\rm{int}} (\cM^{(0)}) +
 \Big[  E (\cM) -V(\cM) +1 \Big] \; ,
\end{equation*}
and $\partial \cM = \partial \cM^{(0)}$.

Starting from $\cM^{(0)}$ we build the maps $\cM^{(s)}$ obtained by eliminating one by one the e--vertices of $\cM$ of coordination one. Choose a univalent e--vertex (hence hooked to a unique e--edge, say of color $c$) in $\cM^{(s)}$ having no cilium. The map $\cM^{(s+1)}$ is obtained by deleting this e--vertex and the e--edge it is adjacent to. The boundary graph is unchanged by this procedure, $ \rho(\cM^{(s)}) = \rho(\cM^{(s+1)}) $, and $D-1$ internal faces are deleted (all the faces of color $c'\neq c$ contained in the e--vertex), hence
\begin{equation}
 F _{\rm{ext}} (\cM^{(s)}) +  F _{\rm{int}} (\cM^{(s)}) + \rho( \partial \cM^{(s)}  ) =
 F _{\rm{ext}} (\cM^{(s+1)}) +  F _{\rm{int}} (\cM^{(s+1)}) + \rho( \partial \cM^{(s+1)}  ) + (D-1) \; .
\end{equation}

If the univalent e--vertex  (hooked to a unique e--edge, say of color $c$) on $\cM^{(s)}$ is ciliated then there are $D$ incoming and $D$ outgoing external strands
at this cilium. Let us denote the cilium by $i$. We build first the map $ \tilde \cM^{(s)}$ by deleting $i$ and all the external strands
which \emph{start and end} at $i$, and reconnecting the remaining external strands incident at $i$ respecting the colors.
The map $\cM^{(s+1)}$ is then obtained
from $ \tilde \cM^{(s)}$ by deleting the resulting univalent e--vertex and the e--edge to which it is hooked.

Going from $\cM^{(s)}$ to $\tilde \cM^{(s)}$ changes the boundary graph: $\partial \cM^{(s)} \neq\partial \tilde\cM^{(s)}  $,
while going from $\tilde \cM^{(s)}$ to $\cM^{(s+1)}$ preserves it.
There are several cases:
\begin{itemize}
 \item The black and white vertices associated to the cilium $i$ in $\partial \cM^{(s)}$ belong to two different connected components of $\partial \cM^{(s)} $. Then the number of connected components of the boundary graph decreases by $1$, $ \rho(\cM^{(s)})=\rho(\tilde \cM^{(s)})+1$. At the same time, a new face is created for every color $c'\neq c$ (this new face is contained in the e--vertex of interest). For the color $c$ (of the e--edge hooked to the e--vertex), at most one face can be deleted, thus:
 \begin{equation}
   F _{\rm{ext}} (\cM^{(s)}) +  F _{\rm{int}} (\cM^{(s)}) + \rho( \partial \cM^{(s)}  )
   \le F _{\rm{ext}} (\tilde \cM^{(s)}) +  F _{\rm{int}} ( \tilde \cM^{(s)}) + \rho( \tilde \partial \cM^{(s)}  ) +1 -(D-1) +1 \; .
 \end{equation}
 \item The black and white vertex associated to the cilium $i$ in $\partial \cM^{(s)}$ belong to the same connected component of
 $\partial \cM^{(s)} $, but not all of the external strands starting at $i$ end at $i$. Then the number of connected components
 of the boundary graph can only increase, $ \rho(\cM^{(s)}) \le \rho(\tilde \cM^{(s)})$. The faces of color $c'\neq c$ containing the
 external strands starting and ending at $i$ become internal. The other faces of color $c'\neq c$ remain external.
 If the external strand of color $c$ starting at $i$ ends also at $i$, the face containing it survives. If not, the number of faces
 of color $i$ can at most decrease by $1$. Thus
 \begin{equation}
   F _{\rm{ext}} (\cM^{(s)}) +  F _{\rm{int}} (\cM^{(s)}) + \rho( \partial \cM^{(s)}  )
   \le F _{\rm{ext}} (\tilde \cM^{(s)}) +  F _{\rm{int}} ( \tilde \cM^{(s)}) + \rho( \tilde \partial \cM^{(s)}  ) + 1
 \end{equation}
 \item all the external strands starting at $i$ end at $i$. Then the number of connected components
 of the boundary graph decreases by $1$, $ \rho(\cM^{(s)}) = \rho(\tilde \cM^{(s)}) + 1$, but none of the faces
 is affected, hence:
 \begin{equation}
   F _{\rm{ext}} (\cM^{(s)}) +  F _{\rm{int}} (\cM^{(s)}) + \rho( \partial \cM^{(s)}  )
   \le F _{\rm{ext}} (\tilde \cM^{(s)}) +  F _{\rm{int}} ( \tilde \cM^{(s)}) + \rho( \tilde \partial \cM^{(s)}  ) + 1 \; .
 \end{equation}
\end{itemize}
 When going from $\tilde \cM^{(s)}$ to $\cM^{(s+1)}$,  $D-1$ faces are deleted. Taking into account that $D\ge 3$,
\begin{equation}
 F _{\rm{ext}} (\cM^{(s)}) +  F _{\rm{int}} (\cM^{(s)}) +  \rho( \partial \cM^{(s)}  )
 \le  F_{\rm{ext}} (\cM^{(s+1)}) +   F _{\rm{int}} (\cM^{(s+1)}) +  \rho( \partial \cM^{(s+1)} ) +  D \; .
\end{equation}
  Eliminating all the e--vertices we obtain the map $\cM^{(s_f)}$ with $s_f = V(\cM)-1$, having only one e--vertex and exactly $D$ faces. The final e--vertex can be ciliated or not,
hence we obtain the bound
\begin{align*}
 F _{\rm{ext}} (\cM^{(0)}) + F _{\rm{int}} (\cM^{(0)}) \le
 D +  \rho( \partial \cM^{(s_f)}  ) - \rho(\partial\cM^{(0)}) + (D-1) (V(\cM)-1) + \begin{cases}
                                           p(\cB) -1 \quad \text{ if ciliated} \\
                                           p(\cB) \quad \text { if not}
                                          \end{cases} \; ,
\end{align*}
hence, taking into according that $\partial \cM^{(0)} = \partial \cM = \cB$ in both cases
\begin{align*}
&  F _{\rm{ext}} (\cM^{(0)}) +  F _{\rm{int}} (\cM^{(0)}) \le D + (D-1) ( V(\cM)-1 ) + p(\cB) - \rho(\cB) \Rightarrow \crcr
&  F _{\rm{int}} (\cM)  \le  D + (D-1) ( V(\cM)-1 ) - (D-1) p(\cB) - \rho(\partial\cB) +
 \Big[  E (\cM) -V(\cM) +1 \Big] \; .
\end{align*}
\qed

This lemma proves in particular in the sense of perturbation theory the scaling behavior in the equation \eqref{eq:propbound}:
\begin{align}\label{eq:bun}
& K \Big( \cB; N,z \Big) = \sum_{v\ge 0 }
    \sum_{ \genfrac{}{}{0pt}{}{\cM, \partial \cM = \cB} { V(\cM) =v } } z^{   E (\cM)}
   \frac{ N^{ - (D-1)p(\cB) -   E (\cM)(D-1) +  F _{\rm{int}}(\cM) } }{  N^{  D - 2 (D-1) p(\cB) -\rho(\cB)  } } \;, \crcr
& \left| \frac{ N^{ - (D-1)p(\cB) -   E (\cM)(D-1) +  F _{\rm{int}}(\cM) } }{  N^{  D - 2 (D-1) p(\cB) -\rho(\cB)  } } \right| \le N^{  -(D-2)  \Big[   E (\cM) -V(\cM) +1 \Big]   } \; ,
\end{align}
where the sum runs over maps $\cM$ with $v$ \emph{unlabeled} vertices (canceling the $1/v!$ factor), and $D\ge 3$ and $ E (\cM) -V(\cM) +1 \ge 0$ for a connected map.

\subsubsection{Leading order of the 4-point cumulants in the $1/N$--expansion}

The relevance of the intermediate field representation is now transparent. Indeed, equation \eqref{eq:bun} teaches us that (in the sense of perturbation theory):
\begin{itemize}
 \item the functions $ K \Big( \cB; N,z \Big)$ are finite for all $N$ and admit a large--$N$ limit.
 \item the leading order of $  K \Big( \cB; N,z \Big)$ is given by trees $\cM$ such that $\partial\cM = \cB$ (in particular $\cM$ must have $p(\cB)$ cilia).
 \item the next to leading order is given by trees decorated by a loop edge, and is suppressed by $N^{-(D-2)}$.
 \item the first $q$ orders in the $1/N$ series are given (at most) by trees decorated with up to $q$ loop edges.
\end{itemize}

From now on we concentrate on the 4-point contributions $K \Big( \cB_{4,\cC}; N,z \Big)$. They are represented by maps with two cilia, $\{i_1,i_2\}$.
At leading order only trees with two cilia contribute. If all the edges in the tree connecting the two ciliated vertices have the same color $c$, the
boundary graph of $\cM$ is $\cB_{4,\{c\}}$. If not, the boundary graph of $\cM$ is $\cB_{4,\emptyset}$. Thus the last statement of equation \eqref{eq:4cumulz}:
\begin{equation}
 K \Big(\cB_{4, \cC } ; N,z  \Big)\Big{|}_{|\cC|\ge 2} =  \mathcal{O}\left( \frac{1}{N^{D-2}} \right) \; ,
\end{equation}
is proven.

The other two statements are obtained as follows. Recall that $T(z) = \frac{1-\sqrt{1-4Dz}}{2Dz}$ is the physical solution of the equation $ 1 -T(z) + Dz T(z)^2=0$,
and counts rooted plane trees with a weight $Dz$ per edge (i.e. tress with a weight $z$ per edge and an arbitrary color $c\in \{1,\dotsc,D\}$).

The graph $\cB_{4,\{c\}}$ is obtained from trees such that the path between $i_1$ and $i_2$ is formed by edges of the same color.
The simplest example is the tree with only two vertices separated by an edge of color $c$. Any other tree contributing at leading order
is obtained by inserting a (possibly empty) tree with colored edges at any one of the four corners of the vertices $i_1$ or $i_2$ and
inserting $d$ additional intermediary vertices on the path between $i_1$ and $i_2$, each equipped with two corners on which arbitrary trees
are inserted, thus:
\begin{equation}
 K \Big(\cB_{4, \{ c \}} ; z  \Big) =  z T(z)^4 \sum_{d=0}^{\infty} [zT(z)^2]^d = \frac{ zT(z)^4 }{ 1- zT(z)^2 } \; .
\end{equation}

For $K \Big(\cB_{4, \emptyset} ; z  \Big)$ the simplest tree has two edges of different colors hooked to $i_1$ and $i_2$ joined at an intermediary bi valent vertex.
Denoting $d$ the number of additional vertices inserted on the path between $i_1$ and $i_2$ and taking into account that only paths in which not all edges
have the same color contribute we have:
\begin{equation}
  K \Big(\cB_{4, \emptyset} ; z  \Big) =   T(z)^4 \Bigl[  Dz \sum_{d=0}^{\infty} [DzT(z)^2 ]^d -  Dz \sum_{d=0}^{\infty} [zT(z)^2]^d \Bigr] =
  \frac{D(D-1)z^2 T(z)^6}{ \Big(1- DzT(z)^2\Big) \Big(1-  zT(z)^2 \Big)} \;,
\end{equation}
which reproduces eq. \eqref{eq:4cumulz} taking into account that $ 1- DzT(z)^2 = 2-T  = T \sqrt{1-4Dz}  $.

\subsubsection{Reduced maps}

The explicit resummation we performed for the leading order in the previous subsection can be extended to all orders in the $1/N$
series and ultimately leads to the double scaling limit of tensor models.

\begin{figure}
\centering
\includegraphics[scale = 1]{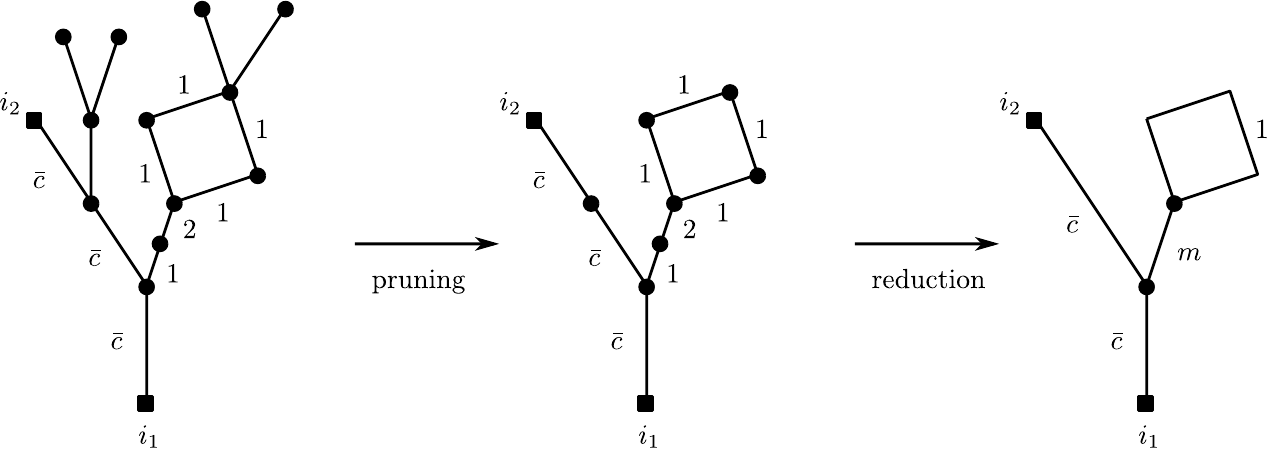}
\caption{\label{fig:prunered} The process of pruning and reduction.}
\end{figure}

It emerges that we can partition the maps $\cM$, with $\partial\cM =\cB$ into classes, each possessing a canonical representative $\overline \cM$, which we
call the \emph{pruned, reduced map} (or simply, reduced map).  There are infinitely many maps $\cM$ in the original sum, which are related through pruning and reduction to the same reduced map $\overline \cM$.
Moreover, the amplitude for the entire class can be resummed and thus assigned to this representative. This process is illustrated in Figure \ref{fig:prunered}.

Let us start with a map $\cM$.
\begin{itemize}
  \item[--]\emph{Pruning} is the iterative removal of non-ciliated e--vertices of degree one.  This procedure effectively removes tree--like sub-structures from the
  map $\cM$.  From the point of view of the original $(D+1)$--colored graph, pruning is equivalent to the iterative removal of elementary melons.

  \item[--]\emph{Reduction} is the removal of all non-ciliated e--vertices of degree two, which from the point of view of the original $(D+1)$--colored graphs, is equivalent to the contraction of certain
chains of $(D-1)$--dipoles. This procedure effectively replaces chains of bivalent vertices by new edges, which we call \emph{bars}. Those bars come in two types: \textit{i}) \emph{Type--$c$ bars} represent chains of e--vertices of degree two where the connecting e--edges all have the same color $c$; and \textit{ii}) \emph{Type-$m$ bars} ($m$ for multicolored) represent chains of e--vertices of degree two, where the connecting e--edges
have at least two different colors.
\end{itemize}
A type--$m$ bar is a sequence of type--$c$ bars connected by vertices of degree two and at least one change of color. Any vertex of the reduced map, except possibly the ciliated ones, therefore has degree at least three.

It is easy to show \cite{DSQuartic} that all the maps $\cM$ associated to the reduced map $\overline \cM$ possess the same scaling in $N$.
The scaling exponent of a map in eq. \eqref{eq:bun}:
\begin{equation*}
-   E (\cM)(D-1) +  F _{\rm{int}}(\cM) \;,
\end{equation*}
is clearly invariant under the deletion of e--vertices of degree one with no cilia and of the e--edges adjacent to them (as exactly $(D-1)$ internal
faces are formed only by this e--vertex). Also, exactly $(D-1)$ internal faces are formed by an e--vertex of degree two with no cilium and
adjacent to two e--edges of the same color.

Type--$c$ bars bring the same scaling as regular e--edges of color $c$, i.e. $N^{-(D-1)}$. However, packing up chains of such bars into type--$m$ bars changes the scaling with an extra $N^{-1}$. Thus, a type--$m$ bar comes with $N^D$. The faces and boundary of the reduced map $\overline \cM$ are defined as before, but taking into account that $\overline \cM_c$ is obtained by deleting not only all the bars of colors different from $c$, but also \emph{all the multicolored bars}. We denote $E^m(\overline \cM)$ the number of multicolored bars, and $E^u(\overline\cM) = \sum_{c=1}^D E^c(\overline\cM) $ the total number of type--$c$ bars of $\overline \cM$.

In addition to its scaling with $N$, a reduced map has a $z$-dependent amplitude. Following the process of pruning and reduction, it is found that this amplitude is evaluated via local weights assigned in the following way:
\begin{itemize}
 \item[--] corners are dressed with the LO full 2-point function $T(z)$, reflecting the summation of arbitrary tree--like structures,
 \item[--] type--$c$ bars represent chains of bubbles $\cB_{4,\{c\}}$ with the same color, hence get the weight
\begin{equation*}
z\sum_{k\geq 0} [z T(z)^2]^k = \frac{z}{1-z T(z)^2}\;,
\end{equation*}
 \item[--] type--$m$ bars represent chains of bubbles $\cB_{4,\{c\}}$ with at least one change of colors, hence the weight
\begin{equation*}
Dz\sum_{k\geq0} [DzT(z)^2]^k - Dz\sum_{k\geq0} [zT(z)^2]^k = \frac{D(D-1)\,z^2 T(z)^2}{(1 - DzT(z)^2)\ (1 - zT(z)^2)}\;.
\end{equation*}
\end{itemize}

\begin{figure}[htb]
    \centering
    \begin{minipage}{0.45\linewidth}
\centering
\includegraphics[scale = 1]{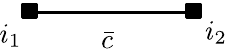}
\caption{\label{fig:simplestc}The leading reduced map for $ K \Big( \cB_{4,\{c\}} ; N,z \Big)  $.}
    \end{minipage}
    \begin{minipage}{0.45\linewidth}
\centering
\includegraphics[scale = 1]{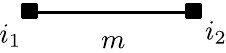}
\caption{\label{fig:simplestm}The leading reduced map for $ K \Big( \cB_{4,\emptyset } ; N,z \Big)  $.}
    \end{minipage}
  \end{figure}

The perturbative expansion of $ K \Big( \cB; N,z \Big) $ can be reorganized in terms of reduced maps $\overline \cM$ with unlabeled vertices:
\begin{align}\label{eq:dsbun}
K \Big( \cB; N,z \Big) = &  \sum_{v\ge 0 }
    \sum_{ \genfrac{}{}{0pt}{}{\overline \cM, \partial \overline \cM = \cB} { V( \overline \cM) =v } }
   \frac{ N^{ - (D-1)p(\cB) -  D E^m(\overline \cM)  - (D-1) E^u(\overline \cM)
      +  F _{\rm{int}}(\overline \cM) } }{  N^{  D - 2 (D-1) p(\cB) -\rho(\cB)  } } \\
 &   ÃÂ T(z)^{p(\cB) + 2 \bigl[ E^m(\overline \cM)  +  E^u(\overline \cM)   \bigr] }
 \Bigl(  \frac{z}{1-zT(z)^2} \Bigr)^{ E^u(\overline \cM)   }  \Bigl(  \frac{D(D-1)z^2T(z)^2}{(1 - DzT(z)^2) (1 - zT(z)^2)} \Bigr)^{ E^m(\overline \cM) } \nonumber \; .
\end{align}
The leading $1/N$ terms for $ K \Big( \cB_{4,\{c\}} ; N,z \Big)  $ and $  K \Big( \cB_{4, \emptyset} ; N,z \Big) $ we computed in the previous sections are exactly
the contributions of the reduced maps in figures \ref{fig:simplestc} and \ref{fig:simplestm}.

\subsubsection{The double scaling limit}

Each type--$m$ bar comes with a factor $1/(1-DzT(z)^2) = 1/(T(z)\sqrt{1-4Dz})$ which diverges as $z\to z_c=1/4D$. Therefore the amplitude of a reduced map has a singular part of the form $(1-4Dz)^{-\frac{ E^m(\overline \cM) }{2}}$ close to criticality. We henceforth look for the most singular reduced maps at each fixed order in $1/N$, by maximizing the number of multicolored bars at that order.

Consider a reduced map $\overline\cM$ with boundary $\cB = \partial \overline\cM$. No face goes along a multicolored edge, neither internal nor external.
We delete all the multicolored bars. The reduced map $\overline\cM$ splits into several connected components.
We denote $\overline\cM(\nu)$, $\nu=1,\dotsc,r$ the connected components which contain ciliated vertices, and
$\overline\cM(\mu)$, $\mu=1,\dotsc,q$ those which do not contain any ciliated vertex. Remark that these connected components \emph{are not} reduced maps, as they can contain vertices of degree two. However, they are edge--colored maps.

As no face goes along the multicolored bars, the boundary graph $\cB = \partial \overline\cM$ also splits
into several connected components $\cB(\nu) = \partial \overline\cM(\nu)$ and $\cB$ is the disjoint union of $\cB(\nu)$.
The type--$c$ bars and internal faces are partitioned between the $\overline\cM(\nu)$s and $\overline\cM(\mu)$s, hence:
\begin{multline} \label{SplitMultiUni}
 -  D E^m(\overline \cM) - E^u  (\overline\cM )(D-1) +  F _{\rm{int}}  (\overline\cM ) = -  D E^m(\overline \cM) +
  \sum_{\nu=1}^r \Bigg[ - E^u \Big(\overline\cM(\nu) \Big)(D-1) +  F _{\rm{int}} \Big(\overline\cM(\nu) \Big) \Bigg] \\
 + \sum_{\mu=1}^q \Bigg[ - E^u \Big(\overline\cM(\mu) \Big)(D-1) +  F _{\rm{int}} \Big(\overline\cM(\mu) \Big) \Bigg] \; .
\end{multline}

\begin{figure}
\includegraphics[scale=.5]{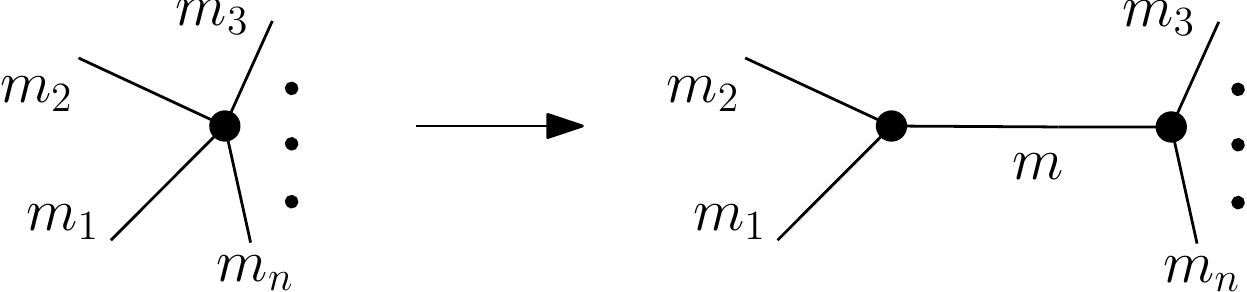}
\caption{\label{fig:MoreSingularBinaryTree} From the left to the right drawing, the scaling with $N$ is preserved but a type--$m$ bar is created. One can proceed until the initial e--vertex has become a binary tree.}
\end{figure}

\begin{figure}
\includegraphics[scale=.5]{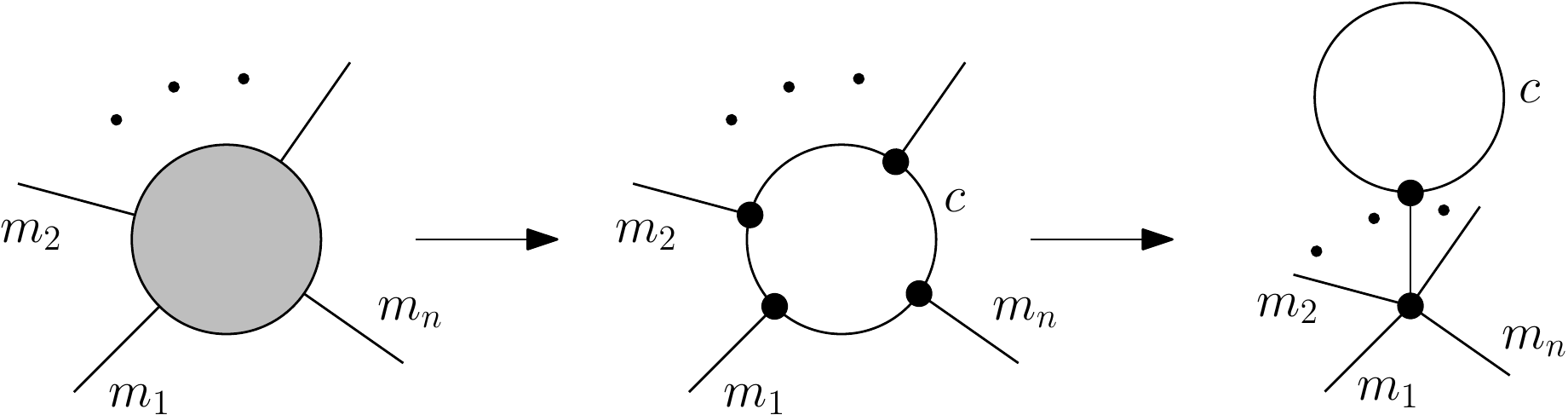}
\caption{\label{fig:MoreSingularCherry} From the left to the middle drawing, one changes the gray blob representing $\overline\cM(\mu)$ with a unicolored loop, which can not decrease the exponent of $N$. From the middle to the right, the scaling with $N$ is preserved but a type--$m$ bar is created.}
\end{figure}

The components $ \overline\cM(\mu)$, $\mu=1,\dotsc,q$, not containing any cilium are treated as follows. Either:
\begin{itemize}
\item $ \overline\cM(\mu) $ \emph{is a tree}, hence $- E^u \Big(\overline\cM(\mu) \Big)(D-1) +  F _{\rm{int}} \Big(\overline\cM(\mu) \Big) =D $.
    There are two cases.
\begin{itemize}
 \item Either $ \overline\cM(\mu) $ has a unique e--vertex, incident to at least three multicolored bars,
 \item or $ \overline\cM(\mu) $ has more than one e--vertex. Then $ \overline\cM(\mu) $ is incident to at least four multicolored bars.
       The reduced map $ \overline\cM$ has the same scaling in $N$ and the same singular behavior as the map where
       $ \overline\cM(\mu) $ has been contracted to a unique e--vertex.
\end{itemize}
Moreover, when $ \overline\cM(\mu) $ is an e--vertex of degree at least four, one can always build a reduced map with the same scaling in $N$, and strictly more type--$m$ bars, by splitting the e--vertex into a binary tree whose edges are type--$m$ bars, as in the fig. \ref{fig:MoreSingularBinaryTree}. Indeed, as no faces go all along a type--$m$ bar, an e--vertex incident to only type--$m$ bars closes exactly $D$ faces. This way, adding both a vertex and a type--$m$ bar does not affect the scaling while bringing an additional power of $1/\sqrt{z_c-z}$.
Therefore the non-ciliated components $\overline \cM(\mu)$ which are trees have to be e--vertices of degree three.

\item Or $ \overline\cM(\mu) $ \emph{is not a tree} and it is incident to at least one multicolored bar.
     In this case, from lemma \ref{lem:bun}:
     \begin{equation*}
    - E^u \Big(\overline\cM(\mu) \Big)(D-1) +  F _{\rm{int}} \Big(\overline\cM(\mu) \Big)  \le D -
   (D-2) \Big[   E^u \Big(\overline\cM(\mu) \Big) -V\Big(\overline\cM(\mu)\Big) +1 \Big] \le D - (D-2) \; ,
    \end{equation*}
    and this bound is saturated by the unicolored loop. It follows that $\overline\cM$ scales at most like the same map where $ \overline\cM(\mu)$ has been replaced with a unicolored loop.

Moreover, if $ \overline\cM(\mu) $ is a unicolored loop incident to more than one type--$m$ bar, then one can always build a reduced map with the same scaling in $N$ but with strictly more type--$m$ bars (hence more singular). Ones detaches the loop and attaches the bars to a common vertex which is then connected to the loop through a new type--$m$ bar (see fig. \ref{fig:MoreSingularCherry}). If $\overline \cM(\mu)$ has $n$ incident bars, the scaling is $N^{2-nD}$ in both cases, but one gets a new singular factor in the second case.
\end{itemize}

Following \cite{DSQuartic}, we call a type--$c$ loop hooked to a single type--$m$ bar a \emph{cherry}, represented like $\begin{array}{c} \includegraphics[scale=.8]{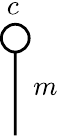} \end{array}$.

We now analyze the connected components $\overline\cM(\nu)$, $\nu=1,\dotsc,r$, which contain the cilia. If several type--$m$ bars are incident to the same component $\overline\cM(\nu)$, one can build a reduced map with the same scaling in $N$ but more type--$m$ bars by a process similar to the one applied on the components $\overline \cM(\mu)$ which are not trees. One indeed detaches all the type--$m$ bars incident to $\overline\cM(\nu)$ but one, say $m_1$, and reconnects them on a new e--vertex created along $m_1$. This splits $m_1$ into two type--$m$ bars, enhancing the singular behavior at criticality. It has to be mentioned that upon detaching all but one of the type--$m$ bars hooked to $\overline\cM(\nu)$, the degree of some e--vertices in $\overline\cM(\nu)$can drop down to two, hence the resulting drawing is not a reduced map. This however is not a problem, as the reduced maps in which those e--vertices of degree two are appropriately exchanged for bars do exist and scale as advertised, hence strictly dominate $\overline\cM$.

Thus, for any boundary graph $\cB$, the leading singular behavior when $z\to z_c$ is captured by the reduced maps whose non-ciliated unicolored components are either vertices of degree three or cherries, and each ciliated unicolored component is incident to exactly one type--$m$ bar. We denote ${\bf V}^3$ and ${\bf V}^{\rm{cherry}}$ the number of e--vertices of degree 3 and the number of cherries. We need the following combinatorial relations:
\begin{itemize}
 \item $ {\bf V}^{\rm{cherry}} +  {\bf V}^3 = q $,
 \item as the type--$m$ bars must connect all the unicolored components in a connected way, we also get
$E^m(\overline \cM)  = q + r -1 + l = {\bf V}^{\rm{cherry}} +  {\bf V}^3 + r -1+l $, for some non-negative integer $l$,
\item as all the cherries are hooked to one type--$m$ bar, all the vertices of degree three to three bars, and all the (ciliated) components $\overline\cM(\nu)$ to one bar, we also have $2 E^m(\overline \cM) = {\bf V}^{\rm{cherry}} +  3 {\bf V}^3 + r  $.

There is however one exception to this relation, namely when $r=1$ and ${\bf V}^{\rm{cherry}}  = {\bf V}^3 =0$, there exists a reduced map with $E^m(\overline \cM)  =0$. It does not diverge at criticality (goes to a constant), and one needs to check its scaling with $N$ separately.
\end{itemize}

From the above three relations, we extract ${\bf V}^3$ and $E^m(\overline\cM)$ as a function of ${\bf V}^{\rm{cherry}},l,r$:
\begin{equation}
\begin{gathered}
{\bf V}^3 =  {\bf V}^{\rm{cherry}} +   r -2+2l \\
E^m(\overline \cM) = 2  {\bf V}^{\rm{cherry}} + 3 l  + 2r -3 \; .
\end{gathered}
\end{equation}

Moreover, we can rewrite the exponent of $N$ due to non-ciliated unicolored components in \eqref{SplitMultiUni} as:
\begin{equation}
  \sum_{\mu=1}^q \bigg[ - E^u \Big(\overline\cM(\mu) \Big)(D-1) +  F _{\rm{int}} \Big(\overline\cM(\mu) \Big) \bigg] =
   D {\bf V}^3+ 2 {\bf V}^{\rm{cherry}} \; ,
\end{equation}
hence the most singular terms in \eqref{eq:dsbun} behave like:
\begin{multline}\label{eq:precis}
\frac{  N^{ -D + p(\cB) (D-1) + \rho(\cB)} N^{-D (  {\bf V}^{\rm{cherry} } +  {\bf V}^3 + r -1+l  ) +  D {\bf V}^3  + 2 {\bf V}^{\rm{cherry}} +
  \sum_{\nu=1}^r \Big[ - E^u \big(\overline\cM(\nu) \big)(D-1) +  F _{\rm{int}} \big(\overline\cM(\nu) \big) \Big] } }
  { (1-4Dz)^{-\frac{ 2  {\bf V}^{\rm{cherry}} + 3 l  + 2r -3  }{2} } } \\
 = \frac{  N^{   p(\cB) (D-1) + \rho(\cB)} N^{- (D-2) {\bf V}^{\rm{cherry} } -D r - D l   +
  \sum_{\nu=1}^r \Big[ - E^u \big(\overline\cM(\nu) \big)(D-1) +  F _{\rm{int}} \big(\overline\cM(\nu) \big) \Big] } }
  { (1-4Dz)^{\frac{ 2  {\bf V}^{\rm{cherry}} + 3 l  + 2r -3  }{2} } } \; .
\end{multline}
Further, as every $\overline\cM(\nu)$ contains a cilium, the lemma \ref{lem:bun} applies, hence every term is bounded by:
\begin{equation}\label{eq:util}
\frac{
 N^{- (D-2) {\bf V}^{\rm{cherry} } - D l   - (D-2)
  \sum_{\nu=1}^r \Big[  E^u \big(\overline\cM(\nu) \big)   - V \big(\overline\cM(\nu) \big)  +1 \Big]  }
}{ (1-4Dz)^{\frac{ 2  {\bf V}^{\rm{cherry}} + 3 l  + 2r -3  }{2} } } \; .
\end{equation}

We are finally in the position to address the double scaling limit of cumulants in the quartic model. The most singular contributions are selected by maximizing $ 2  {\bf V}^{\rm{cherry}} + 3 l $ while keeping $ (D-2) {\bf V}^{\rm{cherry} } + D l $ fixed (which is a linear program similar to the one used for the double scaling limit in \cite{ColoredSchemes}). For $D<6$ the dominant singular behavior is obtained by setting $l$ to zero and introducing the new coupling $ x = N^{D-2}(\frac{1}{4D} - z) $ to be held fixed, leading to the generic double scaling behavior
\begin{equation}
   N^{(D-2)\frac{2r-3}{2}  - (D-2)
  \sum_{\nu=1}^r \Big[  E^u \Big(\overline\cM(\nu) \Big)   - V \Big(\overline\cM(\nu) \Big)  +1 \Big]  } \; \;  f(\cB;N,x)
\end{equation}
where $f(\cB;N,x)$ as a function of $N$ is bounded by a constant.

We now apply this formula to the 2-- and 4--point cumulants.

{\bf The 2-point cumulant.} The 2-point function is represented by reduced maps with a unique cilium, hence $r=1$. Furthermore, the unique connected component containing the ciliated vertex can be chosen to have no loops (i.e. it is formed only by the ciliated vertex). Separating the contribution of the reduced map with only the ciliated vertex (which as we already mentioned must be evaluated separately), we get the double scaling ansatz:
\begin{equation}
 K \Big( \cB_2; N,x  \Big) = f^{(0)}(\cB_2;N,x) + N^{-\frac{D-2}{2}} f^{( -\frac{D-2}{2} )}(\cB_2;N,x) \;,
\end{equation}
in agreement with our ansatz \eqref{DSAnsatz}.

\begin{figure}[htb]
    \centering
    \begin{minipage}{0.4\linewidth}
\centering
\includegraphics[scale = 1]{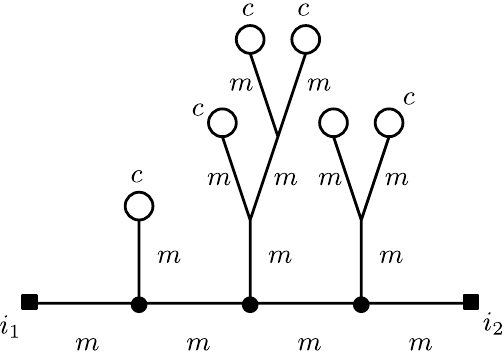}
\caption{\label{fig:optimalm} Leading contribution to $ K \Big( \cB_{4,\emptyset}; N,x  \Big)$.}
    \end{minipage}
    \begin{minipage}{0.4\linewidth}
\centering
\includegraphics[scale = 1]{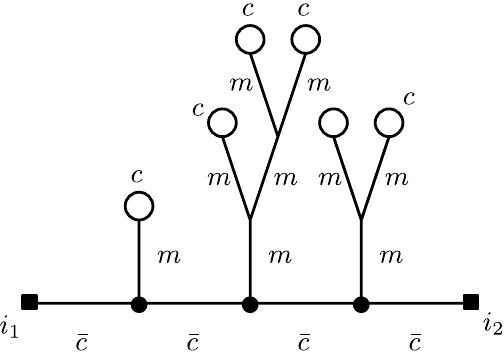}
\caption{\label{fig:optimalc}Leading contribution to $ K \Big( \cB_{4,\{c\} }; N,x  \Big)$.}
    \end{minipage}
  \end{figure}

{\bf The 4-point cumulant.} For $\cB_{4,\emptyset}$, the leading double scaling contributions come from the reduced maps in the figure \ref{fig:optimalm} having $r=2$, while for $\cB_{4,\{c\}}$ the leading double scaling contributions come from the reduced maps in the figure \ref{fig:optimalc} having $r=1$ (and one must remember to treat the contribution of the map with only one edge of color $c$ separately). The maps contributing to $\cB_{4,\cC}$ with $|\cC|\ge 2$ have $r=1$, but $ \partial \overline\cM(\nu) = \cB_{4,\cC}$ imposes that $ \cM(\nu)$ possesses loop edges.
Thus:
\begin{align}
&  K \Big( \cB_{4,\emptyset}; N,x  \Big) = N^{\frac{D-2}{2}} f^{(\frac{D-2}{2})}(\cB_{4,\emptyset};N,x) \;, \crcr
&   K \Big( \cB_{4,\{c\} }; N,x  \Big) = f^{(0)}(\cB_{4, \{c\} };N,x)  +  N^{- \frac{D-2}{2}} f^{ (-\frac{D-2}{2} ) }(\cB_{4, \{c\} };N,x) \; ,\crcr
&  K \Big( \cB_{4,\cC}; N,x  \Big) = \mathcal{O} \Bigl( \frac{1}{N^{D-2}} \Bigr) \; ,
\end{align}
which is precisely \eqref{eq:4cumulx}.

\subsection{From the quartic model to a generic model} \label{ssec:FromQuarticToGeneric}

Although the above proof is restricted to the quartic model, its result can be easily extended to a generic model whose action has bubbles $\{\cB_i\}_{i\in I}$ that are melonic (and symmetrized on their colorings). This is actually expected from universality (in the sense used in statistical mechanics). Indeed, universality means that changing the details of a model (here the building blocks of the graphs) does not change its critical properties.

Let us see to what extent the quartic model and its generic extension coincide:
\begin{description}
\item[Graph structure]
The first thing we shall show is that all the graphs of a generic model can be mapped to graphs of the quartic model. The apparent difference between those two types of graph is that they are built from different bubbles (subgraphs with colors $1,\dotsc,D$). However, any melonic bubble can be obtained as the boundary graph of a gluing of quartic bubbles via propagators. This has already been observed in \cite{TreeAlgebra} and we shall therefore not give too many details.

The main idea behind this fact is that, given a bubble $\cB$, it is easy to add a $(D-1)$--dipole onto any of its edges, using the propagator and an appropriate quartic bubble.  Indeed, say we want to add a $(D-1)$--dipole onto the edge of color $c$ that is incident to the white vertex $v$ in $\cB$. To do so, a propagator is introduced to connect $v$ to a black vertex $\overline{v}$ of $\cB_{4,\{c\}}$. The boundary graph is obtained by cutting out $v$ and $\overline{v}$ as well as the propagator, and reconnecting then the open edges of the same colors,
\begin{equation*}
  \begin{array}{c} \includegraphics[scale=.6]{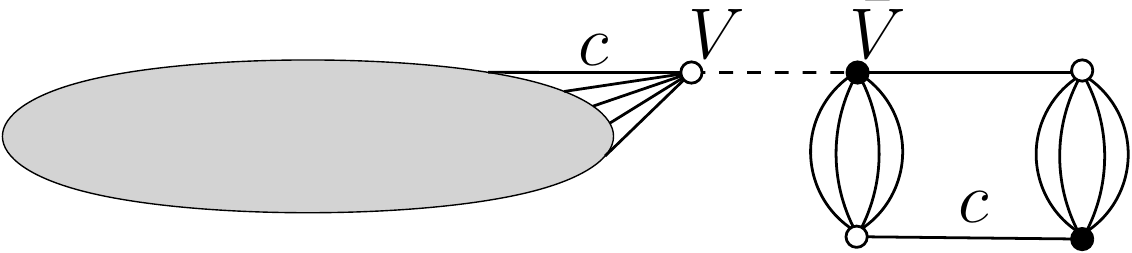} \end{array}
  \qquad \underset{\text{Boundary graph}}{\rightarrow} \qquad
  \begin{array}{c} \includegraphics[scale=.6]{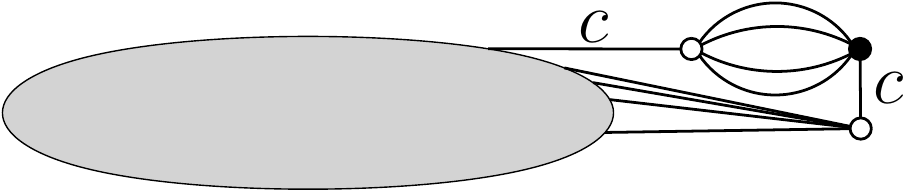} \end{array}
\end{equation*}
Note that this is the same as removing $v$ and gluing instead the right hand side of \eqref{OpenB4}. Importantly, this map can be interpreted the other way around: each $(D-1)$-dipole of external color $c$ can be replaced by a regular edge of color $c$ and a propagator connecting the white vertex to the quartic bubble $\cB_{4,\{c\}}$.

Since any melonic bubble on $p+1$ black vertices can be obtained through the insertion of a $(D-1)$-dipole on a melonic bubble on $p$ black vertices, a simple inductive argument proves the claim: any melonic bubble with $p$ black vertices is thus represented by a gluing of $p-1$ quartic bubbles. This way the Feynman graphs of an arbitrary model are realized as a sub-family of graphs within the quartic model. (Obviously, the quartic model generates arbitrary melonic bubbles as boundary graphs.)
\item[Scaling in $N$]
If $\mathcal{G}$ is a Feynman graph in the generic model, we denote a representative in the quartic model by $Q(\mathcal{G})$. Then, $\mathcal{G}$ and $Q(\mathcal{G})$ turn out to have the same scaling. This stems from the fact that the weight of a melonic bubble reprensented as a gluing of quartic ones is the same as if the bubble were in the action. Indeed, in the latter case, each bubble $\cB_i$, $i\in I$, brings a factor $N^{D-1}$. On the other hand, we need $p_i-1$ quartic bubbles to represent it, each of them also coming up with a factor $N^{D-1}$. There are however $p_i-2$ propagators, each of them with the weight $N^{-(D-1)}$, which eventually leaves us with an overall $N^{D-1}$.

\item[$z$--dependence]
Furthermore, the amplitudes of $\mathcal{G}$ and $Q(\mathcal{G})$ have the same functional dependence on the coupling $z$. A bubble $\cB_i$, $i\in I$, among those of the action comes with a weight $z^{p_i-1}$. Since $p_i-1$ quartic bubbles are required to represented it, each of them bringing a single $z$, we get $z^{p_i-1}$ as expected for the representant of $\cB_i$.
\item[Combinatorial factors]
The only departure between the two families $\{\mathcal{G}\}$ and $\{Q(\mathcal{G})\}$ are combinatorial factors that weight the graphs when they are summed as contributions to some expectation. Those combinatorial factors come from the Feynman expansion (that is, from the expansion of the exponential in the original integral and the counting the number of times a given graph is obtained). But this is where universality intervenes. Such a microscopic change coming from the use of different building blocks (in other words, the bubbles $\cB_i$ are given an internal structure made up of quartic bubbles) may affect the value of the critical coupling but do not change the scaling exponents. Therefore, a family of Feynman graphs of the form $\{Q(\mathcal{G})\}$ which is irrelevant (i.e. suppressed as some powers of $1/N$) in the double scaling limit of the quartic model cannot be a relevant family $\{\mathcal{G}\}$ in the generic model.
\end{description}

\section{Concluding remarks and perspectives}

In this paper, we have used the Schwinger--Dyson equations to both re-derive the double scaling limit of the quartic model and also, with a succinct universality argument, to extend it to models with generic melonic interaction bubbles. More precisely, we have obtained the doubly scaled 2-point function in the equation \eqref{2PtDoublyScaled}. In addition to this result, we have used a new strategy which combine the Schwinger--Dyson equations with combinatorial scale arguments:
\begin{itemize}
 \item[--] expand the expectations onto sums of products of cumulants,
 \item[--] study the (simple or double) scaling of the cumulants to identify the relevant boundary graphs at each order,
 \item[--] solve a linear (in the regular $1/N$--expansion case) or quadratic (in the double scaling case, at least for the 2-point function) system of Schwinger--Dyson equations on those relevant cumulants.
\end{itemize}
We have thus seen at play the hierarchy of cumulants beyond the leading order. The latter is indeed Gaussian, then at NLO one observes 4-point cumulants and so on. Remarkably, the double scaling of the 2-point function only requires to go up to 4-point cumulants, and this is why only two SD equations were needed instead of the full tower. It will be interesting to probe more equations and higher cumulants, in particular in the double scaling regime.

Let us reiterate, given that the doubly--scaled series is summable in dimensions lower than six, this study (complementing \cite{ColoredSchemes} and \cite{DSQuartic}) may be seen as a preliminary study for some multiple scaling limit mechanism (obtained from reiterating the double scaling procedure). This multiple scaling limit offers the hope to escape the branched polymer phase of the theory, emerging into a new continuous random space with hopefully more attractive physical features. This multiple scaling limit thus seem to us an interesting direction for future work.

Another perspective for future work is the generalization of the approach exhibited in this paper to the multi-orientable (MO) random tensor model \cite{MO}, for which a large--$N$ expansion has recently been found and the leading-order \cite{DRT} and the next-to-leading order \cite{RT} have been identified from a combinatorial and topological point of view. A yet more thorough analysis is required, however, as the MO model generates a family of stranded graphs which is larger than the set of regular edge--colored graphs, and for which observables have to be appropriately redefined.

A further perspective is the application of such techniques to tensor models whose propagators break the unitary invariance. Numerous studies have been recently dedicated to the renormalization of such models.
Thus, by introducing a non-trivial index dependence on the propagator (of the type $1/p^2$), a first, just renormalizable, four-dimensional tensor model, the so-called Ben Geloun-Rivasseau model, was
proposed in \cite{BGR,BGR1}. A series of studies \cite{ren,ren1,ren2} has followed this breakthrough, leading, for example, to the striking result of UV asymptotic freedom of renormalizable
tensor models. Moreover, this type of result was generalized within the  Group Field Theory framework \cite{COR,COR1,COR2}. Finally, let
us also mention that methods related to lattice gauge theories of permutation groups have been successfully applied to count
invariants for tensor models \cite{AIHPD}. We hope that SD techniques could be useful for those models too. For instance, when the propagator breaks the unitary invariance of the action, Ward identities become interesting, and in the context of non--commutative field theories, they can be successfully combined with SD equations to study the phase diagram of the model \cite{GW}.

\section*{Acknowledgements}
Adrian Tanasa is partially supported by the ANR JCJC CombPhysMat2Tens grant and by the grant PN 09 37 01 02.
Razvan Gurau, James P. Ryan and Adrian Tanasa ackowldge the E. Schro\"odinger Institute for the work conditions offered to them during the "Combinatorics, Geometry and Physics" programme.

\newpage

\appendix

%
\section{The leading and next-to-leading orders in the $1/N$--expansion}
\label{app:4pdom}

We recall the expansion in cumulants of the expectation of a generic trace--invariant \eqref{eq:momentexpansion}:
\begin{equation}\label{eq:expcumul}
\begin{aligned}
\frac{1}{N}\langle \tr_{\cB}(\bT, \overline{\bT}) \rangle &=\frac{1}{N}
\sum_{\vec a^{v} \vec b^{\bar v}}  \delta^{\cB}_{\vec a^v, \vec b^{\bar v}}
\left \langle
\left(  \prod_{\bar v \in \cV(\cB) } \bar \bT_{\vec b^{\bar v}} \right)
 \left( \prod_{v\in \cV(\cB)} \bT_{\vec a^{v}} \right) \right \rangle \\
 & = \frac{1}{N}\sum_{\vec a^{v} \vec b^{\bar v}}  \delta^{\cB}_{\vec a^v, \vec b^{\bar v}}
 \left(\sum_{ \pi  } \prod_{\cV_\alpha \in \pi}
W^{(2p_\alpha)}_{\vec a^{v^{(\alpha)}_1} \dotsc \vec a^{v^{(\alpha)}_{p_\alpha}}, \vec b^{ \bar v^{(\alpha)}_1 } \dotsc \vec b^{\bar v^{(\alpha)}_{p_\alpha}}} \Big(N, z,\{t_i\} \Big)  \right)\; .
\end{aligned}
\end{equation}
where $\pi$ denotes a partition of the vertex set $\cV(\cB)$ into bipartite subsets
$\cV_\alpha=\{v^{(\alpha)}_1,\dotsc,v^{(\alpha)}_{p_\alpha}, \bar v^{(\alpha)}_1,\dotsc,\bar v^{(\alpha)}_{p_\alpha}\}$, for $\alpha=1,\dotsc,A$, (hence $\sum_{\alpha=1}^A p_\alpha = p(\cB)$). Moreover the cumulants admit an expansion onto invariants:
\begin{equation}\label{eq:graphcumul}
\begin{gathered}
W^{(2p)}_{\vec a^1 \dotsc \vec a^p, \vec b^1 \dotsc \vec b^p}\Big(N, z,\{t_i\} \Big)
   = \sum_{\cB', p(\cB') = p } \bar \delta^{\cB}_{\vec a^v,\vec b^{\bar v}} \; W \Big(\cB' ; N,z,\{t_i\}  \Big)  \; , \\
K \Big( \cB';N,z,\{y_i\} \Big) = \frac{W \Big(\cB' ; N,z,\{t_i\}  \Big)}{  N^{  D - 2(D-1) p(\cB') -\rho(\cB') } } \; ,
\qquad \lim_{N\to \infty} K \Big( \cB';N,z,\{y_i\} \Big) = K \Big( \cB',z,\{y_i\} \Big) \; ,
\end{gathered}
\end{equation}
where $p(\cB')$ denotes the number of black vertices of $\cB'$ and $\rho(\cB')$ denotes the number of connected components.

Substituting \eqref{eq:graphcumul} into \eqref{eq:expcumul} we obtain a sum indexed by a partition $\pi$, and a particular set of invariants labeled by
$\cB'_\alpha$, coming from the cumulants and associated to parts of $\pi$, $\alpha=1,\dotsc,A$,
\begin{equation}
\frac{1}{N}\langle \tr_{\cB}(\bT, \overline{\bT}) \rangle =\frac{1}{N} \sum_\pi \sum_{\{\cB'_\alpha, p(\cB'_\alpha)=p_\alpha\}_{\alpha=1,\dotsc,A}} \left[
\sum_{\{\vec a^v,\vec b^{\bar v}\}} \delta^{\cB}_{\vec a^v, \vec b^{\bar v}}\ \prod_{\alpha=1}^A \bar \delta^{\cB'_\alpha}_{\vec a^{v^{(\alpha)}}, \vec b^{\bar v^{(\alpha)}}}\ W(\cB'_\alpha;N,z,\{t_i\}) \right]
\end{equation}
Each term in this sum is a contraction of the trace-invariant operator $\delta^{\cB}_{\vec a^v, \vec b^{\bar v}}$ with the product over the parts of $\pi$ of
trace-invariant operators $ \bar \delta^{\cB'_\alpha}_{\vec a^{v^{(\alpha)}}, \vec b^{\bar v^{(\alpha)}}} $ coming from the cumulants, multiplied by a product of $W(\cB'_\alpha;\dots)$ factors.
Every vertex of $\cB$ appears in exactly one bubble of the family $(\cB'_\alpha)_\alpha$, and $p(\cB)=\sum_\alpha p(\cB'_\alpha)$. One naturally associates to each term in this sum a $(D+1)$-colored graph
$\cG$ obtained by drawing the $D$--colored graphs $\cB$ and $(\cB'_\alpha)_\alpha$, and connecting the vertices of $\cB$ with their images in the bubbles $\cB'_\alpha$ by edges of color $0$.
In order to maintain bipartiteness, one then flips all the vertices of $\cB$. The graph $\cG$ is called a \emph{doubled graph} \cite{Universality},
and the expectation of $\tr_{\cB}(\bT, \overline{\bT})$ is a sum over all doubled graphs which have $\cB$ as a subgraph.

The scaling with $N$ of a term with associated doubled graph $\cG$ is determined by the explicit scalings of the cumulants and by the number of independent
sums in the contractions of the trace-invariant operators. The independent sums are immediately read off the doubled graph $\cG$: indeed, one
obtains a free sum for every face of color $0c$ of $\cG$.

We study below the leading and next to leading contributions to the expectation of melonic observables.

\subsection{Non--Gaussian contributions to the expectation of $\cB_{4,\{c\}}$}

We first study the non--Gaussian contributions in eq. \eqref{eq:nongauss1}:
\begin{equation}
\frac{1}{N} \sum_{a,b} \Bigl(
   \delta_{a^1_c b^2_c } \delta_{a^2_c b^1_c}\prod_{c_1\neq c} \delta_{a^1_{c_1} b^1_{c_1} } \delta_{a^2_{c_1} b^2_{c_1} } \Bigr)
   W^{(4)}_{ \vec a^1 \vec a^2, \vec b^1 \vec b^2 }\Big(N, z,\{t_i\} \Big) \; .
\end{equation}
Using:
\begin{align}
&  W^{(4)}_{\vec a_1\vec a_2,\vec b_1\vec b_2}  = \sum_{\cC \subset \{1,\dotsc,D\} }
     \delta^{\cB_{4,\cC}}_{\vec a^1 \vec a^2, \vec b^1 \vec b^2 }  \; \;  W \Big(\cB_{4,\cC} ; N,z,\{t_i\}  \Big) \\
& W \Big(\cB_{4,\cC} ; N,z,\{t_i\}  \Big)  = N^{D- \rho(\cB_{4,\cC})-2(D-1)p(\cB_{4,\cC}) } K \Big(\cB_{4,\cC} ; N,z,\{t_i\}  \Big)  \\
&  \bar \delta^{\cB_{4,\cC}}_{\vec a^1 \vec a^2; \vec b^1 \vec b^2 } = \left( \prod_{c\notin \cC} \delta_{a_c^1 b_c^1} \right)
 \left( \prod_{c\notin \cC} \delta_{a_c^2 b_c^2} \right) \left( \prod_{c\in \cC}  \delta_{a_c^1 b_c^2}
  \delta_{a_c^2 b_c^1} \right) + \left( \prod_{c\notin \cC} \delta_{a_c^2 b_c^1} \right)
 \left( \prod_{c\notin \cC} \delta_{a_c^1 b_c^2} \right) \left( \prod_{c\in \cC}  \delta_{a_c^2 b_c^2}
  \delta_{a_c^1 b_c^1} \right) \; ,
\end{align}
we obtain a list of terms (i.e. doubled graphs $\cG$):
\begin{itemize}
 \item the terms with $\cC=\emptyset$:
     \begin{equation}
       \frac{1}{N} \Big( N^{D+1} + N^{ 2D-1} \Big) N^{D-2 -2(D-1)2}  K \Big(\cB_{4,\emptyset} ; N,z,\{t_i\}  \Big) =
       \Bigl( \frac{1}{N^{2(D-1) } } + \frac{1}{N^D} \Bigr)K \Big(\cB_{4,\emptyset} ; N,z,\{t_i\}  \Big)  \; .
     \end{equation}
 \item the terms with $ c \in \cC $:
      \begin{align}
        & \frac{1}{N} \Big( N^{2D - |\cC|+1 } + N^{ D + |\cC|-1 } \Big) N^{D-1 -2(D-1)2}  K \Big(\cB_{4,\cC} ; N,z,\{t_i\}  \Big) \crcr
        & = \Bigl( \frac{1}{N^{D-2}} N^{1-|\cC|} +\frac{1}{N^{D-2}} N^{ -D + |\cC|-1} \Bigr)    K \Big(\cB_{4,\cC} ; N,z,\{t_i\}  \Big)   \; .
     \end{align}
    \item the terms with $\cC \neq \emptyset$, but $c\notin \cC$ scale like:
    \begin{align}
      &  \frac{1}{N} \Big( N^{ D + |\cC|+1 } + N^{2 D - |\cC|-1 } \Big) N^{D-1 -2(D-1)2} K \Big(\cB_{4,\cC} ; N,z,\{t_i\}  \Big) \crcr
      & = \Bigl( \frac{1}{N^{D-2}} N^{-D + 1 + |\cC|} + \frac{1}{N^{D-2}} N^{-1-|\cC|} \Bigr) K \Big(\cB_{4,\cC } ; N,z,\{t_i\}  \Big)   \; .
     \end{align}
\end{itemize}
Thus eq. \eqref{eq:nongauss1} becomes, remembering that $|\cC|\le D/2$,
\begin{align}
 & \Bigl( \frac{1}{N^{2(D-1) } } + \frac{1}{N^D} \Bigr)K \Big(\cB_{4,\emptyset} ; N,z,\{t_i\}  \Big) \crcr
 & + \sum_{\cC \in \{1,\dotsc,D\} , c\in \cC, |\cC|\le D/2 }\Bigl( \frac{1}{N^{D-2}} N^{1-|\cC|} +\frac{1}{N^{D-2}} N^{ -D + |\cC|-1} \Bigr)    K \Big(\cB_{4,\cC} ; N,z,\{t_i\}  \Big) \crcr
&  +\sum_{\cC \in \{1,\dotsc,D\} , c\notin \cC, |\cC| \le D/2 } \Bigl( \frac{1}{N^{D-2}} N^{-D + 1 + |\cC|} + \frac{1}{N^{D-2}} N^{-1-|\cC|} \Bigr) K \Big(\cB_{4,\cC } ; N,z,\{t_i\}  \Big) \;.
\end{align}

\subsection{The expectations of melonic bubbles}

In the rest of this appendix we prove that the expectation of any melonic bubble has an expansion as in eq. \eqref{BiExpNLO}.
The melonic bubbles $\cB$ are obtained by iterated $(D-1)$-dipole insertions. The two vertices of a $(D-1)$-dipole
inserted at some step form a \emph{canonical pair}.
We call a $(D-1)$-dipole inserted at some step in this procedure an \emph{elementary dipole} if no other $(D-1)$-dipole is inserted
on any of its edges.

{\bf Gaussian contributions:}

We first examine the Gaussian contributions to the expectation of $\tr_{\cB}(\bT, \overline{\bT})$, that is the partitions
$\pi$ such that $p_\alpha=1$ for all $\alpha=1,\dotsc,A$. It follows that all the cumulants in the expansion \eqref{eq:expcumul} are 2-point cumulants.
Thus the doubled graphs $\cG$ corresponding to the Gaussian contributions have a subgraph $\cB$, and all their subgraphs $\cB'_\alpha = \cB_2$.
The Gaussian contributions split further into two classes:
\begin{itemize}
 \item either all the parts $\cV_\alpha =\{  v^{(\alpha)}_1, \bar v^{(\alpha)}_1 \} $ are comprised of canonical pairs of vertices.
  In this case the contribution of $\pi$ is:
 \begin{align}
&  \frac{1}{N}  N^{D+ (D-1)(p(\cB)-1)}  \left( \frac{   K \Big(\cB_2; N,z,\{t_i\} \Big)  }{N^{D-1} }\right)^{p(\cB)} = \Big[    K \Big(\cB_2; N,z,\{t_i\} \Big)   \Big]^{p(\cB)} \crcr
&  =  \left[  T(z,\{t_i\}) \right]^{p(\cB)} +
\frac{1}{N^{D-2}} \; p(\cB) \; \left[  T(z,\{t_i\}) \right]^{p(\cB)-1}  K^{\NLO} \Big(\cB_2; N,z,\{t_i\} \Big)   +  \mathcal{O}\left(\frac{1}{N^{D-1}} \right) \; .
\end{align}
\item or some of the sets $\cV_\alpha$ are comprised of two vertices which do \emph{not} form a canonical pair.

   Consider a doubled graph $\cG$ consisting in the invariant $\cB$ and only 2-point invariants $\cB'_\alpha=\cB_2$.
   Let $(x,\bar x)$ be a canonical pair of vertices belonging to an elementary dipole and such that $\cV_\alpha=\{x,\bar x\}$ for some $\alpha$.
   Consider the graph $\cG^{(1)}$ obtained by deleting the vertices $x,\bar x$ and the $D-1$ edges which connect them,
   and reconnecting the remaining two half--edges. As $\cG^{(1)}$ has one fewer 2-point invariant and $D-1$ fewer
   faces that $\cG$, it has the same scaling with $N$.

   Eliminating iteratively canonical pairs of vertices associated to elementary dipoles, we will obtain a doubled graph $\cG^{(s)}$
   having a canonical pair of vertices $(v,\bar v)$ associated to an elementary dipole which do \emph{not} form a part $\cV_\alpha$ for any $\alpha$. Instead we have $\cV_1 = \{ v, \bar x\}$ and $\cV_2 = \{ y ,\bar v \}$ for some vertices $\bar x, y$.
   From eq. \eqref{eq:expcumul}, it is trivial to see that the scaling in $N$ of $\cG^{(s)}$ is suppressed
   by $1/N^{D-2}$ with respect to the scaling of the graph $\tilde \cG^{(s)}$ corresponding to a partition
   in which all other parts $\cV_\beta$ are unchanged, but $\cV_1 = \{v , \bar v \}$ and $\cV_2 =\{ y, \bar x \}$.
   It follows that $\cG$ contributes at most to the order $\frac{1}{N^{D-2}}$. Denoting $k(\cB)$ the number of such doubled graphs
   which actually do contribute at this order we obtain a total contribution:
    \begin{equation}
       \frac{1}{N^{D-2}} k(\cB)  \left[  T(z,\{t_i\}) \right]^{p(\cB)} +   \mathcal{O}\left(\frac{1}{N^{D-1}} \right)
    \end{equation}
\end{itemize}

We have so far obtained the first three terms in eq. \eqref{BiExpNLO} (the leading order and the first two sub--leading corrections).
The fourth term requires more work.

Before proceeding, let us discuss the case when, in a Gaussian pairing $\cG$, at least three parts $\cV_1$, $\cV_2$ and $\cV_3$
are such their two vertices do not form canonical pairs. The elimination of canonical pairs of vertices forming elementary
dipoles which builds the sequence of graphs $(\cG^{(s)})$ can not eliminate any one of the three sets. Furthermore, the passage
from $\cG^{(s)}$ to $\tilde \cG^{(s)}$ concerns only two sets $\cV_\alpha$. It follows that $\tilde \cG^{(s)}$ has at least a set
$\cV_\beta$ (one of the three sets $\cV_1,\cV_2$ or $\cV_3$) whose two vertices do not form a canonical pair, hence the scaling of
$\cG$ is at most $\frac{1}{N^{2(D-2)}}$.

{\bf Non--Gaussian contributions:}

We will show that a generic non--Gaussian contribution is strictly bounded by contributions having exactly
one 4-point cumulant and $(p(\cB)-2)$ 2-point cumulants. We subsequently classify such contributions and identify those which contribute to the
order $\frac{1}{N^{D-2}}$.

Consider the contribution of a doubled graph $\cG$ having a subgraph $\cB'$ chosen among the bubbles $(\cB'_\alpha)_{\alpha=1,\dotsc,A}$ (i.e. coming from some $ K \Big(\cB'_\alpha ; N,z,\{t_i\}  \Big) $), with more than one connected component, $\rho(\cB')> 1$. The graph $\cB'$ brings a scaling:
\begin{equation}
 N^{D - \rho(\cB') - 2(D-1)p(\cB')  } \; .
\end{equation}
The same doubled graph $\cG$ is obtained if the connected components of $\cB'$ come from distinct parts of the partition $\pi$ (i.e. each one from a separate $K$). In this case, the contribution
to the scaling with $N$ is:
\begin{equation}
 N^{ (D -1)\rho(\cB') -2 (D-1) p(\cB') } \; ,
\end{equation}
which is larger by at least a factor $N^D$.

It follows that the corrections at order $\frac{1}{N^{D-2}}$ can only emerge from terms which associate connected bubbles $\cB'_\alpha$ to each part of $\pi$.
In this case, let us denote $B'(\cG)=A$ the number of parts, or equivalently the number of connected subgraphs with colors $1,\dotsc,D$ in $\cG$ minus one (for $\cB$ itself). We also set $F^{0c}(\cG)$ to be the number of faces of colors $0c$ of $\cG$ and $F(\cG)=\sum_{c=1}^D F^{0c}(\cG)$.
The scaling with $N$ of $\cG$ is:
\begin{equation}
\frac{1}{N} N^{F(\cG) + (D-1) B'(\cG)  - 2 (D-1)p(\cB) } \;.
\end{equation}

If $\cG$ is a non--Gaussian contribution, then there exists a subgraph $\cB'_\alpha$, for some $\alpha$, which has more than two vertices.
As long as there exist either two subgraphs $\cB'_\alpha,\cB'_\beta$ both having at least four vertices, or one $\cB'_\alpha$ having at least six vertices,
we proceed as follows. We consider $y$ and $\bar y$ two vertices of $\cB'_\alpha$, connected by at least an edge, say of color
$c_1$. We compare the scaling of $\cG$ with the scaling of the graph $\tilde \cG$, in which the vertices $y$ and $\bar y$ have been separated into
a new, 2-point, connected component $\cB''=\cB_2$, and the rest of the edges incident to $y$ and $\bar y$
in $\cB'_\alpha$ are reconnected respecting the colors. Since the adjacency relations between the edges of colors 0 and $c_1$ and the vertices has not changed, the number of faces of colors $0c_1$ is not affected. As for faces with colors $0c$ for $c\neq c_1$, their number can not decrease by more than 1, so $F(\tilde \cG)+(D-1) \geq F(\cG)$. Clearly, the number of subgraphs increases by at least $1$, $B'(\tilde \cG)\geq B'(\cG)+1$. Thus
\begin{equation}
\frac{1}{N} N^{ F(\cG) + (D-1) B'(\cG)  - 2 (D-1)p(\cB) }
\le \frac{1}{N} N^{F(\tilde\cG) + (D-1) + (D-1) ( B'(\tilde\cG) -1 ) - 2 (D-1)p(\cB) }
= \frac{1}{N} N^{F(\tilde\cG)  +  (D-1)  B'(\tilde\cG) - 2 (D-1)p(\cB)   } \; .
\end{equation}

It follows that all non-Gaussian contributions are:
\begin{itemize}
 \item either bounded by contributions with:
     \begin{itemize}
      \item exactly one 6-point cumulant and $(p(\cB)-3)$ 2-point cumulants.
      \item exactly two 4-point cumulants and $(p(\cB)-4)$ 2-point cumulants.
     \end{itemize}
  \item or formed of exactly one 4-point cumulant and $(p(\cB)-2)$ 2-point cumulants.
\end{itemize}

The graphs with one 6-point cumulant or two 4-point cumulants scale at most as $1/N^{2(D-2)}$,
because in each case, by the same construction, one can bound them by graphs $\cG$ representing
Gaussian contributions with at least three sets $\cV_1$, $\cV_2$ and $\cV_3$ whose vertices do not form
canonical pairs (remember that the graphs coming from the cumulants must be connected, otherwise $\cG$ is already
suppressed by at least $N^{-D}$).

Finally, let us discuss the graphs $\cG$ with exactly one 4-point cumulant and $(p(\cB)-2)$ 2-point cumulants. We repeat the above argument, but in the specific case that the 4-point cumulant has the boundary graph $\cB_{4,\cC}$. We choose $y$ and $\bar y$ in it and build the graph $\tilde \cG$ which is a Gaussian contribution, such that the images of $y $ and $\bar y$ in $\cB$ do not belong in a canonical pair. Therefore the amplitude of $\tilde \cG$ is bounded by $N^{-(D-2)}$. In $\cB_{4,\cC}\subset \cG$, $y$ and $\bar y$ are connected by $l=|\cC|$ or $l= D-|\cC|$ edges. From $\cG$ to $\tilde \cG$, the number of faces with the corresponding colors is unchanged, while the number of faces with the other colors can change by at most one. Therefore $F(\tilde \cG) + (D-l)\geq F(\cG)$. The number of boundary components changes by one. This way we get the bound:
\begin{equation}
F(\cG)+(D-1)B'(\cG) \leq F(\tilde \cG) + (D-l) + (D-1)(B'(\tilde \cG)-1) \leq F(\tilde \cG) + (D-1)B'(\tilde \cG) + \bigl(1-l\bigr).
\end{equation}
In the case $|\cC|>1$, we find that $1-l<0$, implying that $\cG$ is at least suppressed by a factor $1/N$ with respect to $\tilde \cG$.

It follows that at order $\frac{1}{N^{D-2}}$ only graphs with one 4-point cumulant, and with corresponding graph $\cB'=\cB_{4,\{c\}}$
for some $c$ can contribute, and it is simple to check that they do. As, moreover, such contributions are at most at order $N^{-(D-2)}$ with respect to a Gaussian pairing, any contribution at the exact order $N^{-(D-2)}$ must be proportional to
\begin{align}
\left[  T(z,\{t_i\}) \right]^{p(\cB)-2}  K \Big(\cB_{4,\{c\}}; z , \{t_i\} \Big) \; ,
\end{align}
which yields the last term in the equation \eqref{BiExpNLO}.



\begin{thebibliography}{99}


\bibitem{review-vincent}
  V. Rivasseau,
  ``The tensor track III,''
  Fortschr. Phys. {\bf 62}, No. 1, 1-27 (2013);
  arXiv:1311.1461.

\bibitem{colored}
  R.~Gurau,
  ``Colored Group Field Theory,''
  Commun.\ Math.\ Phys.\  {\bf 304} (2011) 69
  [arXiv:0907.2582 [hep-th]].

\bibitem{largeN}
   R.~Gurau,
 ``The 1/N expansion of colored tensor models,''
  Annales Henri Poincar\'e {\bf 12} (2011) 829
  [arXiv:1011.2726 [gr-qc]].

\bibitem{largeN1}
     R.~Gurau and V.~Rivasseau,
``The 1/N expansion of colored tensor models in arbitrary dimension,''
  Europhys.\ Lett.\  {\bf 95}, 50004 (2011)
  [arXiv:1101.4182 [gr-qc]].

\bibitem{largeN2}
R.~Gurau,
 ``The complete 1/N expansion of colored tensor models in arbitrary dimension,''
  Annales Henri Poincar\'e {\bf 13}, 399 (2012)
  [arXiv:1102.5759 [gr-qc]].

\bibitem{largeN3}
  V.~Bonzom,
  ``New 1/N expansions in random tensor models,''
  JHEP {\bf 1306}, 062 (2013)
  [arXiv:1211.1657 [hep-th]].

\bibitem{Uncoloring}
  V.~Bonzom, R.~Gurau and V.~Rivasseau,
  ``Random tensor models in the large N limit: Uncoloring the colored tensor models,''
  Phys.\ Rev.\ D {\bf 85}, 084037 (2012)
  [arXiv:1202.3637 [hep-th]].

\bibitem{Heegaard}
  J.~P.~Ryan,
  ``Tensor models and embedded Riemann surfaces,''
  Phys.\ Rev.\ D {\bf 85}, 024010 (2012)
  [arXiv:1104.5471 [gr-qc]].

\bibitem{GR}
R.~Gurau and J.~P.~Ryan,
``Colored Tensor Models - a review,''
  SIGMA {\bf 8}, 020 (2012)
  [arXiv:1109.4812 [hep-th]].

\bibitem{GR2}
R.~Gurau and J.~P.~Ryan,
``Melons are branched polymers,''
  arXiv:1302.4386 [math-ph].

\bibitem{ColoredNLO}
  W.~Kaminski, D.~Oriti and J.~P.~Ryan,
  ``Towards a double-scaling limit for tensor models: probing sub-dominant orders,''
  arXiv:1304.6934 [hep-th].

\bibitem{ColoredSchemes}
  R.~Gurau and G.~Schaeffer,
  ``Regular colored graphs of positive degree,''
  [arXiv:1307.5279 [math.CO]].

\bibitem{DSQuartic}
  S.~Dartois, R.~Gurau and V.~Rivasseau,
  ``Double Scaling in Tensor Models with a Quartic Interaction,''
  JHEP {\bf 1309}, 088 (2013)
  [arXiv:1307.5281 [hep-th]].

\bibitem{Fukuma}
M.~Fukuma, H.~Kawai and R.~Nakayama,
  ``Continuum Schwinger-dyson Equations and Universal Structures in Two-dimensional Quantum Gravity,''
  Int.\ J.\ Mod.\ Phys.\ A {\bf 6} (1991) 1385.

\bibitem{Dijkgraaf}
R.~Dijkgraaf, H.~L.~Verlinde and E.~P.~Verlinde,
  Nucl.\ Phys.\ B {\bf 348} (1991) 435.

\bibitem{Bouwknegt}
P.~Bouwknegt and K.~Schoutens,
  ``W symmetry in conformal field theory,''
  Phys.\ Rept.\  {\bf 223} (1993) 183
  [hep-th/9210010].

\bibitem{Dijkgraaf2}
R.~Dijkgraaf and C.~Vafa,
  Nucl.\ Phys.\ B {\bf 644} (2002) 21
  [hep-th/0207106].
  
\bibitem{Aganagic}
M.~Aganagic, R.~Dijkgraaf, A.~Klemm, M.~Marino and C.~Vafa,
  ``Topological strings and integrable hierarchies,''
  Commun.\ Math.\ Phys.\  {\bf 261} (2006) 451
  [hep-th/0312085].

\bibitem{eynard}
  B. Eynard,
  ``Topological expansion for the 1-Hermitian matrix model correlation functions,''
  JHEP 0411, 031 (2004)
  [hep-th/0407261].

\bibitem{TreeAlgebra}
  R.~Gurau,
  ``A generalization of the Virasoro algebra to arbitrary dimensions,''
  Nucl.\ Phys.\ B {\bf 852}, 592 (2011)
  [arXiv:1105.6072 [hep-th]].

\bibitem{BubbleAlgebra}
  R.~Gurau,
  ``The Schwinger Dyson equations and the algebra of constraints of random tensor models at all orders,''
  Nucl.\ Phys.\ B {\bf 865}, 133 (2012)
  [arXiv:1203.4965 [hep-th]].

\bibitem{SDE}
  V.~Bonzom,
  ``Revisiting random tensor models at large N via the Schwinger-Dyson equations,''
  JHEP {\bf 1303} (2013) 160,
  arXiv:1208.6216 [hep-th].

\bibitem{Universality}
  R.~Gurau,
  ``Universality for Random Tensors,''
  arXiv:1111.0519 [math.PR].

\bibitem{beyondpert}
  R.~Gurau,
  ``The $1/N$ Expansion of Tensor Models Beyond Perturbation Theory,''
  arXiv:1304.2666 [math-ph].

\bibitem{MO}
  A.~Tanasa,
  ``Multi-orientable Group Field Theory,''
  J.\ Phys.\ A {\bf 45} (2012) 165401
  [arXiv:1109.0694 [math.CO]].

\bibitem{DRT}
  S.~Dartois, V.~Rivasseau and A.~Tanasa,
  ``The 1/N expansion of multi-orientable random tensor models,''
  Annales Henri Poincar\'e {\bf 15}, 965 (2014).
  arXiv:1301.1535 [hep-th].

\bibitem{RT}
  M.~Raasakka and A.~Tanasa,
  ``Next-to-leading order in the large N expansion of the multi-orientable random tensor model,''
  arXiv:1310.3132 [hep-th]. Annales Henri Poincar\'e (in press).

\bibitem{BGR}
  J.~Ben Geloun and V.~Rivasseau,
  ``A Renormalizable 4-Dimensional Tensor Field Theory,''
  Commun.\ Math.\ Phys.\  {\bf 318}, 69 (2013)
  [arXiv:1111.4997 [hep-th]].

\bibitem{BGR1}
  J.~Ben Geloun and V.~Rivasseau,
  ``Addendum to 'A Renormalizable 4-Dimensional Tensor Field Theory',''
  Commun.\ Math.\ Phys.\  {\bf 322}, 957 (2013)
  [arXiv:1209.4606 [hep-th]].

\bibitem{ren}
  J.~Ben Geloun,
  ``Renormalizable Models in Rank $d\geq 2$ Tensorial Group Field Theory,''
  arXiv:1306.1201 [hep-th].
\bibitem{ren1}
  J.~Ben Geloun,
  ``Asymptotic Freedom of Rank 4 Tensor Group Field Theory,''
  arXiv:1210.5490 [hep-th].
\bibitem{ren2}
  J.~Ben Geloun and E.~R.~Livine,
  ``Some classes of renormalizable tensor models,''
  J.\ Math.\ Phys.\  {\bf 54} (2013) 082303
  [arXiv:1207.0416 [hep-th]].

\bibitem{COR}
  D.~O.~Samary and F.~Vignes-Tourneret,
  ``Just Renormalizable TGFT's on $U(1)^{d}$ with Gauge Invariance,''
  Communications in Mathematical Physics (2014)
  [arXiv:1211.2618 [hep-th]].
\bibitem{COR1}
  S.~Carrozza, D.~Oriti and V.~Rivasseau,
  ``Renormalization of Tensorial Group Field Theories: Abelian U(1) Models in Four Dimensions,''
  arXiv:1207.6734 [hep-th].
\bibitem{COR2}
  S.~Carrozza, D.~Oriti and V.~Rivasseau,
  ``Renormalization of an SU(2) Tensorial Group Field Theory in Three Dimensions,''
  arXiv:1303.6772 [hep-th].

\bibitem{AIHPD}
 J.~B.~Geloun and S.~Ramgoolam,
 ``Counting Tensor Model Observables and Branched Covers of the 2-Sphere,''
  arXiv:1307.6490 [hep-th].
  Annales de l'Institut Henri Poincar\'e {\bf D} - Combinatorics, Physics and their Interactions 1 (2014) 77-138.

\bibitem{GW}
  H.~Grosse and R.~Wulkenhaar,
  ``Construction of the $\Phi^4_4$-quantum field theory on noncommutative Moyal space,''
  arXiv:1402.1041 [math-ph].

\end{thebibliography}
\end{document}